\documentstyle[11pt]{article}

\newcommand{\mrm}[1]{\rm{#1}}

\newcommand{\ttt}[1]{\tt{#1}}
 
\renewcommand{\a}{\rm{a}}
\renewcommand{\b}{\rm{b}}
\renewcommand{\c}{\rm{c}}
\renewcommand{\d}{\rm{d}}
\newcommand{\e}{\rm{e}}
\newcommand{\f}{\rm{f}}
\newcommand{\g}{\rm{g}}
\newcommand{\hrm}{\rm{h}}

\newcommand{\n}{\rm{n}}
\newcommand{\p}{\rm{p}}

\newcommand{\s}{\rm{s}}
\renewcommand{\t}{\rm{t}}
\renewcommand{\u}{\rm{u}}
\newcommand{\A}{\rm{A}}
\newcommand{\B}{\rm{B}}
\newcommand{\D}{\rm{D}}

\renewcommand{\H}{\rm{H}}

\newcommand{\K}{\rm{K}}

\newcommand{\W}{\rm{W}}
\newcommand{\Z}{\rm{Z}}

\newcommand{\Jpsi}{\rm{J}/\psi}

{\end{list}}
\newcounter{enumct}

{\begin{list}{}{\setlength{\topsep}{0mm} \setlength{\itemsep}{0mm}
\setlength{\parskip}{0mm} \setlength{\parsep}{0mm}
\setlength{\leftmargin}{20mm} \setlength{\rightmargin}{0mm}
\setlength{\labelwidth}{18mm} \setlength{\labelsep}{2mm}}}%
{\end{list}}
{\begin{list}{}{\setlength{\topsep}{0mm} \setlength{\itemsep}{0mm}
\setlength{\parskip}{0mm} \setlength{\parsep}{0mm}
\setlength{\leftmargin}{10mm} \setlength{\rightmargin}{0mm}
\setlength{\labelwidth}{18mm} \setlength{\labelsep}{2mm}}}%
{\end{list}}

\newlength{\captivewidth}
\newcommand{\captive}[1]{\rule{5mm}{0mm}%
\begin{minipage}{\captivewidth}%
\caption[small]{#1}\end{minipage}}

\newlength{\tablinsep}
\newlength{\halfpagewid}
\def\lline{--------------------------------------------------------------------------------------------------------------------------------------------}
\def\sline{------------------------------------------------------------------------------------------------------------------------------------}
 
\input{epsf}

\begin{document}

\sloppy
\setcounter{page}{0}
\thispagestyle{empty}

\begin{flushright}
BNL-63632\\
hep-ph/9701226
\end{flushright}
\vspace{3.5cm}

\begin{center}
{\Huge {\bf VNI 3.1}}
\bigskip

{\LARGE
{\bf MC-simulation program to study
high-energy particle collisions in QCD
by space-time evolution of parton-cascades 
and parton-hadron conversion}
}

\end{center}
\bigskip

\begin{center}

{\Large
{\bf  Klaus Geiger}
}
\medskip

{\it Physics Department,
Brookhaven National Laboratory, Upton, N.Y. 11973, U.S.A.}
\end{center}
\smallskip

\begin{center}
e-mail: klaus@bnl.gov    \\
http://penguin.phy.bnl.gov/ $\widetilde{  }$ klaus    \\
phone:  (516) 344-3791    \\
fax:    (516) 344-2918    \\
\end{center}
\vspace{0.5cm}

\begin{center}
{\large {\bf Abstract}}
\end{center}
\bigskip

\noindent
$VNI$ is a general-purpose Monte-Carlo event-generator, which includes the
simulation of lepton-lepton, lepton-hadron, lepton-nucleus, 
hadron-hadron, hadron-nucleus, and nucleus-nucleus collisions.
On the basis of renormalization-group improved
parton description and quantum-kinetic theory, it uses
the real-time evolution of parton cascades in conjunction with
a self-consistent hadronization scheme that is governed 
by the dynamics itself.
The causal evolution from a specific initial state (determined by
the colliding beam particles) is followed by the time-development
of the phase-space densities of partons, pre-hadronic parton clusters,
and final-state hadrons, in 
position-space, momentum-space and color-space.
The parton-evolution is described in terms of a
space-time generalization of the familiar momentum-space
description of multipl (semi) hard interactions in QCD, involving
$2 \rightarrow 2$ parton collisions, $2\rightarrow 1$ parton fusion
processes, and $1\rightarrow 2$ radiation processes.
The
formation of color-singlet pre-hadronic clusters and their decays into
hadrons, on the other hand,
is treated by using a spatial criterion motivated by confinement and a
non-perturbative model for hadronization. 
This article gives a brief review of the physics underlying $VNI$, 
which is followed by a detailed description of the program itself.
The latter program description emphasizes easy-to-use
pragmatism and explains how to use the program (including
a simple example), annotates input and control parameters, and 
discusses output data provided by it.
\bigskip
\bigskip

\noindent
{\it Program obtainable from:} 
http://rhic.phys.columbia.edu/rhic/vni

\newpage

\tableofcontents

\newpage

\section{PROGRAM SUMMARY}
\label{sec:section1}
\bigskip

\noindent
{\bf Title of program:} VNI-3.1
\bigskip

\noindent
{\bf Computer fo which the program has been designed and
others on which it has been tested:}
IBM RS-6000, Sun Sparc, Hewlett Packard UX A-9000
\bigskip

\noindent
{\bf Operating systems:}
IBM-AIX, Sun-OS, and any other UNIX operating systems, as well as LINUX.
\bigskip

\noindent
{\bf Programming language used:}
Fortran 77
\bigskip

\noindent
{\bf Memory required to execute with typical data:} 2000 kwords
\bigskip

\noindent
{\bf No. of bits in a word:} 32
\bigskip

\noindent
{\bf No. of lines in distributed program:} 
25760 lines of main program, plus 49 and 244 lines for two example programs.
\bigskip

\noindent
{\bf Keywords:} 
Monte Carlo simulation, event generator, 
QCD kinetic theory, parton cascades, parton coalescence,
hadronic final states.
\bigskip

\noindent
{\bf Nature of physical problem:}

\noindent
In high-energy particle collisions certain phase-space regions
can be populated by a large number of quanta, such that
statistical correlations among them
(e.g., in space, momentum, or color) become of essential importance.
Examples are deep-inelastic lepton-hadron scattering and 
hadron-hadron collisions in the region of very small Bjorken-$x$, or,
collisions involving heavy nuclei in the central rapidity region.
In these cases the produced particles evolve in a 
complicated non-equilibrium environment created by the presence
of neighboring ones. The `deterministic' quantum evolution
of particle states due to self-interactions (depending only on the
particle itself), receives a new 
`statistical' kinetic contribution due to mutual interactions
(depending crucially on the local density).
The theoretical basis for addressing the solution for 
the dynamics of such particle systems is a quantum-kinetic
formulation of the QCD equations of motion,
an approximation that combines field-theoretical aspects
associated with the renormalization group (including
well-known resummation techniques) with aspects of
transport theory associated with non-equilibrium multi-particle
dynamics (including important quantum effects beyond the classical level).

\bigskip

\noindent
{\bf Method of solution:}

\noindent
The solution of the underlying quantum-kinetic equations
of motion for non-equilibrium multi-particle QCD 
by Monte-Carlo simulation of collisions allowing for a variety of 
combinations of beam and target particles.
To simulate the real-time evolution of the collision system in
position space and momentum space on the basis of the equations
of motions, the procedure is three-fold:
i) the construction of the initial state including the
decomposition of the beam and target particles
into their partonic substructure, (ii) the evolution
of parton cascades including multiple scatterings, emission- and
fusion-processes, and (iii)  the self-generated
conversion of partons into hadrons using a phenomenological
model for parton-coalescence into pre-hadronic clusters and
subsequent decay into final-state particles.
\bigskip

\noindent
{\bf Restriction on the complexity of the problem:}

\noindent
For very high collision energy ($\sqrt{s} \gg 10$ TeV in
hadronic collisions,  and
$\sqrt{s} \gg 5$ TeV/nucleon in nuclear collisions)
numerical inaccuracies due to repeated Lorentz boosts, etc.,
may accumulate to cause problems.
Although the most concerned parts of the program use double precision,
for extreme energies the code would require a conversion in full 
to double precision format (which is planned in the near future).
\bigskip

\noindent
{\bf Typical running time:}

\noindent
The CPU time for a typical simulation is strongly dependent on the type of
beam and target, the magnitude of collision energy, as well
as on the time interval $\Delta t$ chosen to follow an event in its real-time
evolution.
Examples  are (for $\Delta t = 35$ $fm$):
a) $e^++e^-$ at $\sqrt{s} = 100$ GeV:   10000 events/hour;
b) $p+\bar{p}$ at $\sqrt{s} = 200$ GeV:   5000 events/hour;
c) $p+ Au$ at $\sqrt{s} = 200$ GeV/nucleon:   100 events/hour;
d) $Au+Au$ at $\sqrt{s} = 200$ GeV/nucleon:   1 event/hour;
All of the above quotes are approximate, and 
refer to a typical
133 Mhz or 166 Mhz processor on a modern
Power-Workstation or Power-PC.
\bigskip

\newpage

\section{LONG WRITE-UP}
\label{sec:section2}
\bigskip

\subsection{Introduction}

VNI 
\footnote{
       The three letters VNI do not mean anything profound. VNI is
       pronounced "Vinnie", short for "Vincent Le CuCurullo Con GiGinello", 
       the little guy who likes to hang out with his pals, the quarks 
       ({\it cucu}rullos) and gluons ({\it gigi}nellos). 
       That is QCD in Wonderland, and that is the whole, true story.
}
is the Monte Carlo implementation 
of a relativistic quantum-kinetic approach \cite{ms39,ms42} to
the dynamics of high-energy particle collisions, inspired by the 
QCD parton picture of hadronic interactions \cite{kogut73,dok80,glr83}.
It is a product of several years of development in both
the improving physics understanding of high-energy multiparticle 
dynamics in QCD, 
as well as the technical implementation in the form of a computer simulation
program. The most relevant references
for the following are Refs. \cite{ms0,ms3,msrep,ms37,ms40,ms41}, 
where details of the
main issues, discussed below, can be found. 
The puropse of VNI is to provide  a comprehensive
description of particle collisions involving beams of
leptons, hadrons, or nuclei,
in terms of the space-time evolution of parton cascades and
parton-hadron conversion.
The program VNI is concepted as a useful {\it tool} (and nothing more) to study
the causal development of the collision dynamics in real time from 
a specified initial state of beam and target particles, all the way to the
final yield of hadrons and other observable particles.
The collision dynamics is traced in detail on the microscopic level 
of quark and gluon interactions in the framework of perturbative QCD,
supplemented by a phenomenological treatment of the non-perturbative dynamics,
including soft parton-collisions and parton-hadronization.  
The generic structure of the simulation concept is illustrated in Fig. 1.
\medskip

\begin{figure}
\epsfxsize=450pt
\centerline{ \epsfbox{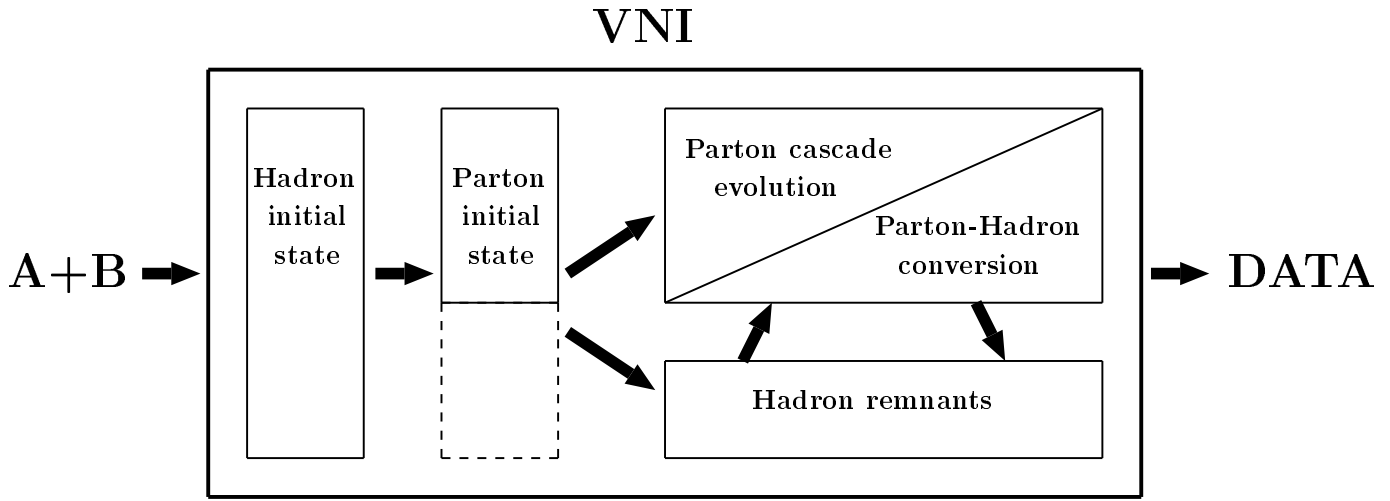} }
\caption{
Generic flow-chart of the simulation concept of VNI:
It starts with
the initial beam target particles $A$ and $B$, decomposing them
(except for leptonic $A$ and/or $B$) 
in their hadronic constituents with partonic substructure, then proceeds through
the parallely evolving stages of parton-cascade, parton-hadron conversion,
and fragmentation of beam/target remnants, and finishes up with
the final particle yield that reflects observables in a detector.
\label{fig:fig1}
}
\end{figure}

The main strength of VNI lies in addressing
the physics of high-density QCD, which  becomes an increasingly popular
object of research, both from the experimental, phenomenological
interest, and from the theoretical, fundamental point of view.
Presently, and in the near future, the collider facilities HERA
($ep$, $eA$?), Tevatron ($p\bar{p}$, $pA$), RHIC and LHC ($p\bar{p}$,
$AA$) are able to probe new regimes of dense quark-gluon matter 
at very small Bjorken-$x$ or/and at large $A$, 
with rather different dynamical properties.
The common feature of high-density
QCD matter that can be produced in these experiments, is an expected
novel exhibition
of the interplay between the high-momentum (short-distance) perturbative
regime and the low-momentum (long-wavelength) non-perturbative physics.
For example, with HERA and Tevatron experiments,
one hopes to gain insight into problems concerning
the saturation of the strong rise
of the proton structure functions at small Bjorken-$x$,
possibly due to color-screening effects that
are associated with the overlappping of a large number of small-$x$
partons.
Another example is the anticipated formation of a quark-gluon plasma in
RHIC and LHC heavy ion collisions, 
where multiple parton rescattering and cascading generates a 
high-density environment, in which the collective 
motion of the quanta imust be taken into consideration.
In this context the most advantageous and novel feature
of VNI
is the space-time cascade picture that provides a potentially powerful
tool to study high-density particle systems in QCD, where 
accounting for the dynamically
evolving space-time structure of the collisions is most important.
The {\it necessity of including space-time variables} in addition to
energy-momentum variables for high-density systems may
be common knowledge in the field of heavy-ion physics, where
the space-time aspect  of many-body transport theory is essential
ingredient to describe nucleus-nucleus collisions, but it is
new to many high-energy particle physicists which only recently
began to acknowleged the necessity to include time and space on top
of the commonly used, pure momentum space description of, e.g.,  lepton and
hadron collisions.
\bigskip

\noindent
{\it
Where does VNI fit into the diverse family of modern
event generators for physics simulation of particle collisions
(where `particle' stands for any beam/target particle from an electron 
to a heavy nucleus)?
}
\smallskip

\noindent
The Monte Carlo models for high-energy particle collisions that are on the 
market, may be crudely divided into two classes:
\begin{description}
\item{(i)}
The first class embodies event generators that restrict to
"clean" reactions  involving lepton or proton beams only,
and which aim at a high-precision description of experimental
tests of QCD's first principles.
Popular examples  are JETSET/PYTHIA \cite{jetset},
HERWIG \cite {herwig}, ARIADNE \cite{ariadne}, LEPTO \cite{lepto},
and ISAJET \cite{isajet}.
The common feature of these event generators is a combination of 
well understood perturbative QCD parton-shower description and
a non-perturbative hadronization prescription to convert
the partonic final state into hadrons.
\item{(ii)}
The second class are event generators that aim to describe
"dirty" reactions involving nuclei, much less based on
first-principle knowledge, but instead rely on phenomenological
models to mimic the unknown details of the underlying physics.
Here the widely used concept is to visualize a nuclear
collision in terms of nucleon-nucleon collisions on the
basis of a constituent valence-quark picture plus
string-excitation and -fragmentation.
Examples for these models are
FRITIOF \cite{fritiof},  DPM \cite{dpm}, 
VENUS \cite{venus}, 
RQMD \cite{rqmd}, and HIJET \cite{hijet}.
Distinct from these is  HIJING \cite{hijing}, 
which also incorporates a perturbative QCD approach to multiple minijet
production, however it does not incorporate a space-time description.
\end{description}
With the exception of HERWIG, all of the
above Monte Carlo generators utilize some form of 
{\it string fragmentation phenomenology} to model 
the non-perturbative hadronization
and final-state particle production.
Most commonly used is the Lund string model \cite{string}.
HERWIG on the other hand is built on a very different 
{\it parton-cluster formation/fragmentation approach}, 
which forms the basis of the hadronization scheme developed in VNI.
\medskip

\noindent
{\it
What are the shortcomings of the above-mentioned Monte Carlo models
with respect to the particle dynamics in finite-density regions
created by high-energy collisions?
}
\smallskip

The high-energy particle physics generators of the first class
lack the inclusion space-time variables in the dynamics
description in both the
perturbative QCD parton evolution and the non-perturbative
process of parton-hadron transition.
These models therefore cannot account  for statistical
interactions due to the presence of a finite density of particles
close-by in space, such as rescattering, absorption of recombination
processes.
Hence, although these models to large extent use QCD's fundamental
quark-gluon  degrees of freedom, important aspects of
parton dynamics and interactions at finite density are left out,
because the particles are assumed to propagate unscathed in free space.

The event generators of the second class, on the other hand, 
mostly do utilize a space-time description, however on the level
of hadronic degrees of freedom (strings, baryons, mesons and resonances)
rather than partonic degrees of freedom.
For ultra-relativistic nucleus-nucleus collisions
the parton approach appears to be more realistic than the 
hadronic or string picture, as it
has been realized that short-range parton interactions play 
a  major role for heavy ion collisions at collider energies of
$\sqrt{s}  \, \lower3pt\hbox{$\buildrel >\over\sim$}\,100$ GeV,
at least during the early and most dissipative stage of the first few $fm$. 
Here copiously produced quark-gluon mini-jets cannot be considered as 
isolated rare events, but must be embedded in complicated multiple 
cascade-type processes.
Thus, the short range character of these  
interactions implies that perturbative QCD can and must be used, and 
that the picture of comparably large distance excitations
of strings or hadronic resonances does not apply in this kinematic regime.
\medskip

In view of this discussion,
the program VNI can be viewed in between the above two classes of
Monte Carlo models:
It provides a kinetic space-time description of parton evolution
by utilizing well-developed techniques for perturbative QCD simulations
at zero density or free space,
as the event generators of the first class. On the other hand,
it applies this concept also to the physics of
finite-density particle systems, e.g., in
collisions involving nuclei, as the event generators of the second class.
  Comparing VNI with the above-mentioned Monte Carlo models, 
the essential differences and partly new aspects are the following:
\begin{description}
\item{a)}
     the aspects of the space-time evolution of the particle distributions
     in addition to the evolution in momentum space \cite{msrep}, 
     and the concepts of quantum kinetic theory and statistical physics
     \cite{ms39,ms42}.
\item{b)}
     the self-consistent interplay of coherent (angular ordered) 
     perturbative 
     parton evolution according to the DGLAP \cite{DGLAP} equations, with the 
     fully dynamical cluster-hadronization according to the 
     phenomenological model of Ref. \cite{ms37}.
\item{c)}
     the microscopic tracing of color degrees of freedom and the effect of
     color-correlations by using explicit color-labels for each parton, 
     which allows to investigate final state interactions in the process 
     of hadron formation \cite{ms40}.
\item{d)}
     the diverse advantages of a stochastic simulation technique with
     which the various particle interaction processes are determined by 
     the dynamics itself, through the local density of particles as they 
     evolve causally in time.
\item{e)}
     the statistical many-particle description for general non-equilibrium
     systems, which allows to study thermodynamic behaviour of the bulk 
     matter \cite{ms2}, such as the evolution of macroscopic energy density, 
     pressure, etc., or the dynamical development of the sytem to 
     thermal/chemical equilibrium in heavy ion reactions.
\end{description}

In summary, the improvement to be expected from VNI
for the physics simulation of high-energy particle collisions 
lies clearly in the `dirty' high-density parton regime, 
where the space-time aspects are most important, and which
currently and in the future is
of central interest
in  experiments  at, e.g.,  HERA, RHIC and LHC.
On the other hand, VNI may also be perceived as a valuable alternative
to high-energy event generators for the
study of `clean', zero-density collisions as $e^+e^-$ annihilation
or $p\bar{p}$ collisions,
where the space-time aspects can provide useful additional insight in
the collision dynamics for experiments at, e.g., LEP or the Tevatron.
\bigskip
\bigskip

\subsection{General concept}

The central element in the physics description implemented in VNI
is the use of
QCD transport theory \cite{msrep} and quantum field kinetics \cite{ms39}
to follow the evolution of a generally mixed multi-particle system
of partons and hadrons
in 7-dimensional phase-space $d^3 r d^3 k dE$.
Included are 
both the parton-cascade development
\cite{dok80,glr83,jetcalc,bassetto}
which embodies the renormalization-group improved evolution of 
multiple parton collisions
including inelastic (radiative) processes,
and the phenomenological
parton-hadron conversion model of Refs.  \cite{ms37,ms40,ms41},
in which the hadronization mechanism is described in terms of 
dynamical parton-cluster formation as
a local, statistical process that depends on the spatial separation and
color
of nearest-neighbor partons, followed by the decay of clusters into
hadrons.

In contrast to the commonly-used momentum-space description,
the microscopic history of the dynamically-evolving particle system
is traced in space-time {\it and} momentum space, so that
the correlations of partons in space,  time, and color can be taken
into account for both the  cascade evolution
and the hadronization mechanism.
It is to be emphasized,
that the interplay
between perturbative and non-perturbative regimes is controlled locally
by the space-time evolution of the mixed parton-hadron system itself
(i.e., the time-dependent local parton density),
rather than by an arbitrary global division
between parton and hadron degrees of freedom 
(i.e., a parametric energy/momentum cut-off).
In particular the parallel evolution of the mixed system
of partons, pre-hadronic clusters, and hadrons, with the relative proportions
determined by the dynamics itself, is a novel feature that
is only possible by keeping track of both space-time and 
energy-momentum variables.

Probably the greatest strength of this approach lies in its
application to the collision dynamics of
complicated multi-particle systems, as for example in  collisions involving
nuclei ($eA$, $pA$ and $AB$), 
for which a causal time evolution in position space
and momentum space is essential: Here statistical,
non-deterministic particle interactions are most important,
which  can only be accounted for by
following the time-evolution of the particle densities in space {\it and}
energy-momentum.
This approach allows to 
study the time evolution of an initially prepared beam/target collision system
in complete phase-space from the instant 
of collisional  contact, through the QCD-evolution of
parton distributions, up to the formation of final hadronic states.
It provides a self-consistent scheme to solve the underlying
equations of motion for the particle densities as determined by the 
microscopic dynamics.
\smallskip

The model as a whole consists of
three major building-blocks, each of which is illustrated schematically
in Fig. 2:
\begin{description}
\item[1.]
The {\it initial state} associated with the incoming
collision partners (leptons, hadrons, or nuclei). Except
for lepton beams,  this involves
the phenomenological construction of hadrons or nuclei in terms of the
partons' phase-space distributions
on the basis of the experimentally measured  
hadron (nucleon) structure functions and
elastic form-factors.
\item[2.]
The {\it parton cascade development} 
with mutual- and self-interactions of the system of quarks and gluons.
This includes multiple inelastic processes, described 
as sequences of elementary $2 \rightarrow 2$ scatterings, $1\rightarrow 2$
emissions and $2 \rightarrow 1$ fusions.
Moreover,  correlations are accounted for between primary virtual
partons, emerging as unscathed remainders from the initial state, and
secondary real partons, materialized or  produced 
through the partonic interactions.
\item[3.]
The {\it hadronization dynamics} of the evolving system
in terms of parton-coalescence to color-neutral clusters
as a local, statistical process that depends on the spatial separation
and color of nearest-neighbor partons.
Each pre-hadronic parton-cluster fragments through isotropic two-body decay
into  primary hadrons, according to the density of
states, followed by the decay of the latter into final 
stable hadrons.
\end{description}

Such a pragmatical division, which assumes complex interference
between the different physics regimes to be negligible, is possible if
the respective dynamical scales are such that
the short-range (semi)hard parton interactions (scattering, radiation,
fusion) of perturbative nature,
and the non-perturbative
mechanism of hadron formation (parton-coalescence and cluster-decay),
occur on well-separated  space-time scales (or momentum scales, by
virtue of the uncertainty principle).
Loosely speaking, the typical momentum scale associated with
parton collisions, radiative emissions, or parton fusion, has to be 
larger than
the inverse `confinement length scale' $\sim1\,fm$ which
separates perturbative and non-perturbative domains.
Further discussion of this condition of validity can be found in, e.g.,
\cite{msrep,dokbook}.

\begin{figure}
\epsfxsize=500pt
\centerline{ \epsfbox{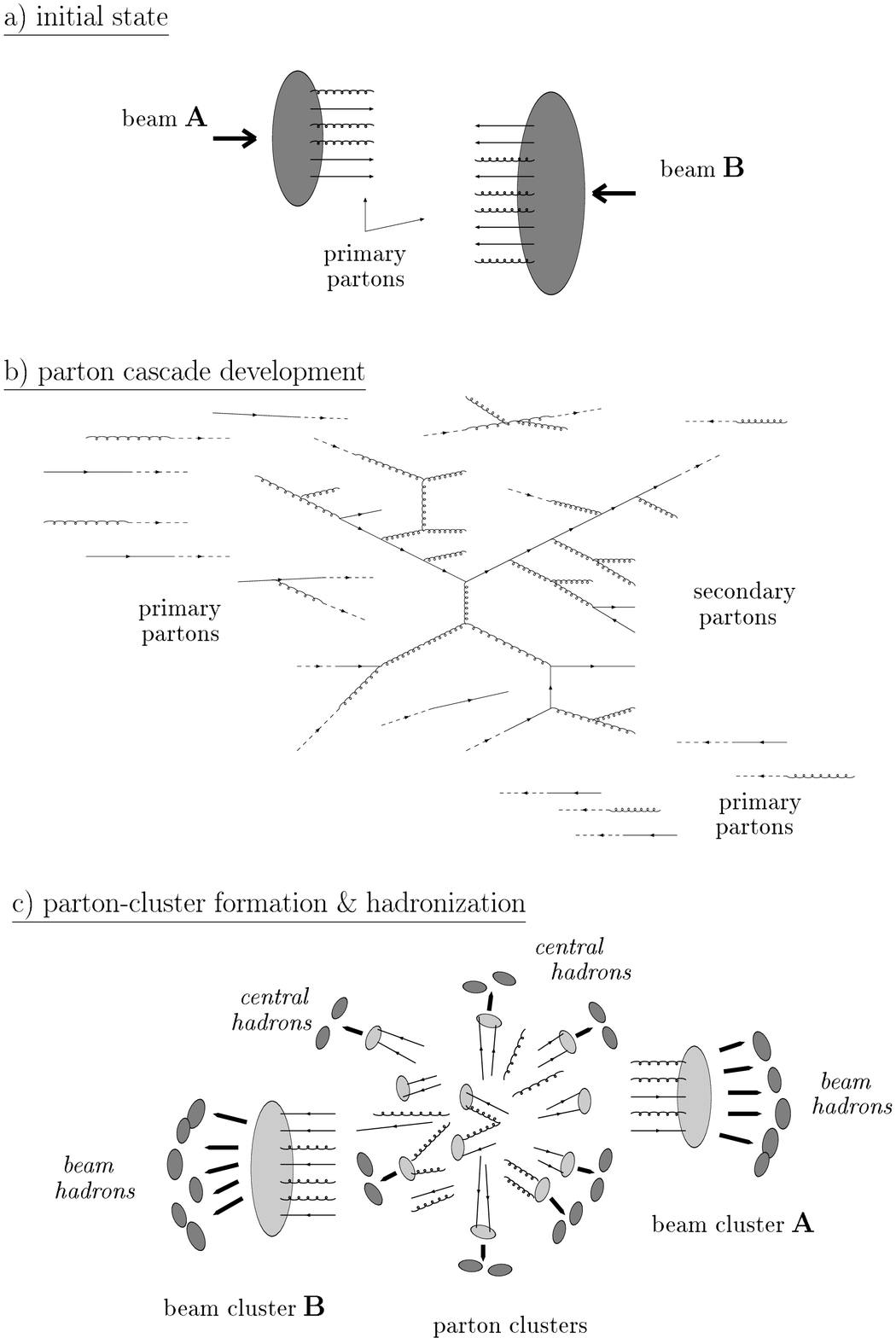} }
\caption{
The three components of the model:
a) the initial state constructed in terms of the parton distribution of
the incoming nuclei;
b) the time-evolution of parton cascades in 7-dimensional phase-space 
c) the formation of color neutral clusters from secondary partons
emerging from cascading, as well of the remnant primary partons from
the initial state, followed by
the fragmentation of the clusters into final hadrons.
\label{fig:fig2}
}
\end{figure}
\bigskip

\subsection{Equations of motion from quantum kinetics for multi-particle
dynamics}

A space-time description of multiparticle systems
in high-energy QCD processes can be derived systematically from
{\it quantum-kinetic theory} on the basis of
QCD's first principles in a stepwise approximation scheme 
(see  e.g., Refs. \cite{ms39,ms42} and references therein).
Applied to the concept sketched in the preceeding subsection,
this framework allows  to
cast the time evolution of the mixed system of
individual partons, composite parton-clusters, and physical hadrons
in terms of a closed set of
integro-differential equations for
the phase-space densities of the different particle excitations.
The definition of these phase-space densities,
denoted by
$F_\alpha$, where $\alpha\equiv p, c, h$
labels the species of partons, pre-hadronic clusters, or hadrons,
respectively, is:
\begin{equation}
F_\alpha(r,k)\;\,\equiv\; \, F_\alpha (t, \vec r; E, \vec k)
\;\,=\;\,
\frac{dN_\alpha (t)}{d^3r d^3k dE}
\;,
\label{F}
\end{equation}
where $k^2 = E^2 -\vec{k}^{\,2}$ can be offshell or on-shell,
as will be discussed in the following subsections.
The densities (\ref{F}) measure the number of particles
of type $\alpha$ at time $t$ with position in $\vec r + d\vec{r}$,
momentum in $\vec k + d\vec{k}$,
and energy in $E + dE$ (or equivalently invariant mass in $k^2 + dk^2$).
The $F_\alpha$ are the quantum analogues of the
classical phase-space distributions, including both off-shell and on-shell
particles, and hence
contain the essential microscopic
information required for a statistical description
of the time evolution of a many-particle system in
complete 7-dimensional phase-space $d^3rd^3kdE$, 
thereby providing the basis for calculating
macroscopic observables.

The phase-space densities (\ref{F}) are determined by the
self-consistent solutions of
a set of {\it transport equations} (in space-time) coupled with
renormalization-group-type {\it evolution equations} (in momentum space).
Referring  to Refs. \cite{ms37,ms39} for details,
these equations can be generically expressed as
convolutions of the densities $F_\alpha$ of particle species $\alpha$,
interacting with  specific cross sections $\hat{I}_j$ for the processes $j$.
The resulting coupled equations for the
space-time development of
the densities of partons $F_{p}$, clusters $F_c$ and
hadrons $F_h$ is a self-consistent set in which the change
of the densities $F_\alpha$ is governed by the balance of
the various possible interaction processes among the particles.
Fig. 3 represents these equations pictorially.
For the densities of {\it partons} 
the {\it transport equation} 
(governing the space-time change with $r^\mu$) and the {\it evolution equation} 
(controlling the change with momentum scale $k^\mu$), read, respectively,
\begin{eqnarray}
k_\mu \frac{\partial}{\partial r^\mu}\; F_p(r,k)
&=&
F_{p''} F_{p'''}\circ 
\left[\frac{}{}
\hat{I}(p''p'''\rightarrow p p') \;+\;\hat{I}(p''p'''\rightarrow p)
\right]
\;-\; 
F_{p} F_{p'}\circ 
\left[\frac{}{}
\hat{I}(pp'\rightarrow p'' p''')\;+\; \hat{I}(pp'\rightarrow p'')
\right]
\nonumber \\
& &
-\;\;
F_{p} F_{p'}\circ \left[\frac{}{} 
\hat{I}(p'p''\rightarrow p)\;-\; \hat{I}(pp'\rightarrow p'')\right]
\;-\;
F_p\,F_{p'}\circ \hat{I}(p p'\rightarrow c)
\label{e1}
\\
k^2  \frac{\partial}{\partial k^2}\; F_p(r,k)
&=&
F_{p'}\circ \hat{I}(p'\rightarrow p p'')\;-\;
F_{p}\circ \hat{I}(p\rightarrow p' p'')
\label{e2}
\;.
\end{eqnarray}
For the densities of {\it pre-hadronic clusters}  and
{\it hadrons}, the evolution equations  are homogeneous
to good approximation,
so that one is left with non-trivial transport equations only,
\footnote{
It is worth noting that eq. (\ref{e2})  embodies the momentum space
($k^2$) evolution of partons through
the renormalization of the phase-space densities $F_p$, determined
by their change $k^2 \partial F_p(r,k)/\partial k^2$
with respect to a variation of the mass (virtuality) scale $k^2$
in the usual QCD evolution framework \cite{dok80,jetcalc,bassetto}.
On the other hand,
for pre-hadronic clusters and hadrons, renormalization effects
are comparatively small, so that their
mass fluctuations $\Delta k^2/k^2$ can be ignored to first
approximation,
implying $k^2 \partial F_c(r,k) /\partial k^2
= k^2 \partial F_h(r,k) / \partial k^2  =0 $.
}
\begin{eqnarray}
k_\mu \frac{\partial}{\partial r^\mu}\; F_c(r,k)
&=&
F_p\,F_{p'}\circ \hat{I}(p p'\rightarrow c)
\;-\;
F_c\circ \hat{I}(c\rightarrow h)
\;\;\;,\;\;\;\;\;\;\;\;\;\;
\;\;\;\;\;\;\;\;\;\;\;\;
\;\;\;\;\;\;\;\;\;\;\;\;
k^2  \frac{\partial}{\partial k^2}\; F_c(r,k)\;\;=\;\;0
\;\;\;\;\;\;\;\;\frac{}{}
\label{e3}
\\
k_\mu \frac{\partial}{\partial r^\mu}\; F_h(r,k)
&=&
F_c \circ\hat{I}(c\rightarrow h)
\;+\;
\left[\frac{}{}
F_{h'}\circ \hat{I}(h'\rightarrow h)
\;-\;
F_h\circ \hat{I}(h\rightarrow h')
\right]
\;\;,\;\;\;\;\;\;
k^2  \frac{\partial}{\partial k^2}\; F_h(r,k)\;\;=\;\;0
\;\;\;\;\;\;\;\;\frac{}{}
\label{e4}
\;.
\end{eqnarray}
In (\ref{e1})-(\ref{e4}),
each convolution $F \circ\hat{I}$ of
the density of particles $F$ entering a particular vertex
${\hat I}$ includes a sum over contributing
subprocesses, and a phase-space integration
weighted with the associated subprocess probability distribution
of the squared amplitude. Explicit expressions are given in
Refs.  \cite{msrep,ms37}.

The terms on the right-hand side
of the transport-  and evolution-equations (\ref{e1})-(\ref{e4})
corresponds to one of the following categories:
\begin{description}
\item[(i)]
parton scattering and parton fusion  through 2-body collisions,
\item[(ii)]
parton multiplication through radiative emission processes
on the perturbative level,
\item[(iii)]
colorless cluster formation through parton recombination
depending on the local color and spatial configuration,
\item[(iv)]
hadron formation through decays of the cluster excitations
into final-state hadrons.
\end{description}

As mentioned before, the
equations (\ref{e1})-(\ref{e4}) reflect a {\it probabilistic
interpretation} of the multi-particle evolution in 
space-time and momentum space
in terms of sequentially-ordered interaction processes $j$,
in which the rate of change of the particle distributions $F_\alpha$
($\alpha=p,c,h$)
in a phase-space element $d^3rd^4k$
is governed by the balance of gain (+) and loss ($-$) terms.
The left-hand side
describes free propagation of a
quantum of species $\alpha$, whereas
on the right-hand side the interaction kernels $\hat{I}$
are integral operators that incorporate the effects of
the particles' self-  and mutual interactions.
This probabilistic  character 
is essentially an effect of time dilation, because in any frame
where the particles move close to the speed of light, the associated
wave-packets are highly localized to short space-time extent, so that
comparatively
long-distance quantum interference effects are generally very small.

\begin{figure}
\epsfxsize=500pt
\centerline{ \epsfbox{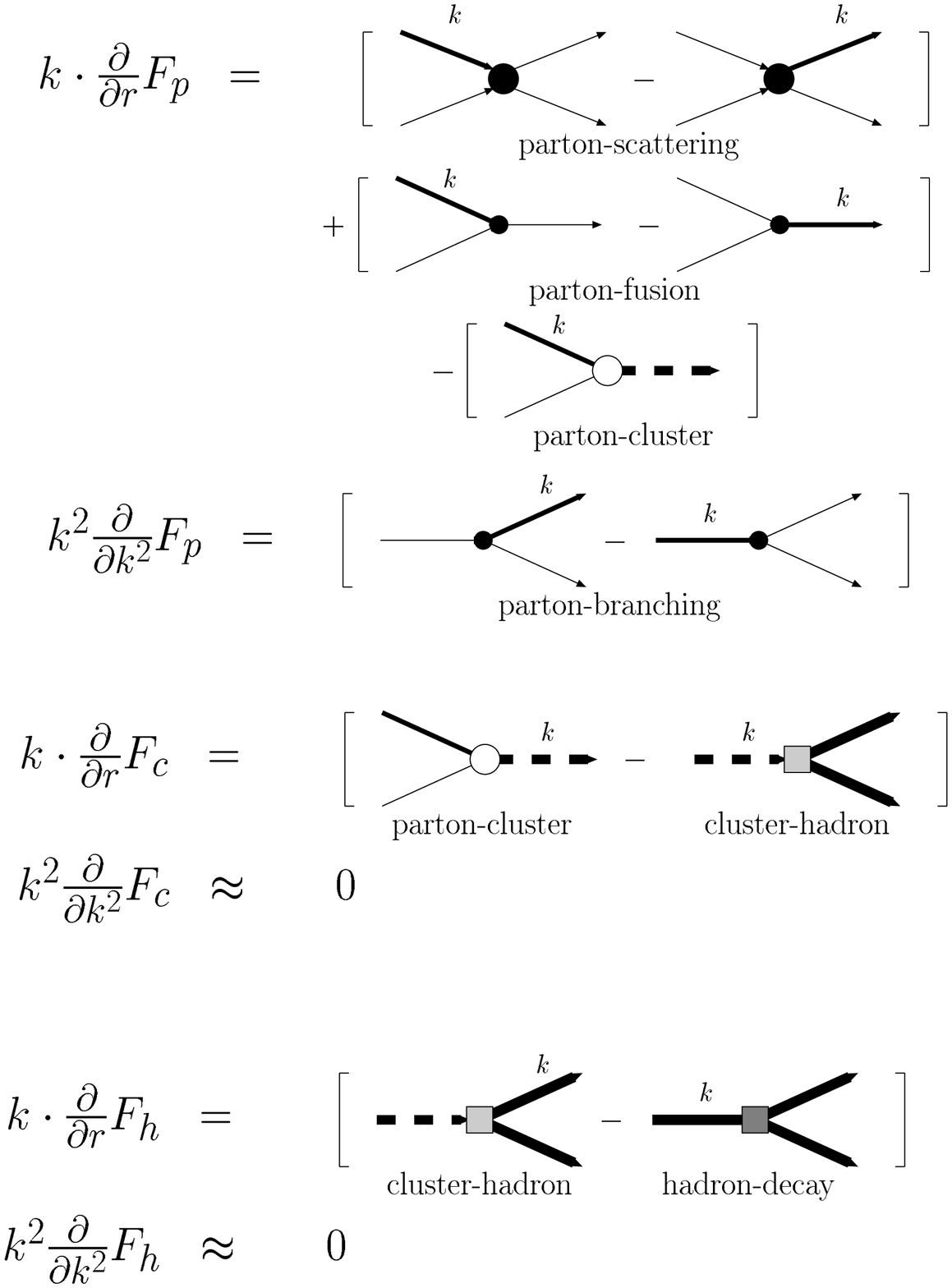} }
\caption{
Graphical representation of the equations (2)-(5) for the
particle phase-space densities $F_p$ of partons, $F_c$ of
pre-hadronic clusters, and $F_h$ of hadrons.
\label{fig:fig3}
}
\end{figure}
\bigskip

\subsection{Scheme of solution in global Lorentz-frame of reference}

In the above kinetic approach to the multi-particle
dynamics,
the probabilistic character of the transport- and evolution equations
(\ref{e1})-(\ref{e4})
allows one to solve for the phase-space densities $F_\alpha(r,k)$ by
simulating
the dynamical development as a Markovian process causally in time.
Because it is an `initial-value problem', one must specify
some physically appropriate initial
condition $F_\alpha(t_0,\vec{r},k)$ at starting time $t_0$, such that
all the dynamics prior to this point is effectively embodied in this
initial form of $F_\alpha$.
The set of equations (\ref{e1})-(\ref{e4}) can
then be solved
in terms of the evolution of the phase-space densities $F_\alpha$
for $t > t_0$ using Monte Carlo methods to
simulate the time development of the mixed system
of partons, clusters, and hadrons
in position and momentum space \cite{ms37,msrep}.

With the initial state specified as discussed below,
the phase-space distribution of particles at $t=t_0 \equiv 0$ can be
constructed and then evolved in small time steps 
$\Delta t =O(10^{-3}\;fm)$ forward throughout the parallely evolving
stages of parton cascade, parton-cluster formation, and cluster-hadron decays,
until stable final-state hadrons and other particles (photons, leptons, etc.) 
are left as freely-streaming particles.
The partons propagate along classical trajectories until they interact,
i.e., collide (scattering or fusion process),
decay (emission process) or recombine to pre-hadronic composite
states (cluster formation).
Similarly, the so-formed pre-hadronic parton-clusters 
travel along classical paths until they convert into
primary hadrons (cluster decay), followed by the hadronic decays
into stable final state particles.
The corresponding probabilities and time scales of interactions are
sampled stochastically from the relevant probability distributions
in the kernels $\hat{I}$ of eq. (\ref{e1})-(\ref{e4}).

It is important to realize, that 
the spatial density and the momentum distribution of the 
particles are intimately connected: The momentum 
distribution continously changes through the interactions and
determines how the quanta propagate in coordinate space.
In turn, the probability for subsequent interactions depends on the 
resulting local particle density. Consequently, the development
of the phase-space densities is a complex
interplay, which - at a given point of time - contains implicitely the
complete preceeding history of the system.
\medskip

It is clear that
the description of particle evolution is
Lorentz-frame dependent, and a suitable reference frame
(henceforth called {\it global frame})
must be chosen (not necessarily the laboratory frame of an experiment).
When computing Lorentz-invariant quantities, such as
cross sections or final-state hadron spectra, the particular choice is
irrelevant, whereas for non-invariant observables, such as energy
distibutions
or space-time-dependent quantities, one must at the end transform
from the arbitrarily-chosen frame of theoretical description to the
actual frame of measurement.

For calculational convenience, it is most suitable
to choose the {\it global center-of-mass ($cm$) frame of the colliding 
beam particles}, with the collision axis in the $z$-direction.
In this global $cm$-frame, the incoming particles $A$, $B$ 
(= lepton, hadron, or nucleus) have four-momenta,
\begin{eqnarray}
& &
\;\;\;\;\;\;\;\;\;\;\;
P^\mu_{A,B} \; = \;\left( E_{A,B}, 0,0,\pm P_{cm}\right)
\nonumber
\\
E_{A,B} &=& \frac{1}{2 \sqrt{s}}\;\left(s \,+\,M_A^2\,\pm\,M_B^2\right)
\;\;\;\;\;\;\;\;\;\;\;\;\;\;\;
s\;=\; E_{cm}^2 = \left(P_A^\mu+P_B^\mu\right)^2
\label{cm}
\\
P_{cm}&=& \frac{1}{2\sqrt{s}}\;
\sqrt{\left[s-(M_A+M_B)^2\right]\; \left[s-(M_A- M_B)^2\right]}
\; ,
\nonumber
\end{eqnarray}
where $M_{A,B}$ are the masses of $A$ and $B$,
and so  incoming particles then carry well-defined fractions of $P_{cm}$,
having only a non-vanishing longitudinal momentum along 
the $z$-axis.
Particularly in the case of nuclei $A$, the daughter nucleons $N_i=1,\ldots A$, 
have momenta, $\vec P_{N_i}= (0,0,\pm P_{cm}/A)$.
\smallskip

In the following, the global $cm$-frame of $A$ and $B$ is assumed to be the
reference frame, with the initial energy-momentum of the collision system
given by (\ref{cm}).
Furtermore the terms `hadron', respectively `nucleon', are used to
distinguish initial states $A+B$ with $A$ and/or $B$ being 
a single hadron, respectively a nucleus with $A$ ($B$) nucleons.

\bigskip

\subsection{Initial state}

If one or both beam particles $A$, $B$ are leptons,
they are considered as  point-like objects
which carry the full beam energy, meaning
that  any QED or QCD substructure of the leptons,
as well as initial-state photon radiation by the leptons, is neglected.
For {\it lepton-lepton annihilation},
it is assumed that the colliding leptons produce a 
time-like $\gamma$ or $Z^0$ boson of invariant mass
$Q^2\equiv +q^2$ at time $t = -Q^{-1}$, so that $t=0$ characterizes the point
when the $\gamma$ ($Z^0$)  decays into a quark-antiquark pair.
Similarly, for {\it lepton-hadron (nucleus) collisions}, the lepton is
emitting a space-like virtual $\gamma$
of invariant mass $Q^2\equiv -q^2$ at time $t = -Q^{-1}$, and hence $t=0$ 
is the point when the $\gamma$ hits the hadron (nucleus).

In the general case of {\it collisions involving hadrons and/or nuclei},
the incoming particles $A$ and $B$ are
decomposed into their parton substructure by
phenomenological construction of the
momentum and spatial distributions of their daughter partons
on the basis of the known hadron (nucleon) structure functions
and elastic hadron (nucleon) form-factors.
In the $cm$-frame, where the two incoming particles
$A, B$ (= hadron, nucleus), are moving close to the speed of light, 
the parton picture is applicable and the parton substructure 
of the hadrons or nucleons
can be resolved with reference to some {\it initial resolution scale} $Q_0$. 
This resolution scale 
generally varies with beam-energy and mass number $A$, $B$, in that it depends
on the typical momentum and spatial density of partons as well as on
their interaction probability.
To be specific, it may be identified with
the statistically estimated expectation value for the 
interaction scale $Q^2$ of all {\it primary} parton-parton collisions 
(that are those in which at least one initial state parton is involved)
\cite{miklos}:
\begin{equation}
Q_0^2 \;\,\equiv\;\,
Q_0^2 (x,P,A) \,=\, A^\alpha \;\left(\frac{1}{\langle p_\perp^2 \rangle_{prim}}\,+\,
\frac{\langle p_\perp^2 \rangle_{prim}\,R_0^2}{2 x P}
\right)^{-1}
\;,
\label{Q0}
\end{equation}
where
$\langle p_\perp^2 \rangle_{prim}$
is the average relative transverse momentum squared generated
by primary parton-parton collisions.
and $R_0 = 1 GeV^{-1}$.
The pre-factor $A^\alpha$ ($0\le\alpha\le 4/3$ a parameter) 
accounts for the nuclear dependence for $A,B > 1$,
and 
$x$ is the parton's momentum fraction of the mother hadron or nucleon
which carries momentum $P$ ($= P_{cm}$ for single hadron, or $=P_{cm}/A$
for a nucleon in a nucleus $A$).
The scale $Q_0$ defines the initial point in momentum space above which the 
system of partons is treated as an ensemble
of incoherent quanta. The  dynamics prior to this point
is contained in the initial parton-structure function of the mother hadron
(nucleon).
Clearly, the convention (\ref{Q0}) yields only an average value 
dominated by the most probable (semi)hard parton collisions with
relatively small momentum transfers of a few GeV. 
However, primary parton collisions with a momentum scale
$Q^2\gg Q_0^2$, which correspond to relatively rare fluctuations, 
are taken into account individually
by the $Q^2$-evolution of the hadron (nucleon) structure functions through 
space-like and time-like radiation processes, discussed later.

The actual form of the  phase-space distribution of partons,
eq. (\ref{F}) is initially to be specified at the reference point $Q_0$.
It is constructed as the following superposition of the 
parton distributions in the individual 
hadrons at $Q_0$ and at time $t=t_0\equiv 0$, the point of collision of $A$ and $B$:
\begin{equation}
\left. F_a (r,k)\right|_{r^0  = t_0}\; = \; \sum_{i=1}^{N_{h}} 
P_a^{N_i} ( k, \vec P; Q_0^2) \cdot R_a^{N_i} ( k, \vec r,\vec R)
\left.\frac{}{}\right|_{r^0=t_0} 
\;\;\;.
\label{Fa0}
\end{equation}
The right hand side is a sum over all $1 \le N_i\le N_{h}$ 
hadrons or nucleons contained in the collision system $A+B$.
Each term in the sum is a  parton phase-space density given by a
convolution of an initial momentum distribution $P_a^{N_i}$ and a spatial
distribution $R_a^{N_i}$. 
The subscript $a = g, q_j, \bar q_j$ ($j= 1,\ldots , n_f)$
labels the species of partons, and $N_i$ refers to the type of the $i$-th
hadron (for nuclei this is just proton or neutron). The four-vectors
$k\equiv k^\mu=(E,\vec k)$ and $r \equiv r^\mu = (t, \vec r)$
refer to the partons. The  3-momenta and 
initial positions of the hadrons or nucleons are
denoted $\vec P$, respectively $\vec R$. All vectors are understood
with respect to the global $cm$-frame, at time $t = t_0 =0$.
The partons' energies
$E \equiv E_a(\vec{k}^{\,2},q^2) = \sqrt{\vec{k}^{\,2} + m_a^2 + q^2}$
take into account their initial space-like virtualities $q^2<0$
which are distributed around $\langle \vert q^2 \vert \rangle = -Q_0^2/4$,
under the constraint that for each hadron in the initial state,
the total invariant mass of the daughter partons must equal the mother 
hadron mass.  This mimics the fact that 
the initial partons are confined inside their parent hadrons or nucleons 
and cannot be treated as free particles, i.e. they have not enough energy to
be on mass shell, but  are off mass shell by a 
space-like virtuality $q^2$ (eq. (\ref{invmass}) below).
\smallskip

As suggestively illustrated in Fig. 2 and discussed in more detail now,
the initial state $A+B$ involving hadrons and/or nuclei,
appears in the global $cm$-frame as two approaching `tidal waves' of
large-$x$ partons (mainly valence quarks), where each
`tidal wave' has an extended tail
of low$-x$ partons (mostly gluons and sea-quarks).
\medskip

\subsubsection{Initial momentum distribution}

\noindent
For each hadron or nucleon the number of partons, the distribution of 
the flavors, their momenta and associated initial space-like virtualities,
are obtained from the function $P_a^{N_i}$ in (\ref{Fa0}).
Denoting  $P\equiv P_{cm}/N_{h}$, with $N_h$ the total number
of hadrons (nucleons) in $A+B$, it is represented in  the form
\begin{equation}
P_a^{N_i} (k,\vec P; Q_0^2)\;=\; 
\left(\frac{x}{\tilde x}\right) \; F_a^{N_i} (x,Q_0^2) \;\, \rho_a^A(x)\;\,
g (\vec k_{\perp}) 
\;\,\delta\left(P_z \,-\,P\right) \; \delta^2\left(\vec P_\perp\right)
\label{PaN}
\end{equation}
with the momentum and energy fractions
$
x = k_z/P
$,
$
\tilde x= E/P= \sqrt{x^2 + (k_\perp ^2 + m_a^2 + q^2)/P^2)}
$
and the normalizations
\begin{eqnarray}
&&
\;\;\;\;\;\;\;\;\;\;
\sum_a \int_0^1 dx \; x F_a^{N_i} (x, Q_0^2) \;= \; 1
\;,\;\;\;\;\;\;\;\;\;\;\;\;\;\;\;\;
\int_0^\infty d^2 k_\perp \;g(\vec k_\perp)  \;=\; 1
\label{norm1}
\\ 
&&
\;\;\;\;\;\;\;\;\;\;
\sum_a \int \, \frac{dk^2\,d^3 k}{(2\pi)^3 2 E}
\; E \; P_a^{N_i} ( k^2,\vec k, \vec P; Q_0^2) \;=\;
n^{N_i} (Q_0^2,\vec P) 
\label{norm3}
\;\; .
\end{eqnarray}

\noindent
The physics behind the ansatz (\ref{PaN}) is the following:
\begin{description}
\item[(i)]
The functions $F_a^{N_i} (x,Q_0^2)$ are the scale-dependent measured
hadron (nucleon) structure functions with $x$ being the fraction
of the parent hadron's or nucleon's longitudinal momentum $P$ carried by the parton.
The transverse momentum distribution $g(k_\perp)$ is specified below.
The factor
$x / \tilde x$ in (\ref{norm1}) is included to form the
invariant momentum integral $\int d^3 k / [(2 \pi)^3 2 E]$ out of
the distribution $P_a^{N_i}$ \cite{feyfld}.
In (\ref{norm3}) the quantity $n^{N_i}$ is the total
number of partons in a given nucleon with momentum $P=|\vec P|$ at $Q_0^2$.
\item[(ii)]
The function $\rho_a^A(x)$ in (\ref{PaN}) takes into account
nuclear shadowing effects
affecting  mainly soft (small $x$) partons in a nucleus
\footnote{This feature is optional in the simulation, by default it is
switched off.}.
These shadowing effects are evident 
in the observations of the European Muon Collaboration \cite{emc} as
a depletion of the nuclear structure functions at small $x$ relative to
those of a free nucleon.
Several mechanisms have been proposed to explain this nuclear shadowing
effect on the basis of the parton model \cite{nucshad4,nucshad2,nucshad3}.
However, here instead the phenomenological approach of Wang and Gyulassy 
\cite{hijing2} is adopted
which is based on the following parametrization \cite{nucshad3}
for the $A$ dependence of the shadowing for both quarks and gluons:
\begin{eqnarray} 
\rho_a^A(x)  &\equiv&
\frac{ {\sl F}_a^{A, N_i} ( x, Q_0^2)}{A \, F_a^{N_i}(x, Q_0^2)}
\;\,=\;\,
1 \,+ \,1.19 \,\left(\frac{}{}\ln A\right)^{1/6} \, \left[
x^3 - 1.5 (x_0 + x_L) x^2 + 3 x_0 x_L x \right]
\\
& &
\;\;\;\;\;\;\;\;\;\;\;\;\;\;\;\;\;\;\;\;\;\;\;\;
\;\;\;\;\;\;\;
- \left[ \;
\beta_A (R_\perp) - 
\frac{1.08 \,( A^{1/3} -1)}{\ln(A + 1)} \, \sqrt{x}\, \right]
\; \exp(-x^2 / x_0^2)
\;\;\;.
\label{rhoA}
\nonumber
\end{eqnarray}
where ${\sl F}_a^{A, N}$ and $F_a^{N}$ represent the structure function of
a nucleus $A$, respectively those of a free nucleon, and
$x_0 = 0.1$, $x_L = 0.7$. The function
$
\beta_A (r) = 0.1 \,\left(\frac{}{}A^{1/3} -1\right) 
\,\frac{4}{3} \,\sqrt{1 - \frac{R_\perp^2}{R_A^2}}
$
takes into account the impact parameter dependence, with $R_\perp$ labeling
the transverse distance of a nucleon from 
its nucleus center and $R_A$ the radius of the nucleus.
\item[(iii)]
The distribution $g(k_\perp)$ specifies
the primordial transverse momenta of partons  according to
a normalized Gaussian 
\begin{equation}
g (\vec k_{\perp})\; = \; \frac{1}{2\pi k_0^2} \; 
\exp \left[-\frac{\vert \vec k_\perp\vert^2}{k_0^2}\right] 
\;,
\label{gpT}
\end{equation}
independent of the type of parton or
hadron (nucleon). 
It takes into account the uncertainty of momentum 
(Fermi motion) due to the fact that the initial
partons are  confined within the nucleons. 
This intrinsic 
$k_{\perp}$ can be observed in Drell-Yan experiments
where it is found that the distribution is
roughly independent of $s$ and $Q^2$ \cite{pTprim}.
As inferred from these analyses, the
width in (\ref{gpT}) is 
$k_0 = 0.42$ GeV, corresponding to a mean
$\langle k_\perp \rangle \approx 0.38$ GeV. 
\end{description}

The scheme (i)-(iii) to sample the flavor and momentum distribution is carried out 
independently for each nucleon subject to the requirement
\begin{equation}
\left( \sum_i E_i \right)^2 - \left( \sum_i k_{x_i}\right) - \left( 
\sum_i k_{y_i}\right)^2 - \left( \sum_i k_{z_i}\right)^2\; = \; 
M_h^2 
\label{invmass}
\end{equation}
where the summation runs over all partons belonging to the same nucleon as determined by
(\ref{norm1}) and (\ref{norm3}),
and $M_h$ is the mother hadron (nucleon) mass.  
With the partons' 4-momenta distributed as outlined above,
the constraint (\ref{invmass}) determines the distribution
in the variable
$q^2 = k^2 - m_a^2 = E^2 -\vec k^2 -m_a^2 < 0$, 
the partons' initial space-like virtualities.
The resulting distribution in $q^2$ 
is a strongly peaked  Gaussian  with a mean value of $\approx Q_0^2/4$.
\medskip

\subsubsection{Initial spatial distribution}

\noindent
The initial spatial distribution of the partons, $R_a^{N_i}$, 
appearing in eq. (\ref{Fa0}), 
depends on the magnitude of their momenta, the
positions of their parent hadrons (nucleons), 
as well as on the spatial substructure
of the latter. It is represented as
\begin{equation}
R_a^{N_i} (\vec p, \vec r, \vec R) \;=\; 
\delta^3\left(\vec R \,-\,\vec R_{AB}\right)\;\,
\left[\frac{}{} \, h_a^{N_i} (\vec r) \;
\,H_{N_i} (\vec R) \, \right]_{boosted} 
\;\;,
\label{RaN}
\end{equation}
where the momentum dependence is here  purely due to
boosting the distributions to the global $cm$-frame. 
The components of the ansatz (\ref{RaN}) have the following meaning:
\begin{description}
\item[(i)]
The incoming beam particles (a single hadron, or a nucleus) 
are assumed to have initial $cm$-positions 
$\vec R_{AB} = (\pm \Delta Z_{AB}/2, \vec B_{AB})$,
corresponding to a chosen impact parameter $\vec B_{AB}=(B_x,B_y)$ 
and a minimum longitudinal separation $\Delta Z_{AB}$.
In the case where $A$ and/or $B$ is not a single hadron, but 
a nucleus,  the individual nucleons are 
assigned positions around the centers of the parent nucleus in its rest-frame,
according to a Fermi distribution for nuclei with mass number $A \geq 12$ and
a Gaussian distribution for nuclei $A < 12$,
\begin{equation}
H_{N_i} (\vec R)\;=\; \left\{ \,
\begin{array}{lr}
\, \frac{1}{4 \pi}\, \left( 1 + \exp \left[ (R-c)/a \right] \right)^{-1}
& \, ( A \geq 12 ) \\
\, \frac{1}{4 \pi}\, \frac{2}{\sqrt{\pi} b} \; \exp \left[ - R^2 / b^2 
\right]
& \, ( A < 12 ) 
\end{array}
\right.
\;\;.
\end{equation}
The parameters are $c = r_0 \, A^{1/3}$, $r_0 = 1.14$ $fm$, $a = 0.545$ $fm$ and
$b = \sqrt{2/3}\, R_A^{ms}$, where $R_A^{ms}$ is the mean square radius
of the nucleus.
\item[(ii)]
Next, the partons are distributed around the centers of their
mother hadrons or nucleons, still in the rest frame of the $A$, respectively $B$,
with an exponential distribution
\begin{equation}
h_a^{N_i} (\vec r)\;=\; 
\, \frac{1}{4 \pi}\,\frac{\nu^3}{8 \pi} \, \exp \left[ - \nu r \right]
\;\;,
\end{equation}
where $\nu = 0.84$ GeV corresponds to the measured 
elastic formfactor of the mother hadron or  nucleon, with a mean square radius
of $R_h^{ms} \equiv \sqrt{12/\nu} = 0.81$ $fm$.
\item[(iii)]
Finally, as indicated by the subscript {\it boosted} in eq. (\ref{RaN}),
the positions of the hadrons or nucleons and their associated valence quarks
are boosted into the global $cm$-frame of the colliding beam particles $A$ and $B$.
The valence quarks then occupy the Lorentz contracted region
$(\Delta z)_v \approx 2 R_A \,M_h/P$, whereas
the sea quarks and gluons are smeared out in the longitudinal
direction by an amount $(\Delta z)_{g,s} \approx 1/k_z < 2 R_A$ ($R_A=r_0 A^{1/3}$)
{\it behind}  the valence quarks. This is an important
feature of the partons when boosting a hadron or nucleus to high rapidities 
\cite{bj76,hwa86,mueller89}.
As a consequence, the parton positions are correlated in longitudinal
direction with their momenta, as required by the uncertainty principle.
This leads to an enhancement of the densities of gluons and sea quarks
with $x< 1/(2R_h M_h)$ proportional to $A^{1/3}$, because such partons from
different nucleons overlap spatially when the nucleons are at the same impact parameter.
\end{description}
\bigskip

\subsection{Parton cascade development}

With the above construction of the initial state,
the incoming beam particles $A$ and $B$ are decomposed in   
their associated parton content at the initial
resolution scale $Q_0^2$ at time $t=t_0\equiv0$, 
and the subsequent dynamical development of the
system for $t > t_0$
can now be traced according to the kinetic equations
(\ref{e1})-(\ref{e4}).
In the kinetic approach, the space-time evolution of  parton densities
may be  described in terms of
parton cascades. Fig. 4 illustrates  a typical parton cascade sequence. 
It is important to realize that, in general,
there can be many such cascade sequences, being internetted
and simultanously evolving (typical for hadron-nucleus or nucleus-nucleus
collisions). 

\begin{figure}
\epsfxsize=450pt
\centerline{ \epsfbox{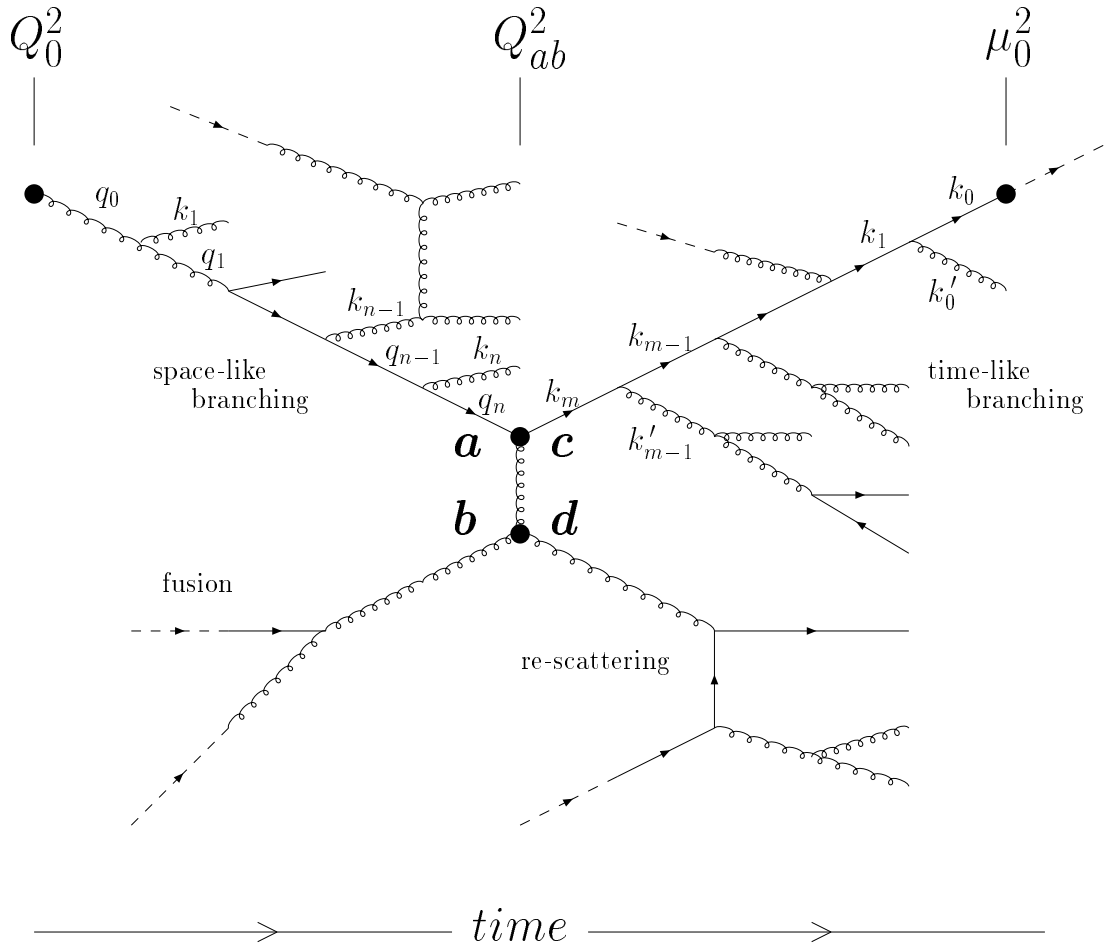} }
\caption{
Schematical illustration of 
a typical parton cascade development initiated by a 
collision of two partons $a$ and $b$.
The incoming primary parton $a$ evolves through
a space-like cascade from the 
initial resolution scale $Q_0^2$ up to
the scale $Q_{a b}^2$ at which it collides with parton $b$.
The outgoing partons $c$ and $d$ both initiate time-like
cascades which are described as a combination of 
multiple branchings (or emissions), fusions (or absorptions),
or secondary scatterings (or rescatterings).
A particular branch in a cascade tree terminates
(locally in space-time), when the partons in that branch 
recombine with neighboring ones to pre-hadronic clusters and 
their subsequent decay into hadrons.
\label{fig:fig4}
}
\end{figure}

Each cascade can be subdivided into elementary
$2 \rightarrow 2$ scatterings, $1 \rightarrow 2$ branchings (emissions), and
$2\rightarrow 1$ fusions (absorptions).
In Fig. 4, a primary parton $a$ that originates from one of the incident
nuclei, collides with another parton $b$ with some momentum transfer
characterized by the scale $Q_{a b}^2$ at the collision vertex.
The parton $a$ has evolved from the initial scale $Q_0^2$, at which it was
resolved in its parent hadron or nucleus, 
up to $Q_{a b}^2$ by successive space-like
branchings. From the scattering of $a$ and $b$ the partons $c$ and $d$ emerge,
both of which can initiate sequences of time-like branchings. These newly
produced partons can themselves branch,
rescatter, or undergo fusion with other partons.

In practice the dynamical interplay of $2 \rightarrow 2$, $2 \rightarrow 1$
and $1 \rightarrow 2$ processes can be simulated
as follows:
Since the time evolution of the parton system is described in
small discrete time steps $\Delta t = O(10^{-3} fm)$, 
the time scale of an interaction
process is compared with $\Delta t$ to decide whether the
interaction occurs within this time interval. 
A virtual parton
$e^\ast$ with momentum $k^\mu$, $k^2\ne 0$,
that is produced via $a + b \rightarrow e^\ast$, has a short
life-time $\tau_{e^\ast}\propto 1/\sqrt{|k^2|}$, 
if its invariant mass $|k^2|$ is large and it is likely
to decay within $\gamma \tau_{e^\ast} < \Delta t$
(the full formulae are derived  in \cite{ms1}).
Therefore, the $2 \rightarrow 2$ process
$a + b \rightarrow e^\ast \rightarrow c + d$ preferably occurs.
On the other hand, if $|k^2|$ is small, corresponding to a long
$\tau_{e^\ast}$, parton $e^\ast$ is likely not to decay within 
the time step $\Delta t$
and the fusion process $a + b \rightarrow e^\ast$ rather happens.
In this case, $e^\ast$ propagates as a quasi-stable particle until in
the following time step it has an increased decay probability \cite{ms1}.
This may then result in the
$1 \rightarrow 2$ decay $e^\ast \rightarrow c + d$. Alternatively,
$e^\ast$ might collide with another parton before its decay or 
may propagate freely until the following time step, and so on.
In this manner the elementary $2\rightarrow 2$, $2\rightarrow 1$,
and $1\rightarrow 2$ processes are treated on equal footing
and double counting is prevented:
either  a $2\rightarrow 2$, or a $2\rightarrow 1$,
and possibly a subsequent $1\rightarrow 2$ process can occur, 
but not both. The relative probability is  determined by
the uncertainty principle, i.e. by relating the momentum scale of 
the process to the time scale as explained.

For the collisional processes $2\rightarrow 2$ and $2 \rightarrow 1$,
the statistical occurence is determined by the 2-body cross-section
in Born apprpoximation, with higher-order inelastic
corrections effectively included in the parton evolution
before and after each collision.
This  parton-evolution is calculated
using the well-known jet calculus \cite{jetcalc,bassetto}
based
on the `modified leading logarithmic approximation' (MLLA)
to the QCD evolution of hard processes \cite{dokbook,dok88}.
Each individual parton-collision factorizes then in an elementary
2-body collision, in which the in- and out-going partons undergo
an ordered sequence
of elementary branchings $a\rightarrow b+c$, accounting for
higher-order radiative corrections to the lowest-order Born approximation.
These  branchingss can be described stochastically as a
Markov cascade in position and momentum space.
One distinguishes  initial-state, {\it space-like} branchings
of the two partons entering a collision vertex,
and final-state, {\it time-like} radiation off the 
collided partons after the collision.

The specific feature of the present approach is that, in addition to the
definite
virtuality and momentum,  each elementary vertex has a certain space and
time
position which is obtained by assuming that the partons in the cascade
propagate on
straight-line trajectories in between their interactions.
In the MLLA framework, the
basic properties of both space-like and time-like showers are
determined by the DGLAP equations \cite{DGLAP}, but with essential
differences in time ordering, kinematics and the treatment of
infrared singularities associated with soft gluon emission.
\medskip

\subsubsection{Elementary parton-parton collisions}

For the elementary  parton scatterings $a + b \rightarrow c + d$,
and fusion processes  $a + b \rightarrow c^\ast$, 
two distinct classes of processes 
are considered: 
\begin{description}
\item[(i)] truly perturbative QCD {\it hard}  parton collisions
with a sufficiently large momentum transfer $p_\perp^2$ or invariant mass
$\hat s$;
\item[(ii)] 
phenomenological treatment of {\it soft} parton collisions
\footnote{The inclusion of soft parton collisions is optional in the program,
and by default is switched off.} 
with low momentum
transfer $p_\perp^2$ or invariant mass $\hat s$.
\end{description}
The motivation for this differentiation of parton-parton collisions
is to regulate the singular behavior of the collision integrals
in (\ref{e1}) which results from the divergence of the 
associated Born-amplitudes squared
for small momentum transfers $Q^2$ (except for $\hat s$ channel
processes such as $q \bar q$ annihilation).
To render the parton-parton cross-sections finite, an invariant 
{\it hard-soft division scale} $p_0^2$ is introduced such that
collisions occurring at a momentum scale $Q^2 \geq p_0^2$
are treated as perturbative (semi)hard collisions,
whereas for those with $Q^2 < p_0^2$
a soft, non-perturbative  interaction is assumed to occur. That is, the total
parton-parton cross-section for a collision
between two partons $a$ and $b$ is represented as
\begin{equation}
\hat \sigma_{a b} ( \hat s ) \;=\;
\sum_{c, (d)} \, \left\{ \,\frac{}{}
\int_{0}^{p^2_0}  d Q^2
\left(
\frac{d \hat \sigma_{a b \rightarrow c (d)}^{soft}}{d Q^2}
\right)
\;+\;
\int_{p^2_0}^{\hat s}  d Q^2
\left(
\frac{d \hat \sigma_{a b \rightarrow c (d)}^{hard}}{d Q^2}
\right)
\, \right\}
\;\;\;,
\label{sigma}
\end{equation}
where
\begin{eqnarray}
Q^2 \,&\equiv& \,Q^2(\hat{s},\hat{t},\hat{u}) \;=\; \left\{
\begin{array}{cl}
p_\perp^2 \simeq  -\hat{t} (-\hat{u})
& \;\mbox{for} \;\hat{t} (\hat{u}) \; \mbox{channel}
\\
m^2 = \hat{s}
& \;\mbox{for} \;\hat{s} \;\mbox{channel}
\end{array}
\right.
\\
\hat s  = (p_a + p_b)^2
& &
,\;\;\;\;\;\;\;\;\;\;
\hat t = (p_a - p_c)^2
\;,\;\;\;\;\;\;\;\;\;\;
\hat u = (p_a - p_d)^2
\end{eqnarray}
That is, the scale $Q^2$ of the collision is set by
$p_\perp^2$ for scatterings
$a+b\rightarrow a+b$, or by $\hat s$ for annihilation
processes and $a+b\rightarrow c+d$ and fusion processes $a+b\rightarrow c$.
The sum over $c$, $(d)$ corresponds to summing over all possible reaction
channels (processes).
The specific value of $p_0$ is generally dependent on beam energy 
$E_{cm}=(P_A+P_B)^2$ of colliding hadrons (nuclei) $A+B$,
as well as on the  mass numbers $A$ and $B$. It is treated as a parameter
that is determined by existing experimental data. A suitable form is
\begin{equation}
p_0\;\,\equiv\;\,
p_0(E_{cm},A,B)\;=\; \frac{a}{4}\;\left(\frac{2 E_{cm}/\mbox{GeV}}
{A+B}\right)^b
\;\;\;\;\;\;\;\;
(a = 2.0 \;\mbox{GeV} , \;\;\;\;\;b= 0.27)
\;.
\end{equation}

The complementation of both hard and soft 
contributions renders the parton cross-section
finite and well defined for {\it all} $Q^2$.
It must be emphasized that the 
effect of soft parton collisions on the global dynamics is not essential
for collisions involving only leptons and/or hadrons, 
but plays an important role
in hadron-nucleus, or nucleus-nucleus collions.
The soft parton collisions
naturally involve comparably small momentum transfer, so that their contribution
to transverse energy production is small, but the effect on soft particle
production is significant.

In accord with the two-component structure of the cross-section (\ref{sigma}),
these parton collisions are distinguished depending on the momentum transfer:
\begin{description}

\item[(i)]
{\bf Hard collisions above} {\boldmath $p_{0}$}:
\smallskip

For the {\it perturbative (semi)hard collisions} above 
$p_{0}$, the momentum scale $Q^2$ is determined by the
corresponding differential cross-sections
\begin{equation}
\frac{d \hat \sigma_{ab \rightarrow cd}^{hard}}
{d  Q^2}
\,=\,
\frac{1}{16 \pi \hat s^2} \,
\left| \overline{M}_{ab \rightarrow cd} \right| ^2 
(\hat s, Q^2) 
\;\;\propto \frac{\pi \alpha_s^2(Q^2)}{Q^2}
\;,
\label{sigh}
\end{equation}
where $| \overline{M} | ^2$ 
is the process-dependent spin- and color-averaged squared matrix element
in Born approximation,  published in
the literature (see e.g. Refs. \cite{sig1} and for 
massive quarks Refs. \cite{sig2}).
\medskip

\item[(ii)]
{\bf Soft collisions below} {\boldmath $p_{0}$}:
\smallskip

In the case of a non-perturbative soft collision between two partons
it is assumed that a very low energy double gluon
exchange occurs. This provides a natural continuation
to the harder collisions above
$p_{0}$ where the dominant 
one gluon exchange processes
$g g \rightarrow g g$, $g q \rightarrow g q$ and
$q q \rightarrow q q$ have the same overall  structure \cite{gustaf82}.
A simple, and physically plausible, form for the soft cross-section
that continues the hard cross-section for $Q^2$ below $p_0$ down to $Q^2 = 0$,
may be modelled by introducing a screening mass-term $\mu^2$ in the
nominator of (\ref{sigh}),
\begin{equation}
\frac{d \hat \sigma_{ab \rightarrow cd}^{soft}}
{d  Q^2}
\;\propto\;
\frac{2\pi \alpha_s^2(p_0^2)}{Q^2 + \mu^2}
\label{sigs}
\end{equation}
where $\mu$ is a phenomenological parameter that 
governs the overall magnitude of the integrated  
$\sigma^{soft} \propto \ln[(p_0^2+\mu^2)/\mu^2]$.
The value of $\mu$ 
is not known precisely, but it can be estimated to be in the
range 0.3 - 1 GeV.
\end{description}

Notice that
both hard and soft scatterings are treated in this approach on completely equal footing.
That is, with the four momenta of the incoming partons known, the momentum
transfer and scattering angle are sampled from the respective
differential cross-sections $d\hat \sigma / d\hat t$
(\ref{sigh}) and (\ref{sigs}), assuming azimuthal symmetry of the
scattering geometry.
\medskip

\subsubsection{Space-like parton evolution}

\noindent
Space-like parton cascades are associated with
QCD radiative corrections due to brems-strahlung emitted 
by primary space-like partons, which are
contained in the initial state hadrons (nuclei),
and which encounter their very
first collision. Thereby such a primary, virtual parton is
struck out of the coherent hadron (nucleus) 
wavefunction to materialize to a real excitation.
A typical space-like cascade is illustrated in the left part of Fig. 4, 
where the parton $a_0$, originating from the initial hadron (nucleus) state, 
undergoes successive space-like branchings $a_0 \rightarrow a_1 a_1'$,
$a_1 \rightarrow a_2 a_2'$, ..., $a_{n-1} \rightarrow a_n a_n'$ to become
the parton $a \equiv a_n$ which then actually collides with another parton $b$.
The branching chain proceeds by increasing the virtualities 
$|q_i^2|$ of the partons $a_i$ in
the cascade, starting from $a_0$ with $|q_0^2|  \simeq Q_0^2/4$
up to $|q^2| \simeq  Q^2$ 
the space-like virtuality of the scattering parton $a$,
where $Q^2 \equiv Q_{ab}^2$ sets the scale of the collision
between partons $a$ and $b$, and hence for the evolution from $Q_0^2$ to $Q^2$.
The emitted partons $a_i'$ on the side branches with momenta $k_i$
on the other hand have, due to energy-momentum
conservation at the branching vertices, time-like virtualities and
each of them can initiate a time-like cascade, discussed afterwards.

The proper inclusion of both collinear, hard and coherent, soft branchings is
achieved by describing the space-like cascade in both space-time {\it and} 
momentum space by using
a so-called {\it angular-ordered} evolution variable 
$\tilde{q}_j^2$ (rather than the virtualities $|q_j^2|$)
\cite{webber88}
\begin{equation}
\tilde{q}_j^2 \;\equiv E_j^2\;\zeta_{j+1}
\;\;,\;\;\;\;\;\;\;\;\;\;\;
\zeta_{j+1} \;=\;\frac{q_0\cdot k_{j+1}}{\omega_0 \;E_{j+1}}
\;\simeq \;
1 - \cos \theta_{0,\;j+1}
\;\;\;\;\;\;\;\;\;(0 \le j \le n)
\;,
\label{tildep}
\end{equation}
where $q_j=(\omega_j,\vec{q}_j)$ and
$k_{j+1}=(E_{j+1},\vec k_{j+1})$ refer to
the $j^{th}$ branching $a_j \rightarrow a_{j+1} a_{j+1}'$ with
momentum assignment
$q_j\rightarrow q_{j+1} k_{j+1}$ (see Fig. 4).
The space-like cascade is then strictly ordered in the variable
$\tilde{q}_{j+1}^2 > \tilde{q}_j^2$, which is equivalent to the
ordering of emission angles,
$\omega_j \theta_{0, \;j+1^\prime} < \omega_{j+1} \theta_{0, \;j+2^\prime}$.

The space-like cascade terminates with parton $a_n \equiv a$ entering
the vertex of collision with parton $b$, that is, $Q_{ab}^2$ in Fig. 4.
The history of parton $a$ is however not known until after
it has collided with parton $b$, because it is this very collision
that causes the cascade evolution of parton $a$.
Therefore one must reconstruct the cascade {\it backwards} in
time starting from the time of the collisions at the vertex $Q_{ab}^2$ and
trace the history of the struck parton $a$ back to 
the initial state at time $t_0 =0$ at which it was 
originally resolved with $Q_0^2$ in its hadron (nucleon) mother.
The method used here is a space-time generalization of the
`backward evolution scheme'  \cite{backevol}.
To sketch the procedure, consider the space-like branching
$q_{n-1} \rightarrow q_n k_n$ which is closest to the 
collision vertex $Q_{ab}^2$  in Fig. 4.
The virtualities satisfy \cite{bassetto}
$\vert q_n^2 \vert > \vert q_{n-1}^2 \vert$, and $q_{n}^2, q_{n-1}^2 <
0$
(space-like) but $k_n^2 > 0$ (time-like).
The relative probability for a branching to occur  between
$\tilde{q}^2$
and $\tilde{q}^2 + d\tilde{q}^2$ is given by
\begin{eqnarray}
d {\cal P}_{n-1,\,n}^{(S)} ( x_{n-1}, x_{n}, \tilde{q}^2;\,\Delta t)
&=&
\frac{d \tilde{q}^2}{\tilde{q}^2}\, \frac{d z}{z}
\,
\frac{\alpha_s\left((1-z) \tilde{q}^2\right)}{2 \pi} \,
\,
\gamma_{n\mbox{-}1 \rightarrow n n^\prime} (z)
\;\;
\left(
\frac{F(r_{n-1}; x_{n-1}, \tilde{q}^2)}
{F(r_n; x_n, \tilde{q}^2)}
\right)
\;\,{\cal T}^{(S)}(\Delta t)
\;,
\label{PS}
\end{eqnarray}
where $x_j = (q_j)_z / P_z$  ($j=n, n-1$) are the
fractions of longitudinal momentum $P_z$ of the initial mother hadron (nucleon),
with $F(r_j;x_j, \tilde{q}^2)\equiv F(r_j,q_j)$
the corresponding parton distributions
defined by (\ref{F}),
and the variables
\begin{equation}
z \;=\;\frac{E_n}{E_{n-1}} \;\simeq \;\frac{x_n}{x_{n-1}}
\;\;\;,\;\;\;\;\;\;\;\;
1\;-\;z \;=\;\frac{E^\prime_n}{E_{n-1}} \;\simeq \;
\frac{x_{n-1}-x_{n}}{x_{n-1}}
\label{tildez}
\end{equation}
specify the fractional energy or longitudinal
momentum of parton $n$ and $n^\prime$, respectively, taken away from
$n-1$.
The function $\alpha_s/(2\pi)\, \gamma (z)$
is the usual DGLAP branching probability \cite{dok80,dokbook},
with $\gamma (z)$ giving the energy distribution in the variable $z$.
The last factor 
${\cal T}^{(S)}(\Delta t)$
in (\ref{PS}) determines the time interval in the global $cm$-frame,
$\Delta t = t_n - t_{n-1}$, that is associated with the branching
process
$a_{n-1}\rightarrow a_n a_n'$.
It accounts for the formation time of $a_n$ from $a_{n-1}$ on
the basis of the uncertainty principle:
$\Delta t = \Delta \omega/|q_n^2|$, 
$\Delta \omega \simeq  (x_n - x_{n-1})\,P_z$.
A very simple form is taken here, 
\begin{equation}
{\cal T}^{(S)}(\Delta t)
\;=\; \delta\left( \frac{x_n - x_{n-1}}{\vert q_n^2\vert}\, P_z
\;-\; \Delta t\right)
\label{delts}
\;.
\end{equation}
The backwards evolution of the space-like branching
$q_{n-1} \rightarrow q_n + k_n$ is expressed in terms of the
probability that parton $a_{n-1}$ did {\it not} branch between the
lower bound $\tilde{q}_0^2$, given by the initial resolution scale
$Q_0^2$, and
$\tilde{q}_n^2 \equiv \tilde{q}^2 \simeq Q_{ab}^2$.
In that case, parton $n$ can {\it not} originate from this branching,
but must have been produced otherwise or already been present in
the initial parton distributions.
This non-branching probability is given by the
{\it Sudakov form-factor for space-like branchings}:
\begin{equation}
S_n ( x_{n}, \tilde{q}^2, \tilde{q}_0^2;\,\Delta t)
\;=\;
\exp
\left\{
\,-\, \sum_{a}
\,
 \int_{\tilde{q}_0^2}^{\tilde{q}^2}
\,
\int_{z_- (\tilde{q}^{\prime})}^{z_+ (\tilde{q}^{\prime})}
\,
d {\cal P}_{n,\,n-1}^{(S)} ( x_n,  z, \tilde{q}^{\prime 2}; \,\Delta t)
\right\}
\;,
\label{slff2}
\end{equation}
where the sum runs over the possible species $a = g, q, \bar q$
of parton $a_{n-1}$.
The upper limit of the $\tilde{p}^2$-integration is set by
$\tilde{q}^2  \, \lower3pt\hbox{$\buildrel <\over\sim$}\,Q_{ab}^2$,
associated with the collision vertex with parton $b$.
The limits $z_\pm$ are determined by kinematics \cite{webber88}:
$z_-(\tilde{q})= Q_0/\tilde{q}$ and
$z_+(\tilde{q})= 1 - Q_0/\tilde{q}$.

The knowledge of the space-like formfactor 
$S_n ( x_{n}, \tilde{q}^2, \tilde{q}_0^2;\,\Delta t)$
is enough to trace
the evolution of the branching closest to the hard vertex backwards
from $q_n^2$ at  $t=t_n$, the time of collision in the global $cm$-frame,
$q_{n-1}^2$ at $t_{n-1}= t_n- x_n/|q_n^2|\,P_z$.
The next preceding branchings  $q_{n-2}\rightarrow q_{n-1}
k_{n-1}$, etc., are then reconstructed in exactly the same manner
with the replacements $t_n \rightarrow t_{n-1}$,
$x_n \rightarrow x_{n-1}$, $q_n^2 \rightarrow q_{n-1}^2$, and so forth,
until the initial point  $q_0^2$ at $t_0 = 0$ is reached.
\medskip

\subsubsection{Time-like parton evolution}

\noindent
Time-like parton cascades are initiated by secondary partons
that  emerge either from the side-branches of a 
preceding space-like or time-like cascade,
or directly from a scattering or fusion process.
Consider the time-like cascade
initiated by the parton $c$ in the right part of Fig. 4, 
with momentum $k = k_n$.
This parton has been produced in the collision
$a +b \rightarrow c + d$ with a time-like off-shellness 
$k_n^2 \simeq Q_{ab}^2$.

Again an {\it angular-ordered} (rather than virtuality-ordered) 
evolution in space-time {\it and} momentum-space
of the cascade is employed to incorporate interference effects of soft
gluons emitted along the time-like cascade tree of Fig. 4,
$c \equiv c_m \rightarrow c_{m-1} c_{m-1}'$, ..., 
$c_{1} \rightarrow c_0 c_0'$.
In contrast to (\ref{tildep}), the time-like version of the angular
evolution variable  is \cite{webber86}
\begin{equation}
\tilde{k}_j^2 \;\equiv E_j^2\;\xi_{j-1}
\;\;,\;\;\;\;\;\;\;\;\;\;\;
\xi_{j-1} \,=\,
\frac{k_{j-1} \cdot k^\prime_{j-1}}{E_{j-1} E^\prime_{j-1}} \,\simeq \,
1 - \cos \theta_{(j\mbox{-}1), (j\mbox{-}1)^\prime}
\;\;\;\;\;\;\;\;\;(m \ge j \ge 1)
\;.
\label{tildek}
\end{equation}
so that the  time-like cascade can be described by
a $\tilde{k}^2$-ordered (rather than $k^2$-ordered) evolution,
which corresponds to an angular ordering with decreasing emission angles
$\theta_{j, j^\prime} > \theta_{(j\mbox{-}1), (j\mbox{-}1)^\prime}$.

Proceeding analogously to the space-like case (c.f. (\ref{PS})),
the probability $d {\cal P}_{m,\,m-1}^{(T)}$ for the
first branching after the $\gamma q$ vertex,
$k_m \rightarrow k_{m-1} k_{m-1}^\prime$
with $k^2_{m-1}, k^{\prime \,2}_{m-1}$,
is given by the space-time extension \cite{ms39,msrep}
of the usual DGLAP probability distribution \cite{dok80,dokbook},
\begin{equation}
d {\cal P}_{m,\, m-1}^{(T)} (z,\tilde{k}^2;\,\Delta t) \,=\,
\frac{d \tilde{k}^2}{\tilde{k}^2}\, dz
\;
\frac{\alpha_s( \kappa^2 )}{2 \pi} \,
\gamma_{m \rightarrow (m\mbox{-}1),(m\mbox{-}1)^\prime} (z)
\;\,
{\cal T}^{(T)} (\Delta t)\,
\; ,
\label{PT}
\end{equation}
where
${\cal T}^{(T)} (\Delta t)$ is the probabilty that parton $m$ with
virtuality
$k_m^2$ and  corresponding proper lifetime $\tau_m \propto
1/\sqrt{k_m^2}$
 decays within a time interval $\Delta t$,
\begin{equation}
{\cal T}^{(T)} (\Delta t)
\;=\; 1\;  - \;\exp \left( - \frac{\Delta t}{t_m(k)}\right)
\label{deltt}
\;.
\end{equation}
The actual lifetime of the decaying parton $m$ in the global $cm$-frame
is then $t_m(k) = \gamma/\tau_m(k)$, where
$t_q(k) \approx 3 E/(2 \alpha_s k^2)$ for quarks and
$t_g(k) \approx  E/(2 \alpha_s k^2)$ for gluons \cite{ms3}.
As before,
$F_j$ denotes the local density of parton species $j=m,m-1$, and
$\alpha_s/(2\pi) \gamma(z)$ is the DGLAP branching kernel
with energy distribution $\gamma(z)$.
The probability (\ref{PT}) is formulated in terms of the  energy
fractions carried by the daughter partons,
\begin{equation}
z \;=\; \frac{E_{m-1}}{E_m}
\;\;\;,\;\;\;\;\;\;\;\;
1 - z \; = \; \frac{E_{m-1}^\prime}{E_m}
\;\;\; ,
\end{equation}
with the virtuality $k_m$ of  the parton $m$  related to $z$ and $\xi$
through
$k_m^2 = k_{m_1}^2 + k_{m-1}^{\prime \, 2} + 2 E_m^2 z ( 1 - z ) \xi$,
and the argument $\kappa^2$ in the running coupling $\alpha_s$ in
(\ref{PT})
is \cite{webber88}
$\kappa^2= 2 z^2 ( 1 - z )^2  E_m^2  \xi  \simeq k_\perp^2$.

The branching probability (\ref{PT}) determines the distribution
of emitted partons in both coordinate and momentum space, because
the knowledge of four-momentum and lifetime (or $\Delta t$ between
successive branchings)  give the spatial positions of the
partons, if they are assumed to propagate on straight paths between
the vertices.
The probability that parton $m$ does {\it not} branch between
$\tilde{k}^2$ and
a minimum value $\tilde{k}^2_0 \equiv \mu_0^2$
is given by the exponentiation of (\ref{PT}),
yielding the {\it Sudakov form-factor for time-like branchings}:
\begin{equation}
T_m (\tilde{k}^2,\tilde{k}_0^2; \, \Delta t)
\;=\;
\exp
\left\{
\,-\, \int_{\tilde{k}_0^2}^{\tilde{k^2}}
\, \sum_a
\, \int_{z_{-}(\tilde{k}^\prime)}^{z_{+}(\tilde{k}^\prime)}
\;\,
d {\cal P}_{m,\,m-1}^{(T)} (z, \tilde{k}^{\prime\,2};\,\Delta t)
\right\}
\;,
\label{tlff2}
\end{equation}
which is summed over the species $a = g , q, \bar q$
of parton $m-1$.
The integration limits $\tilde{k}_0^2$ and $z_\pm$
are determined by the requirement that the branching must terminate
when the partons enter the non-perturbative regime and begin
to hadronize.
This condition can be parametrized by
the confinement length scale $L_c = O(1 \,fm)$  with
$\tilde{k_0}^2  \, \lower3pt\hbox{$\buildrel >\over\sim$}\, L_c^{-2}
\equiv \mu_0^2$, and
$
z_{+} ( \tilde{k}_m ) =
1 -z_{-} ( \tilde{k}_m ) =
\mu_0/\sqrt{4 \tilde{k}_m^2}
$,
so that for
$z_{+} ( \tilde{k}_0^2)=
z_{-} ( \tilde{k}_0^2) = 1/2$
the phase space for the branching vanishes.

The time-like form factor 
$T_m (\tilde{k}^2,\tilde{k}_0^2; \, \Delta t)$
determines the four-momenta and
positions of the partons of a particular emission vertex
as sketched above for the first branching
from $k_m^2$ at  $t=t_m$, the time of production of parton
$c$ in the global $cm$-frame,
to $k_{m-1}^2$ at $t_{m-1}= t_m+ E_m/|k_m^2|$.
Subsequent branchings are described
completely analogously by replacing $t_m\rightarrow t_{m-1}$,
$k_m^2 \rightarrow k_{m-1}^2$, etc..
Hence $T(\tilde{k}^2,\tilde{k}_0^2; \, \Delta t)$
generates a time-like cascade
as sequential branchings starting from $t=0$ at the
hard vertex forward in time, until the partons eventually hadronize
as discussed below.
\bigskip

\subsection{Cluster formation and hadronization}

In view of lack of knowledge about the details of confinement dynamics 
and the non-perturbative hadronization mechanism, one  must
rely at present on  model-building.
In the present approach, the cluster-hadronization scheme of Ref. 
\cite{ms37,ms40,ms41} is employed.
This phenomenological scheme is inspired by the Marchesini-Webber model 
\cite{webber84}, however it works in space-time plus color-space.
On the other hand it is very different from
commonly used string-fragmentation models such as the Lund-model \cite{string}.

In VNI,
both the cluster-formation from the collection
of quarks and gluons at the end of the perturbative phase
and the subsequent cluster-decay into final hadrons
consist of two components:
\begin{description}
\item[1.]
The recombination of the {\it secondary time-like partons},
their conversion into colorless
{\it parton clusters} and the subsequent
decay into secondary hadrons.
\item[2.]
The recombination of the {\it primary space-like partons} that remained
spectators throughout the collision development into {\it beam clusters}
and the fragmentation of these clusters.
\end{description}

The important assumption here is that the process of hadron formation
depends only on the local space-, time-, and color-structure of the
parton system,
so that the hadronization mechanism can be modelled as the formation of
color-singlet clusters of partons as independent entities
(pre-hadrons),
which subsequently decay into hadrons.
This concept is reminiscent of the `pre-confinement' property
\cite{preconf} of parton evolution,
which is the tendency of the produced partons
to arrange themselves in color-singlet clusters with limited
extension in both position and momentum space, so that it is
suggestive to suppose that these clusters are the basic units out of
which hadrons form.
\medskip

\subsubsection{Cluster formation}

\noindent
{\bf (i)  Parton clusters:}
\smallskip

Parton clusters are formed from secondary partons, i.e. those
that have been produced by the hard interaction and the
parton shower development.
The coalescence of these secondary  partons to color-neutral clusters
has been discussed in detail in Refs. \cite{ms37,ms40}.
Throughout the dynamically-evolving
parton cascade development,  every parton and its nearest
spatial neighbour are considered as as potential candidates for
a 2-parton cluster, which, if {\it color neutral}, plays the role
of a `pre-confined' excitation in the process of hadronization.
Within each single time step, the probability for parton-cluster
conversion is determined for each nearest-neighbor pair by
the requirement that the total color charge of the two partons must
give a composite color-singlet state (if necessary by accompanying
gluon emission), and the condition that
their {\it relative spatial distance} $L$ exceeds the critical
{\it confinement length scale} $L_c$.
The scale $L$ is defined as
the Lorentz-invariant distance $L_{ij}$ between parton $i$
and its nearest neighbor $j$:
\begin{equation}
L\;\equiv\; L_{ij} \;= \;
\mbox{min} (\Delta_{i 1}, \ldots , \Delta_{i j}, \ldots , \Delta_{i n})
\;,
\label{L}
\end{equation}
where $\Delta_{ij} = \sqrt{ (r_{ij})_0^2 + 
(r_{ij})_x^2 + (r_{ij})_y^2 + (r_{ij})_z^2}$, 
with $r_{ij}^\mu = r_i^\mu - r_j^\mu$,
and the probability for the coalescence of the two partons $i$, $j$ to
form
a cluster is modelled by a distribution of the form
\begin{equation}
\Pi_{ij\rightarrow c}\;\propto \; \left(\frac{}{}
1\,-\, \exp\left(-\Delta F\;L_{ij}\right)
\right)
\;\,\simeq \;\,
1\;-\;\exp\left(\frac{L_0-L_{ij}}{L_c-L_{ij}} \right)
\;\;\;\;\;\mbox{if $L_0 \;<\;L_{ij}\;\le\;L_c$}
\;,
\label{Pi2}
\end{equation}
and $\Pi_{ij\rightarrow c}= 0\; (1)$ if $L_{ij} < L_0$ ($L_{ij} > L_c$).
Here $\Delta F$ is the local change in the free energy
of the system that is associated with the
conversion of the partons to clusters,
and the second expression on the right side is a parametrization
in terms of $L_0 = 0.6$ $fm$ and $L_c = 0.8$ $fm$
that define the transition regime.
As studied in Ref. \cite{ms40}, the aforementioned color constraint,
that only colorless 2-parton configurations may produce a cluster,
can be incorporated by allowing
coalescence for any pair of color charges, as determined by the
space-time separation $L_{ij}$ and the probability (\ref{Pi2}),
however, accompanied by the additional emission
of a gluon or quark that carries away any unbalanced net color
in the case that the two coalescing partons are not in a colorless
configuration.
\medskip

\noindent
{\bf (ii)  Beam clusters:}
\smallskip

If one or both of the colliding beam/target particles $A$ and $B$ were
a hadron or a nucleus, the one, respectively two beam clusters
are formed from the spectator partons that represent
the receding beam/target remnants of the original particles $A$ and $B$.
More precisely,
the remaining fraction of the longitudinal momentum and energy that has
not materialized or been redirected 
and harnessed during the course of the collision, is
carried by those primary partons of the initial-state hadrons or nuclei, which
remained spectators throughout. In the present approach
these partons maintain their originally assigned momenta
and their space-like virtualities. Representing the  beam remnants
of $A$ and/or $B$, they
may be pictured as the coherent relics of  the original
hadron (or nucleus) wavefunctions.
Therefore the primary virtual partons must be treated differently
than the secondary partons which are real excitations that contribute
incoherently to the hadron yield.
In the global $cm$-frame,  the primary partons are
grouped together to form a  massive beam cluster with its four-momentum
given by the sum of the parton momenta and its position given by
the 3-vector mean of the  partons' positions.
\bigskip

\subsubsection{Hadronization of clusters}

\noindent
{\bf (i) Parton clusters:}
\smallskip

For  the decay of each parton cluster into final-state hadrons,
the scheme presented in detail in Refs. \cite{ms37}
is employed:
If a cluster is too light to
decay into a pair of hadrons, it is taken to represent
the lightest single meson that corresponds to its
partonic constituents. Otherwise, the cluster
decays isotropically in its rest frame into
a pair of hadrons, either mesons or baryons, whose combined
quantum numbers correspond to its partonic constituents.
The corresponding  decay probability is chosen to be
\begin{equation}
\Pi_{c\rightarrow h}\;=\;
\;{\cal T}_c(E_c,m_c^2) \;
\;\,{\cal N}
\int_{m_h}^{m_c}
\frac{dm}{m^3}\;\exp\left(-\frac{m}{m_0}\right)
\;,
\label{pi3}
\end{equation}
where ${\cal N}$ is a normalization factor,
and the integrand is a Hagedorn spectrum \cite{hagedorn} that
parametrizes quite well
the density of accessible hadronic states below $m_c$ which are
listed in the particle data tables, and $m_0 = m_{\pi}$.
In analogy to (\ref{deltt}), ${\cal T}_c$ is a
life-time factor giving the probability
that a cluster of mass $m_c^2$ decays  within
a time interval $\Delta t$ in the global $cm$-frame,
\begin{equation}
{\cal T}_c(E_c,m_c^2)\;=\; 1\;-\;
\exp\left( - \frac{\Delta t}{t_c(E_c,m_c^2)}\right)
\;,
\label{lft2}
\end{equation}
with the Lorentz-boosted life time
$t_c= \gamma_c \tau_c\simeq E_c/m_c^2$.
In this scheme, a particular cluster
decay mode is obtained from (\ref{pi3})
by summing over all possible decay channels,
weighted with the appropriate spin, flavour, and
phase-space factors, and then choosing the actual decay mode
acording to the relative probabilities of the channels.
\medskip

\noindent
{\bf (ii) Beam clusters:}
\smallskip

The fragmentation of the beam clusters containing the spectator partons
mimics in the present model what is commonly termed the `soft underlying event',
namely, the emergence of those final-state hadrons that are
associated with the non-perturbative physics which underlies the
perturbatively-accessible dynamics of the hard interaction
with parton shower fragmentation.

In the spirit of the Marchesini-Webber model \cite{webber88},  
a version (suitably modified
for the present purposes) of the  soft hadron production model
of the UA5 collaboration \cite{UA5}, which is based on a parametrization
of the CERN $p\bar p$ collider data for minimum-bias hadronic
collisions.
The parameters involved in this model are set to give a good
agreement with those data.

Soft hadron production is known to be a universal mechanism
\cite{Mdiff} that is
common to all high-energy collisions that involve beam hadrons in the
initial state,
and that depends essentially on the total energy-momentum of
the fragmenting final-state beam remnant.
Accordingly, one may assume that the fragmentation of the final-state
beam cluster depends solely on its invariant mass $M$, and that it
produces a charged-
particle multiplicity with a binomial distribution \cite{UA5},
\begin{equation}
P(n) \;=\; \frac{\Gamma(n+k)}{n! \Gamma(k)} \;
\frac{ (\overline{n}/k)^n}{(1 + \overline{n}/k)^{n+k}}
\label{ndist}
\;,
\end{equation}
where the mean charged multiplicity
$\overline{n}\equiv\overline{n}(M^2)$
and the parameter $k\equiv k(M^2)$
depends on the invariant cluster mass
\footnote{
Notice that in this model $M$ fluctuates statistically, as a result
of fluctuations of the initial-state parton configuration in
the incoming hadrons (or nuclei), as well as
due to the fluctuating number of remnant partons during the space-time
evolution.
Hence the distribution (\ref{ndist}) and the mean multiplicity
(\ref{nk})
vary from event to event. This is in contrast to the original UA5 model,
in which the fixed beam energy $\sqrt{s}/2$ controls the energy
dependence of
soft hadron production.
}
according to the
following particle data parametrization \cite{UA5},
\begin{equation}
\overline{n}(M^2) \;=\;  10.68 \; (M^2)^{0.115} \;-\; 9.5
\;\;\;\;\;\;\;\;\;
k(M^2) \;=\;  0.029 \; \ln(M^2) \;-\; 0.064
\label{nk}
\;.
\end{equation}
Adopting the scheme of Marchesini and Webber \cite{webber88}, the
fragmentation of a beam cluster of mass $M$ proceeds then as
follows:
First, a particle multiplicity $n$ is chosen from (\ref{ndist}), and the
actual charged particle multiplicity is taken to be $n$ plus the
modulus of the beam cluster charge.
Next, the beam cluster is split into
sub-clusters $(q_1 \bar{q}_2), (q_2 \bar{q}_3), \ldots$ ($q_i = u, d$),
which are
subsequently hadronized in the beam cluster rest frame,
in the same way as the parton clusters described above.
To determine the sub-cluster momenta,  the following mass distribution
is assumed,
\begin{equation}
P(M) \;=\; c \; (M-1) \; \exp \left[ -a (M-1) \right]
\label{mdist}
\;,
\end{equation}
with $c$ a normalization constant and $a = 2$ GeV$^{-1}$, resulting
in an average value of $\langle M \rangle \approx 1.5$ GeV.
The transverse momenta are taken from the distibution
\begin{equation}
P(p_\perp) \;=\; c^\prime \; p_\perp \;
\exp \left[ -b \sqrt{p_\perp^2 +M^2}\right]
\label{pTdist}
\;,
\end{equation}
with normalization $c^\prime$ and slope parameter
$b = 3$ GeV$^{-1}$, and
the rapidities $y$ are drawn from a simple flat distribution
$P(y) \propto const.$ with an extent of
0.6 units  and Gaussian tails with 1 unit standard deviation at
the ends.
Finally, all  hadronization products of the sub-clusters are
boosted from the rest frame of the original beam cluster back into the
global $cm$-frame.
\bigskip
\bigskip

\section{PROGRAM DESCRIPTION}
\label{sec:section3}
\bigskip

\subsection{The package VNI-3.1}

The  program package VNI-3.1
is a completely new written code, using only fragments of its precedessors 
VNI-1.0 (Ref. \cite{ms0}) and VNI-2.0 (Ref. \cite{ms3}). 
The current program is a 
first release of a long term project that aims at unifying the simulation 
of a wide class of high energy QCD processes, ranging from elementary 
particle physics processes, e.g., $e^+e^-$-fragmentation, deep inelastic 
$ep$-scattering, via hadronic collisions such as $pp$ ($p\bar{p}$), 
to reactions involving (heavy) nuclei, 
e.g. deep inelastic $eA$ scattering, $pA$, or $AA$ ($AB$) collisions.
The types of collision processes that are available 
and are discussed in more detail
in Section 3.4 below, include
collisions involving
$e^\pm,\mu^\pm$ for leptons, $p,n,\pi,\Lambda,\Sigma\ldots$ for hadrons,
and
$\alpha, \H\e, {\rm O}, \ldots,
\A\u, {\rm P}\b, {\rm U}$ for nuclei.
Some of these collision processes have of course been 
investigated before; here 
they are revisited within the space-time picture of 
parton cascades and cluster dynamics in order to provide an alternative 
method for Monte Carlo simulation and also to check the consistency with 
the wide literature on both experimental measurements and theoretical 
predictions. 

\subsection{Overall structure of the program}

  The program package VNI is written entirely in Fortran 77, and should 
run on any machine with such a compiler.  The program VNI adopts the 
common block structure, particle identification codes, etc.,
from the Monte Carlos of the Lund family, e.g.,
the JETSET/PYTHIA programs \cite{jetset}. 
In addition, various program elements 
of the latter are used in the sense of a library, 
however modified to incorporate 
the additional aspects of the space-time and color label information. 
VNI also employs certain (significantly modified) parts of the 
HERWIG Monte Carlo \cite{herwig}, used for the 
final decays of clusters into hadronic resonances and stable particles.
It is important to stress that the correspondence of 
VNI with JETSET/PYTHIA and HERWIG is rather loose and should not 
lead to naive identification, in particular since VNI is taylored for a 
larger set of degrees of freedom that is necessary to describe the 
space-time dynamics and the information on color correlations. 
Nevertheless, by truncating the additional information of the particle 
record of VNI, a direct interface to the particle records of the Lund 
Monte Carlo JETSET/PYTHIA, and to the program HERWIG, is actually 
straightforward. Instructions are given below in  Section 3.10.
\smallskip

  The package VNI-3.1 consists of the following files:

\begin{description}
\item{(i)}  the documentation {\bf vni-3.1.ps} (this one)
\item{(ii)}  the program VNI-3.1 in 7 parts: {\bf vni1.f -- vni7.f}
\item{(iii)}  two include files {\bf vni1.inc} and {\bf vni2.inc}
\item{(iv)}   two example programs {\bf vnixpl1.f} and {\bf vnixpl2.f }.
\item{(vi)}   a simple, portable histogram package {\bf vnibook.f }.
\end{description}

  The seven components {\it vni1.f} $-$ {\it vni7.f} of Fortran source code
contain all the subroutines, functions, and block data, as categorized
above. All default settings and particle data are automatically loaded 
by including the two include files {\it vni1.inc} and {\it vni2.inc}.
The example programs {\it vnixpl1.f} (a generic one) and {\it vnixpl2.f} 
(a more application-oriented one), as well as the histogram package
{\it vnibook.f} are not integral part of the program, i.e. are 
distributed supplementary as useful guiding tools.
\medskip

  The program source code is organized in 7 parts, in which subroutines
and functions that are related in their performance duties, are grouped 
together
(a summary list of all components with a brief description of their 
purpose is given in Appendix E):

\begin{description}
\item[vni1.f:]
            Main steering routines VNIXRIN (initialization), VNIXRUN 
           (event generation), and VNIXFIN (finalization), plus other
	    general-duty subroutines.
           VNIXRIN and VNIXFIN are only to be called once, while
	    VNIXRUN
           is to be called anew for each event, in which the 
           particle system is evolved in discrete time steps and 
           7-dimensional phase-space.
\item[vni2.f:]
           Initialization routines. They include, e.g. the process
           dependent event-by-event initialization of the chosen type 
           of collision process, and various other slave routines 
           associated with the initial set up of an event.
\item[vni3.f:]
           Package for parton distributions (structure functions). A
           portable and self-contained collection of routines with
           a number of structure function parametrizations, plus a
           possible interface to the CERN PDFLIB, plus various slave 
           routines. 
\item[vni4.f:]
           Evolution routines. These form the heart of the parton cascade/
           cluster-hadronization simulation, and include 
           both (process-dependent) 
           particle evolution routines and  (process-independent)
           universal routines for space-like and time-like shower evolution, for
           parton scatterings, parton recombination to clusters, and 
           cluster decays to hadrons and other final state particles.
\item[vni5.f:]
           Diverse utility routines and functions performing the 
           slave-work for the part of perturbative parton evolution.
\item[vni6.f:]
           Diverse utility routines and functions performing the 
           slave-work for the part of cluster fragmentation to hadrons.
\item[vni7.f:]
           Event study and analysis routines for accumulating, analysing
           and printing out statistics on observables.
\end{description}
\medskip

\subsection{Special features and machine dependence}

Important to remark are the following special features, being 
not strictly common in standard Fortran 77:

\begin{description}
\item[1.]
  The program  uses  of a {\it machine-dependent timing routine},
called $VNICLOC$, which measures the elapsed CPU time 
\footnote{
The usage of the time measurement is not 
necessary for the performance of the program as-a-whole, although it 
is a nice convenience.}.
Two alternatives are provided: 
a) the $TIMEX$ routine of the CERN library, in which case the 
   latter must be linked with the program, or,
b) the $MCLOCK$ routine which is specific to IBM-AIX compilers. 
If neither of these options are desired, or does not work in the users 
local environment, the lines involving 
calls for $TIMEX$ ($MCLOCK$) calls can easily commented out.
\footnote{
The subroutine $VNICLOC$ is listed as the very first one in {\it vni1.f}.
}.
\item[2.]
All common block variables and dimensions are defined globally within
the {\it include files  vni1.inc and vni2.inc}.
This greatly simplifies for instance possible modifications 
(e.g., enlarging or decreasing) of global common block arrays.
Consequently, all subroutines contain INCLUDE statements for these include
files, and therefore
when compiling and linking the program, the files 
{\it vni1.inc} and {\it vni2.inc} must be properly accessible.
\item[3.]
  {\it Single precision} is normally assumed throughout, since most 
machines are 32 bit processors. For applications at extremely high 
energies, single precision for any real variable starts to become a 
problem. The main source of numerical precision-loss arises from 
multiple Lorentz boosts of particles throughout the simulation. 
Therefore the relevant parts of the code are  written in 
double precision. All double precision variables have as first 
character D.  As additional precaution, particle energies are recalculated 
from momentum and mass after each time step during the evolution of 
an event. 
\item[4.]
{\it Implicit variable-type assignment} is assumed, such that variables
or functions beginning with letters I - M are of {\it integer}$\ast$4
type, and variables or functions with first letters A - H and O - Z
are of type {\it real}$\ast$4 (single precision), except for
occasional {\it real}$\ast$8 (double precision) occurrences
as stated in item 3.
\item[5.]
{\it  SAVE statements} have been included in accordance with the Fortran
standard. Since most ordinary machines take SAVE for granted, this
part is not particularly well tried out, however. Users on machines
without automatic SAVE are therefore warned to be on the lookout for
any variables which may have been missed.
\end{description}
\medskip

\subsection{The main subroutines}

  There is a minimum of three routines that a user must know: 
\begin{description}
\item[1.] VNIXRIN for the overall initialization of a sample of 
          collision events, 
\item[2.] VNIXRUN for the subsequent actual generation of each event,
\item[3.] VNIXFIN for finishing up simulation.
\end{description}
These three routines, which are briefly descibed below, 
are to be called in that order by a user's program as is exemplified by the
example programs {\it vnixpl1.f} and {\it vnixpl2.f} 
(see also Section 3.10).
\smallskip


\begin{verbatim}
SUBROUTINE VNIXRIN(NEVT,TFIN,FRAME,BEAM,TARGET,WIN)
\end{verbatim}

\noindent
{\it Purpose}: to initialize the overall simulation procedure.
\medskip

\begin{description}
\item[{\boldmath $NEVT$} :] 
    integer specifying the number of collision events to be 
    simulated for a selected physics process.
\item[{\boldmath $TFIN$} :] 
    final time (in $fm$) in the global Lorentz frame up to which each
    collision event is followed.
\item[{\boldmath $FRAME$} :] 
    a character variable used to specify the global Lorentz frame  of the
    experiment. Uppercase and lowercase letters may be freely mixed.
\\
{\bf = 'CMS' :} colliding beam experiment in CM frame, with beam momentum
        in +$z$ direction and target momentum in $-z$ direction.
\\
{\bf = 'FIXT':} fixed target experiment, with beam particle momentum
        pointing in $+z$ direction.
\\
{\bf = 'USER' :} full freedom to specify frame by giving beam
momentum in \tt{PSYS(1,1)}, \tt{PSYS(1,2)} and \tt{PSYS(1,3)} and target
momentum in \tt{PSYS(2,1)}, \tt{PSYS(2,2)} and \tt{PSYS(2,3)} in
common block \tt{VNIREC}. Particles are assumed on the mass shell,
and energies are calculated accordingly.
\\
{\bf = 'FOUR' :} as \tt{'USER'}, except also energies should be
specified, in \tt{PSYS(1,4)} and \tt{PSYS(2,4)}, respectively. The
particles need not be on the mass shell; effective masses are 
calculated from energy and momentum. (But note that numerical
precision may suffer; if you know the masses the option \ttt{'FIVE'}
below is preferrable.)
\\
{\bf = 'FIVE' :} as \tt{'USER'}, except also energies and masses
should be specified, i.e the full momentum information in
\tt{PSYS(1,1) - PSYS(1,5)} and \tt{PSYS(2,1) - PSYS(2,5)} should be given for
beam and target, respectively. Particles need not be on the mass
shell. Space-like virtualities should be stored as $-\sqrt{-m^2}$.
Four-momentum and mass information must match. 
\item[{\boldmath $BEAM$, }] {\boldmath $TARGET$} : 
     character variables to specify beam and target particles.
    Uppercase and lowercase letters may be freely mixed. An
    antiparticle may be denoted either by "$\tilde{}$" (tilde) or 
    "$\bar{}$" (bar) at the end of
    the name. It is also possible to leave out the charge for neutron
    and proton. 
\\
{\bf = 'e--'   :} electron $e^-$.
\\
{\bf = 'e+'   :} positron $e^+$.
\\
{\bf = 'mu--'  :} muon $\mu^-$.
\\
{\bf = 'mu+'  :} antimuon $\mu^+$.
\\
{\bf = 'gamma':} real photon $\gamma$ (not yet implemented).
\\
{\bf = 'pi+'  :} positive pion $\pi^+$.
\\
{\bf = 'pi--'  :} negative pion $\Pi^-$.
\\
{\bf = 'pi0'  :} neutral pion $\pi^0$.
\\
{\bf = 'n0'   :} neutron $n$.
\\
{\bf = 'n{\boldmath $\tilde{}$}0'  :} antineutron $\overline{n}$.
\\
{\bf = 'p+'   :} proton $p$.
\\
{\bf = 'p{\boldmath $\tilde{}$}--'  :} antiproton $\overline{p}$.
\\
{\bf = 'Lambda0':} $\Lambda$ baryon.
\\
{\bf = 'Sigma+' :} $\Sigma^+$ baryon.
\\
{\bf = 'Sigma--' :} $\Sigma^-$ baryon.
\\
{\bf = 'Sigma0' :} $\Sigma^0$ baryon.
\\
{\bf = 'Xi--'   :} $\Xi^-$ baryon.
\\
{\bf = 'Xi0'   :} $\Xi^0$ baryon.
\\
{\bf = 'Omega--':} $\Omega^-$ baryon.
\\
{\bf = 'Alpha' :} $^2_{1}\alpha$ particle.
\\
{\bf = 'He' :}  $^4_2\H\e$ (Helium) nucleus.
\\
{\bf = 'Ox' :}  $^{16}_8{\rm O}$ (Oxygen) nucleus.
\\
{\bf = 'Su' :}  $^{32}_{16}{\rm S}$ (Sulfur) nucleus.
\\
{\bf = 'Ag' :}  $^{108}_{47}\A\g$ (Silver) nucleus.
\\
{\bf = 'Au' :}  $^{197}_{79}\A\u$ (Gold) nucleus.
\\
{\bf = 'Pb' :}  $^{208}_{82}{\rm P}\b$ (Lead) nucleus.
\\
{\bf = 'Ur' :}  $^{238}_{92}{\rm U}$ (Uranium) nucleus.
\item[{\boldmath $WIN$} :] 
    related to energy of system (in GeV). Exact meaning depends on FRAME, 
    i.e. for
    FRAME='CMS' it is the total energy $\sqrt{s}$ of system, and for
    FRAME='FIXT' it is the  absolute 3-momentum $P = \sqrt{\vec{P}^2}$ of 
    beam particle.
\end{description}
\bigskip

\begin{verbatim}
SUBROUTINE VNIXRUN(IEVT,TRUN)
\end{verbatim}
\medskip

\noindent
{\it Purpose}: to generate one event of the type specified by the VNIXRIN. 
    This is the main routine, which administers the overall
    run of the event generation and calls a number of other routines 
    for specific tasks.

\begin{description}
\item[{\boldmath $IEVT$} :] 
    integer labeling the current event number.
\item[{\boldmath $TRUN$} :] 
    time (in $fm$) $TRUN\le TFIN$ in the global Lorentz frame up to which 
    the space time evolution
    of the current event $IEVT$ is carried out. If $TRUN < TFIN$, then
    the particle record contains the history up to this time,
    and repeated calls resume from $TRUN$ of the previous call 
    (c.f. Section 3.10).
\end{description}
\medskip

\begin{verbatim}
SUBROUTINE VNIXFIN()
\end{verbatim}

\noindent
{\it Purpose}:
    to finish up overall simulation procedure, to write data 
    analysis results to files, evaluate CPU time, close files, etc..
    This routine must be called after the last event has been
    finished (c.f. Section 3.10).

\bigskip
\bigskip

For a deeper understanding of the 
physics routines and their connecting structure,
I refer to Appendix A for a brief description.
Generally, for 
each of the included collision processes, there is an initialization 
routine that sets up the initial state, and an evolution 
routine  that carries out the time evolution of the particle 
distributions in phase-space, starting from the initial state. 
Within the evolution routines, a number of 
universal slave routines (i.e. independent of the process under consideration)
perform the perturbative parton evolution in terms of space-like branchings, 
time-like branchings, and parton collisions, as well as
the cluster hadronization. 
\bigskip

  Finally, the program provides some useful
pre-programmed event-analysis routines which collect, analyze and print out information
on the result of a simulation, and give the user immediate access
to calculated particle data, spectra, and other observables. 
A more detailed description of their duties, and how
to call them can be found in Appendix B.
\newpage

\begin{table}[ptb]
\captive{Available collision processes.
\protect\label{t:ipro} } \\
\vspace{1ex}
\begin{center}
\begin{tabular}{|rc|cl|}
\hline\hline
    IPRO  &&&             type of collision process
\\
\hline\hline
&&& \\
     1    &&&       $l^+l^- \rightarrow  \gamma/Z_0  \rightarrow 2-jets \rightarrow hadrons$
\\
     2    &&&       $l^+l^- \rightarrow Z_0 \rightarrow W^+W^- \rightarrow 4-jets \rightarrow hadrons \;\;\;\;\;\;$
\\
     3    &&&       $\gamma + h \rightarrow jets \rightarrow hadrons$      
\\
     4    &&&       $\gamma + A \rightarrow jets \rightarrow hadrons$     
\\
     5    &&&       $l + h  \rightarrow  l + jets + X  \rightarrow  hadrons$ 
\\
     6    &&&       $l + A  \rightarrow  l + jets + X  \rightarrow  hadrons$ 
\\
     7    &&&       $h + l  \rightarrow  l + jets + X  \rightarrow  hadrons$ 
\\
     8    &&&       $h + h'  \rightarrow  jets + X  \rightarrow  hadrons$ 
\\
     9    &&&       $h + A  \rightarrow  jets + X  \rightarrow  hadrons$ 
\\
    10    &&&       $A + l  \rightarrow  l + jets + X  \rightarrow  hadrons$
\\
    11    &&&       $A + h  \rightarrow  jets + X  \rightarrow  hadrons$ 
\\
    12    &&&       $A + A'  \rightarrow  jets + X  \rightarrow  hadrons$ 
\\ &&& 
\\ \hline\hline
\end{tabular}
\end{center}
\end{table}

\begin{table}[ptb]
\captive{Available beam and target particles.
\protect\label{t:bt} } \\
\vspace{1ex}
\begin{center}
\begin{tabular}{|c|c|c|}
\hline\hline
  $\;\;\;$ leptons $l$ $\;\;\;$ & $\;\;\;$ 
hadrons $h$ $\;\;\;$ & $\;\;\;$ nuclei $A$ $\;\;\;$
\\
\hline\hline
&& \\
$ e^\pm $ & $\pi^\pm, \pi^0$ &  $^2_{1}\alpha$
\\
$ \mu^\pm $ & $n, \overline{n}$ &  $^4_2\H\e$ 
\\
$  $ & $p, \overline{p}$ &  $^{16}_8{\rm O}$
\\
$  $ & $\Lambda^0$ &  $^{32}_{16}{\rm S}$
\\
$  $ & $\Sigma^\pm, \Sigma^0$ &  $^{108}_{47}\A\g$
\\
$  $ & $\Xi^-, \Xi^0$ &  $^{197}_{79}\A\u$ 
\\
$  $ & $\Omega^-$ &  $^{208}_{82}{\rm P}\b$ 
\\
$  $ & $  $  &  $^{238}_{92}{\rm U}$ 
\\ && 
\\ \hline\hline
\end{tabular}
\end{center}
\end{table}

\subsection{The physics processes}

The program is structured to incorporate different classes of particle 
collision processes in a modular manner. The 
general classes that are part of the program are summarized in
Table 1.
All of the generic collision processes in Table 1
 allow a further specification 
of beam and/or target particle, .e.g., $l = e, \mu$ for leptons, 
$h = p, n, \pi, \ldots$ for hadrons, or $A =\alpha, He, O, 
\ldots$  for nuclei. They are summarized in Table 2.
\medskip

As explained before, the
actual evolution of any beam/target-particle collision
is simulated
on microscopic level of the system of partons, clusters and hadrons, and 
their interactions. For the parton interaction processes, a selection 
of elementary hard/soft $2\rightarrow 2$ scatterings and $2\rightarrow 1$ 
fusions are included 
in VNI, and future extensions are planned. In addition 
there are the standard space-like and time-like $1\rightarrow 2$ branching 
processes, which, when combined with the elementary tree processes, 
provide the usual parton cascade method of including
higher order, real and virtual, corrections to the 
elementary tree processes. The parton-cluster
formation and hadronization scheme includes a number of 2-parton
recombination processes and the decay of the formed clusters into
final state hadron. 

\begin{table}[ptb]
\captive{Elementary partonic subprocesses.
\protect\label{t:isub} } \\
\vspace{1ex}
\begin{center}
\begin{tabular}{|lc|cc|}
\hline\hline
$ \;\;\;$ Class &    ISUB $\;\;\;\;\;$ &&             type of subprocess $\;\;\;\;$
\\
\hline\hline
&&& \\
  a) $2 \rightarrow> 2$ : 
&                  1  &&  $ q_i q_j \rightarrow  q_i q_j$
\\
&                  2  &&  $ q_i \bar{q}_i \rightarrow  q_k \bar{q}_k    $    
\\
&                  3  &&  $ q_i \bar{q}_i \rightarrow  g g         $   
\\
&                  4  &&  $ q_i \bar{q}_i \rightarrow  g \gamma     $  
\\
&                  5  &&  $ q_i \bar{q}_i \rightarrow  \gamma \gamma $ 
\\
&                  6  &&  $ q_i g \rightarrow  q_i g         $ 
\\
&                  7  &&  $ q_i g \rightarrow  q_i \gamma    $ 
\\
&                  8  &&  $ g g  \rightarrow  q_k \bar{q}_k      $ 
\\
&                  9  &&  $ g g  \rightarrow  g g          $ 
\\
&                 10  &&   soft  scattering 
\\
&&& \\
    b) $2 \rightarrow 1$ :
&                 11  &&  $ q_i \bar{q}_i \rightarrow  g^\ast$
\\
&                 12  &&  $ q_i g \rightarrow  q_i^\ast$
\\
&                 13  &&  $ g g  \rightarrow  g^\ast$
\\
&&& \\
    c) $1 \rightarrow 2$ (space-like) :
&                 21  &&  $ g^\ast \rightarrow  q_i q_i$
\\
&                 22  &&  $ q_i^\ast \rightarrow  q_i g$
\\
&                 23  &&  $ g^\ast \rightarrow  g g$
\\
&&& \\
    d) $1 \rightarrow 2$ (time-like) :
&                 31  &&  $ g \rightarrow  q_i \bar{q}_i$
\\
&                 32  &&  $ q_i \rightarrow  q_i g$
\\
&                 33  &&  $ g \rightarrow  g g$
\\ 
\hline\hline
\end{tabular}
\end{center}
\end{table}
 
It is possible to select a combination of partonic subprocesses to 
simulate. For this purpose, all subprocesses are numbered according
to an $ISUB$ code. The list of allowed codes is given below.
In the following $g$ denotes a gluon, $q_i$ represents a quark of 
flavour $i=1,\ldots,n_f$, i.e. for $n_f=6$ this translates
 to $d, u, s, c, b, t$. 
A corresponding antiquark 
is denoted $\bar{q}_i$. The notation $\gamma$ is for a real photon, i.e. on shell.  
An asterix $\ast$ denotes an off-shell parton.
\bigskip

\subsection{The particle record}

  Each new event generated is in its entity stored in the commonblock
VNIREC, which thus forms the event record. Here each particle that
appears at some stage of the time evolution of the system, will
occupy one line in the matrices. The different components of this line
will tell which particle it is, from where it originates, its
present status (fragmented/decayed or not), its momentum, energy and
mass, and the space-time position of its production vertex.
The structure of the particle record VNIREC follows closely the
one of the Lund Monte carlos \cite{jetset,ariadne,lepto},
employing the same overall particle classification of
particle identification, status code, etc.,
however with important differences
concerning color- and space-time
degrees of freedom..

  Note: When in the following reference is made to certain switches 
and parameters MSTV or PARV, these are described in Sec. 3.7 below. 

\begin{verbatim}
     PARAMETER (NV=100000)
     COMMON/VNIREC/N,K(NV,5),L(NV,2),P(NV,5),R(NV,5),V(NV,5)
\end{verbatim}

\noindent
{\it Purpose}: 
Contains the event record, i.e. the complete list of all 
         partons and particles in the current event.
\medskip

\begin{description}
\item[{\boldmath $N$} :]  number of lines in the $K$, $L$, $P$, $R$, and $V$ 
    arrays occupied by the current
    event. $N$ is continuously updated as the definition of the original
    configuration and the treatment of parton cascading, cluster fragmentation
    and hadron production proceed.
    In the following, the individual parton/particle number, running
    between 1 and $N$, is called $I$.
    The maximum $N$ is limited by the dimension $NV$ of the arrays.
\item[{\boldmath $K(I,$}1) :] status code ($KS$), which gives the current status of the
    parton/particle stored in the line. The ground rule is that codes
    1 - 10 correspond to currently existing partons/particles, while
    larger codes contain partons/particles which no longer exist.
\\
{\bf = 0 :} empty line.
\\
{\bf = 1 :} an undecayed particle or an unfragmented parton.
\\
{\bf = 2 :} an unfragmented parton, which is followed by more partons in 
        the same color-singlet parton subsystem.
\\
{\bf = 3 :} an unfragmented parton with special color-flow information
        stored in $K(I,4)$ and $K(I,5)$, such that color connected partons 
        need not follow after each other in the event record.
\\
{\bf = 6 :} a final state hadron resulting from decay of clusters
       that formed from materialized (interacted) partons.
\\
{\bf = 7 :} a final state hadron resulting from soft cluster decay of 
     beam/target remnants formed from left-over (non-interacted)
     initial state partons.
\\
{\bf = 11 :} a decayed particle or a fragmented parton, c.f. = 1.
\\
{\bf = 12 :} a fragmented jet, which is followed by more partons in the 
        same color-singlet parton subsystem, c.f. = 2.
\\
{\bf = 13 :} a parton which has been removed when special color-flow
        has been used to rearrange a parton subsystem, cf. = 3.
\\
{\bf = 16 :} an unstable and removed hadron resulting from decay of clusters
       that formed from materialized (interacted) partons, c.f. = 6.
\\
{\bf = 17 :} an unstable and removed hadron 
     resulting from soft cluster decay of 
     beam/target remnants formed from left-over (non-interacted)
     initial state partons, c.f. = 7.
\\
{\bf {\boldmath $<$} 0 :}
     Particle entries with negative status codes (e.g. -1, -6, -7) parallel
     in their meaning those from above, but refer to particles
     which are only virtually present and become active only after
     a certain formation time, upon which their status code is set to
     the appropriate positive value.
\\
\item[{\boldmath $K(I,$}2) :] 
    Flavor code ($KF$) for partons, hadrons and electromagnetic 
    particles included in the current version of VNI.  The particle code 
    is summarized in Tables 3 -- 9. It is based on
    the flavor and spin classification of particles, following the 1988
    Particle Data Group numbering conventions. Nuclei are attributed
    a non-standard code that is internal to VNI, given by 
    $KF = 1000 \cdot N_{neutrons} + N_{protons}$. 
    A negative $KF$ code, where existing, 
    always corresponds to the antiparticle of the one listed in the Tables
    3 -- 9.
\item[{\boldmath $K(I,$}3) :] 
    line number of parent particle or jet, where known, else 0.
    Note that the assignment of a particle to a given jet of a jet
    system is unphysical, and what is given there is only related to
    the way the event was generated.
\item[{\boldmath $K(I,$}4) :] 
    Special color-flow information (for internal use only) of the 
    form $K(I,4) =$ MSTV(2)$\cdot  ICFR + ICTO$, where $ICFR$ and $ICTO$ give 
    the line numbers of the partons {\it from}
     which the color comes and {\it to} where it goes, respectively. 
\item[{\boldmath $K(I,$}5) :] 
    Special color-flow information (for internal use only) of 
    the form $K(I,5) =$ MSTV(2)$\cdot JCFR + JCTO$, where 
    $JCFR$ and $JCTO$ give 
    the line numbers of the partons {\it from} which the anticolor comes 
    and {\it to} where it goes, respectively. 
\\
\item[{\boldmath $L(I,$}1) :] 
    color label ($L=1,\ldots, NC$) of a parton, where $NC$ is number of
    colors specified by the parameter MSTV(5). Is = 0 for antiquarks 
    and all non-colored particles. 
\item[{\boldmath $L(I,$}2) :] 
    anticolor label ($L=1,\ldots, NC$) of a parton. Is = 0 for quarks 
    and all non-colored particles.
\\
\item[{\boldmath $P(I,$}1) :] 
    $p_x$, momentum in the $x$ direction in GeV.
\item[{\boldmath $P(I,$}2) :] 
     $p_y$, momentum in the $y$ direction in GeV.
\item[{\boldmath $P(I,$}3) :] 
     $p_z$, momentum in the $z$ direction in GeV.
\item[{\boldmath $P(I,$}4) :] 
     $E$, energy, in GeV.
\item[{\boldmath $P(I,$}5) :] 
     $m = p^2$, mass in GeV. For partons with space-like virtualities, 
    i.e. where $Q^2 = - m^2 > 0$, its value is 
    $P(I,5) = -\sqrt{|m^2|} = -Q$.
\\
\item[{\boldmath $R(I,$}1) :] 
    current $x$ position in frame of reference in 1/GeV.
\item[{\boldmath $R(I,$}2) :] 
    current $y$ position in frame of reference in 1/GeV.
\item[{\boldmath $R(I,$}3) :] 
    current $z$ position in frame of reference in 1/GeV.
\item[{\boldmath $R(I,$}4) :] 
    time (in 1/GeV) of active presence in the system, 
    i.e. $t_{pres}= t - t_{prod}$. 
    Is set equal to PARW(2) at time of production $t_{prod}$ of the
    particle and then increased
    by PARW(2) in each timestep until the particle fragments or decays.
    If $ \langle$ 0, giving the remaining formation time for virtually produced,
    but not yet real particles. 
\item[{\boldmath $R(I,$}5) :] 
    number and type of interactions of a particle, and is equal to
    MSTV(3)$^2\cdot NCO +$ MSTV(3)$\cdot NSB + NTB$, 
    where -- for partons -- $NCO$ is 
    the number of 2-body collisions, $NSB$ of space-like branchings, and 
    $NTB$ of time-like branchings, undergone up to current time. For 
    clusters, hadrons and unstable particles, $NSB$ is zero and $NTB$ 
    counts the number of 2-body decays.
\\
\item[{\boldmath $V(I,$}1) :] 
    $x$ position of production vertex, in 1/GeV.
\item[{\boldmath $V(I,$}2) :] 
    $y$ position of production vertex, in 1/GeV.
\item[{\boldmath $V(I,$}3) :] 
    $z$ position of production vertex, in 1/GeV.
\item[{\boldmath $V(I,$}4) :] 
    time $t$ of production, in 1/GeV.
\item[{\boldmath $V(I,$}5) :] 
    encodes origin of production as 
    $10\cdot$MSTV(4)$\cdot IMO + 10\cdot ITM + OR$, 
    where $IMO$ is the direct mother, $ITM$ the time of production, and $OR$ 
    the generation of particle $IP$ in a cascade tree, i.e. the genetical 
    origin with respect to the production vertex of its original 
    `Ur'-mother. (This keeps track of the production history even if the 
    decayed mother has been removed and allows to follow the genealogy
    of the system.)
\end{description}
\medskip

\noindent
{\bf  Special cases:} 
\smallskip

  For the simulation of collisions involving hadrons or nuclei in 
the initial state, additional initial state information on the 
status and specific variables of "primary partons" of the incoming 
hadron (nucleus) are stored as follows. Note however, that after
the first interaction of a "primary parton", this initial state
information is lost and the assignments for the components of 
the arrays $K,R,V$ are the same as above.
\begin{description}
\item[{\boldmath $K(I,$}4) :] 
    $KF$ flavour code of nucleus from which a primary parton 
    originates.
\item[{\boldmath $K(I,$}5) :] 
    $KF$ flavour code of hadron (nucleon), which is mother of 
    the primary parton. (For hadrons $K(I,4)$ and $K(I,5)$ are equal). 
\item[{\boldmath $R(I,$}5) :] 
    Is equal to $-|x|$, where $x$ is the longitudinal momentum 
    fraction of primary parton that it carries off its mother hadron. 
\item[{\boldmath $V(I,$}4) :] 
    encodes location $IB$ in the particle record of a primary 
    sea-quark's brother, i.e. the sea-antiquark belonging to the same
    vacuum polarization loop,  as 
    MSTV(4)$^2\cdot IB + $MSTV(4)$\cdot ITM + OR$,
    analogous to  $V(I,5)$ above. For primary valence quarks and gluons 
    it is 0.
\end{description}
{\it  Additional remark}: The arrays $K(I,3)-K(I,5)$ and the arrays $P, R$, 
and $V$ may temporarily take special meaning other than the above, 
for some specific internal use.
\bigskip

In Section 3.10 below, a typical event listing of the
particle record is printed, which 
serves as an example for the organization of the 
particle record, exhibiting the information for 
{\it momentum space}:
It lists the particles 
contained in VNIREC at a certain point during the evolution of an event, 
where $KS$ and $KF$ give status and flavor codes, $C$ and $A$ the color and 
anticolor index, $P$ the energy-momentum-mass in GeV
for each particle that appeared at some point in the event history.
An analogous  listing for  the {\it position space}
particle record (not printed here) contains the spatial
coordinates and time of active presence for each particle, as well as 
the particle's rapidity and transverse momentum
with respect to the $z$-axis (jet-axis, or beam-axis).
Note: both listings 
can  be obtained at any point by calling the routine VNILIST,
described in Appendix B.

\begin{table}[thb]
\captive{Quark, lepton and gauge boson codes.
\protect\label{t:codeone} } \\
\vspace{1ex}
\begin{center}
\begin{tabular}{|c|c|c||c|c|c||c|c|c|@{\protect\rule{0mm}{\tablinsep}}}
\hline \hline
KF & Name & Printed & KF & Name & Printed & KF & Name & Printed \\
\hline\hline
    1 & $\d$ & \ttt{d}   &  11 & $\e^-$ &  \ttt{e-}  & 21 & $\g$ & \ttt{g} \\  
    2 & $\u$ & \ttt{u}   &  12 & $\nu_{\e}$ &  \ttt{nu\_e} & 22 & $\gamma$ & \ttt{gamma} \\
    3 & $\s$ & \ttt{s}   &  13 & $\mu^-$ &  \ttt{mu-}  & 23 & $\Z^0$ & \ttt{Z0} \\
    4 & $\c$ & \ttt{c}   &  14 & $\nu_{\mu}$ & \ttt{nu\_mu} & 24 & $\W^+$ & \ttt{W+} \\
    5 & $\b$ & \ttt{b}   &  15 & $\tau^-$ & \ttt{tau-}  & 25 & $\H^0$ & \ttt{H0} \\
    6 & $\t$ & \ttt{t}   &  16 & $\nu_{\tau}$ &  \ttt{nu\_tau}   & 26 & &  \\
    7 & & & 17 & & & 27 & & \\
    8 & & & 18 & & & 28 & & \\
    9 & & & 19 & & & 29 & &\\
   10 & & & 20 & & & 30 & &\\
\hline \hline
\end{tabular}
\end{center}
\end{table}
\begin{table}[bhp]
\captive{Special codes.
\protect\label{t:codetwo} }   \\
\vspace{1ex}
\begin{center}
\begin{tabular}{|c|c|c|@{\protect\rule{0mm}{\tablinsep}}}
\hline \hline
KF & Printed & Meaning \\
\hline \hline
   90 &  \ttt{CMsoft}  & Center-of-mass of beam/target remnant system before
	 soft fragmentation  \\
   91 &  \ttt{cluster}  & Cluster from coalescence of secondary partons  \\
   92 &  \ttt{Beam-REM}   & Cluster from remnant primary partons of beam particle\\
   93 &  \ttt{Targ-REM}   & Cluster from remnant primary partons of target particle\\
   94 &  \ttt{CMshower} & Four-momentum of time-like showering system  \\
   95 &  \ttt{SPHEaxis} & Event axis found with \ttt{VNISPHE}  \\
   96 &  \ttt{THRUaxis} & Event axis found with \ttt{VNITHRU}  \\
   97 &  \ttt{CLUSjet}  & Jet (cluster) found with \ttt{VNICLUS}  \\
   98 &  \ttt{CELLjet}  & Jet (cluster) found with \ttt{VNICELL}  \\
   99 &      &  \\
  100 &           &   \\
\hline \hline
\end{tabular}
\end{center}
\end{table}

\newpage
 
{\footnotesize
\begin{table}[thb]
\captive{Meson codes, part 1.
\protect\label{t:codethree} }  \\
\vspace{1ex}
\begin{center}
\begin{tabular}{|c|c|c||c|c|c|@{\protect\rule{0mm}{\tablinsep}}}
\hline \hline
KF & Name & Printed & KF & Name & Printed \\
\hline \hline
  211 & $\pi^+$ & \ttt{pi+}        & 213 & $\rho^+$ & \ttt{rho+}  \\
  311 & $\K^0$ & \ttt{K0}          & 313 & $\K^{*0}$ & \ttt{K*0}  \\
  321 & $\K^+$ & \ttt{K+}          & 323 & $\K^{*+}$ & \ttt{K*+}  \\
  411 & $\D^+$ & \ttt{D+}          & 413 & $\D^{*+}$ & \ttt{D*+}  \\
  421 & $\D^0$ & \ttt{D0}          & 423 & $\D^{*0}$ & \ttt{D*0}  \\
  431 & $\D_s^+$ & \ttt{D\_s+}     &
433 & $\D_{\s}^{*+}$ & \ttt{D*\_s+}  \\
  511 & $\B^0$ & \ttt{B0}          & 513 & $\B^{*0}$ & \ttt{B*0}  \\
  521 & $\B^+$ & \ttt{B+}          & 523 & $\B^{*+}$ & \ttt{B*+}  \\
  531 & $\B_s^0$ & \ttt{B\_s0}     &
533 & $\B_{\s}^{*0}$ & \ttt{B*\_s0}  \\
  541 & $\B_c^+$ & \ttt{B\_c+}     &
543 & $\B_{\c}^{*+}$ & \ttt{B*\_c+}  \\
  111 & $\pi^0$ & \ttt{pi0}        & 113 & $\rho^0$ & \ttt{rho0}  \\
  221 & $\eta$ & \ttt{eta}         & 223 & $\omega$ & \ttt{omega}  \\
  331 & $\eta'$ & \ttt{eta'}       & 333 & $\phi$ & \ttt{phi}  \\
  441 & $\eta_{\c}$ & \ttt{eta\_c} & 443 & $\Jpsi$ & \ttt{J/psi}  \\
  551 & $\eta_{\b}$ & \ttt{eta\_b} &
553 & $\Upsilon$ & \ttt{Upsilon}  \\
  661 & $\eta_{\t}$ & \ttt{eta\_t} & 663 & $\Theta$ & \ttt{Theta}  \\
  130 & $\K_{\mrm{L}}^0$ & \ttt{K\_L0} & & &  \\
  310 & $\K_{\mrm{S}}^0$ & \ttt{K\_S0} & & &  \\
\hline \hline
\end{tabular}
\end{center}
\end{table}
} 

\vspace{-0.75cm}

{\footnotesize
\begin{table}[thb]
\captive{Meson codes, part 2.
\protect\label{t:codefour} }  \\
\vspace{1ex}
\begin{center}
\begin{tabular}{|c|c|c||c|c|c|@{\protect\rule{0mm}{\tablinsep}}}
\hline \hline
KF & Name & Printed & KF & Name & Printed \\
\hline \hline
10213 & $\b_1$ & \ttt{b\_1+}              &
10211 & $\a_0^+$ & \ttt{a\_0+}  \\
10313 & $\K_1^0$ & \ttt{K\_10}            &
10311 & $\K_0^{*0}$ & \ttt{K*\_00}  \\
10323 & $\K_1^+$ & \ttt{K\_1+}            &
10321 & $\K_0^{*+}$ & \ttt{K*\_0+}  \\
10413 & $\D_1^+$ & \ttt{D\_1+}            &
10411 & $\D_0^{*+}$ & \ttt{D*\_0+}  \\
10423 & $\D_1^0$ & \ttt{D\_10}            &
10421 & $\D_0^{*0}$ & \ttt{D*\_00}  \\
10433 & $\D_{1 \s}^+$ & \ttt{D\_1s+}      &
10431 & $\D_{0 \s}^{*+}$ & \ttt{D*\_0s+}  \\
10113 & $\b_1^0$ & \ttt{b\_10}            &
10111 & $\a_0^0$ & \ttt{a\_00}  \\
10223 & $\hrm_1^0$ & \ttt{h\_10}            &
10221 & $\f_0^0$ & \ttt{f\_00}  \\
10333 & $\hrm'^0_1$ & \ttt{h'\_10}          &
10331 & $\f'^0_0$ & \ttt{f'\_00}  \\
10443 & $\hrm_{1 \c}^0$ & \ttt{h\_1c0}      &
10441 & $\chi_{0 \c}^0$ & \ttt{chi\_0c0}  \\  \hline
20213 & $\a_1^+$ & \ttt{a\_1+}            &
  215 & $\a_2^+$ & \ttt{a\_2+}  \\
20313 & $\K_1^{*0}$ & \ttt{K*\_10}        &
  315 & $\K_2^{*0}$ & \ttt{K*\_20}  \\
20323 & $\K_1^{*+}$ & \ttt{K*\_1+}        &
  325 & $\K_2^{*+}$ & \ttt{K*\_2+}  \\
20413 & $\D_1^{*+}$ & \ttt{D*\_1+}        &
  415 & $\D_2^{*+}$ & \ttt{D*\_2+}  \\
20423 & $\D_1^{*0}$ & \ttt{D*\_10}        &
  425 & $\D_2^{*0}$ & \ttt{D*\_20}  \\
20433 & $\D_{1 \s}^{*+}$ & \ttt{D*\_1s+}  &
  435 & $\D_{2 \s}^{*+}$ & \ttt{D*\_2s+}  \\
20113 & $\a_1^0$ & \ttt{a\_10}            &
  115 & $\a_2^0$ & \ttt{a\_20}  \\
20223 & $\f_1^0$ & \ttt{f\_10}            &
  225 & $\f_2^0$ & \ttt{f\_20}  \\
20333 & $\f'^0_1$ & \ttt{f'\_10}          &
  335 & $\f'^0_2$ & \ttt{f'\_20}  \\
20443 & $\chi_{1 \c}^0$ & \ttt{chi\_1c0}  &
  445 & $\chi_{2 \c}^0$ & \ttt{chi\_2c0}  \\  \hline
30443 & $\psi'$     & \ttt{psi'}     &   &   &  \\
30553 & $\Upsilon'$ & \ttt{Upsilon'} &   &   &  \\
\hline \hline
\end{tabular}
\end{center}
\end{table}
}
 
\begin{table}[thp]
\captive{Baryon codes.
\protect\label{t:codefive} }  \\
\vspace{1ex}
\begin{center}
\begin{tabular}{|c|c|c||c|c|c|@{\protect\rule{0mm}{\tablinsep}}}
\hline \hline
KF & Name & Printed & KF & Name & Printed \\
\hline \hline
      &  &                            &
 1114 & $\Delta^-$ & \ttt{Delta-} \\
 2112 & $\n$ & \ttt{n0}                     &
 2114 & $\Delta^0$ & \ttt{Delta0} \\
 2212 & $\p$ & \ttt{p+}                     &
 2214 & $\Delta^+$ & \ttt{Delta+} \\
      &  &                            &
 2224 & $\Delta^{++}$ & \ttt{Delta++} \\
 3112 & $\Sigma^-$ & \ttt{Sigma-}           &
 3114 & $\Sigma^{*-}$ & \ttt{Sigma*-} \\
 3122 & $\Lambda^0$ & \ttt{Lambda0}         & & &  \\
 3212 & $\Sigma^0$ & \ttt{Sigma0}           &
 3214 & $\Sigma^{*0}$ & \ttt{Sigma*0} \\
 3222 & $\Sigma^+$ & \ttt{Sigma+}           &
 3224 & $\Sigma^{*+}$ & \ttt{Sigma*+} \\
 3312 & $\Xi^-$ & \ttt{Xi-}                 &
 3314 & $\Xi^{*-}$ & \ttt{Xi*-} \\
 3322 & $\Xi^0$ & \ttt{Xi0}                 &
 3324 & $\Xi^{*0}$ & \ttt{Xi*0} \\
      &  &                            &
 3334 & $\Omega^-$ & \ttt{Omega-} \\
 4112 & $\Sigma_{\c}^0$ & \ttt{Sigma\_c0}   &
 4114 & $\Sigma_{\c}^{*0}$ & \ttt{Sigma*\_c0} \\
 4122 & $\Lambda_{\c}^+$ & \ttt{Lambda\_c+} & & &  \\
 4212 & $\Sigma_{\c}^+$ & \ttt{Sigma\_c+}   &
 4214 & $\Sigma_{\c}^{*+}$ & \ttt{Sigma*\_c+} \\
 4222 & $\Sigma_{\c}^{++}$ & \ttt{Sigma\_c++} &
 4224 & $\Sigma_{\c}^{*++}$ & \ttt{Sigma*\_c++} \\
 4132 & $\Xi_{\c}^0$ & \ttt{Xi\_c0}         & & &  \\
 4312 & $\Xi'^0_{\c}$ & \ttt{Xi'\_c0}       &
 4314 & $\Xi_{\c}^{*0}$ & \ttt{Xi*\_c0}  \\
 4232 & $\Xi_{\c}^+$ & \ttt{Xi\_c+}         & & &  \\
 4322 & $\Xi'^+_{\c}$ & \ttt{Xi'\_c+}        &
 4324 & $\Xi_{\c}^{*+}$ & \ttt{Xi*\_c+}  \\
 4332 & $\Omega_{\c}^0$ & \ttt{Omega\_c0}   &
 4334 & $\Omega_{\c}^{*0}$ & \ttt{Omega*\_c0}  \\
 5112 & $\Sigma_{\b}^-$ & \ttt{Sigma\_b-}     &
 5114 & $\Sigma_{\b}^{*-}$ & \ttt{Sigma*\_b-}  \\
 5122 & $\Lambda_{\b}^0$ & \ttt{Lambda\_b0} & & &  \\
 5212 & $\Sigma_{\b}^0$ &\ttt{Sigma\_b0}   &
 5214 & $\Sigma_{\b}^{*0}$ & \ttt{Sigma*\_b0}  \\
 5222 & $\Sigma_{\b}^+$ & \ttt{Sigma\_b+}   &
 5224 & $\Sigma_{\b}^{*+}$ &  \ttt{Sigma*\_b+}  \\
\hline \hline
\end{tabular}
\end{center}
\end{table}
\vspace{-0.5cm}
\begin{table}[bhp]
\captive{Nucleus codes.
\protect\label{t:codesix} }  \\
\vspace{1ex}
\begin{center}
\begin{tabular}{|c|c|c||c|c|c|@{\protect\rule{0mm}{\tablinsep}}}
\hline \hline
KF & Name & Printed & KF & Name & Printed \\
\hline \hline
& & & & &\\ 
 1001 & $^2_1\alpha$ & \ttt{alpha}  & 61047 & $^{108}_{47}\A\g$ & \ttt{ag} \\
& & & & &\\ 
 2002 & $^4_2\H\e$ & \ttt{he} & 118079 & $^{197}_{79}\A\u$ & \ttt{au} \\
& & & & &\\ 
 8008 & $^{16}_8{\rm O}$ & \ttt{ox} & 126082 & $^{208}_{82}{\rm P}\b$ & \ttt{pb} \\
& & & & &\\ 
16016 & $^{32}_{16}{\rm S}$ & \ttt{su}   & 146092 & $^{238}_{92}{\rm U}$ & \ttt{ur} \\
& & & & &\\ 
 \hline \hline
\end{tabular}
\end{center}
\end{table}

\newpage

\subsection{The general input and control parameters}
\label{sec:section5}
\bigskip

The VNIPAR common block contains the status code and parameters
that regulate the performance of the program. All of them are
provided with sensible default values, so that a novice user can
neglect them, and only gradually explore the full range of
possibilities. 

\begin{verbatim}
COMMON/VNIPAR/MSTV(200),PARV(200),MSTW(200),PARW(200)
\end{verbatim}
 
\noindent
{\it Purpose:} to give access to status code and parameters that
regulate the performance of the program. If the default values,
denoted below by (D=\ldots), are not satisfactory, they must in
general be changed before the VNIXRIN call. Exceptions, i.e.
variables that can be changed for each new event, are denoted by
(C).
\bigskip
\bigskip
 
\begin{center}
{\bf MSTV(200), PARV(200): control switches and physics parameters}
\end{center}
\bigskip

\noindent
\lline

\noindent{\bf 1. General:}

\noindent
\lline

\begin{description}

\item[ MSTV(1)]  : (D=100000) number of lines available in the commonblock
    VNIREC. Should always be changed if the dimensions of the
    $K,P,R,V$ arrays are changed.
\item[ MSTV(2)]  : (D=25000) is used to trace internal color flow information 
    in the array $K$.
\item[ MSTV(3)]  : (D=10000) is used to store number and type of interactions
    of a particle in the array $R$ of VNIREC.
\item[ MSTV(4)]  : (D=1000) is used to trace origin of a particle's production
    in the array $V$ of VNIREC.
\item[ MSTV(5)]  : (D=3) number of colors $N_c$ included in the simulation.
\item[ MSTV(6)]  : (D=6) number of quark flavors $N_f$ included.
\item[ MSTV(7)] : (D=2) energy and momentum conservation options. Since 
    both the initial generation of initial hadronic or nuclear parton
    system and the final cluster-hadron decay scheme require occasional 
    reshuffling of 4-momentum, an energy and/or momentum imbalance can 
    occur. This can be corrected, if desired, by renormalizing particle 
    energies and momenta.\\
    {\bf = 0 :} no explicit conservation of any kind.\\
    {\bf = 1 :} particles share energy imbalance compensation according
        to their energy (roughly equivalent to boosting event to CM
        frame).\\
    {\bf = 2 :} as =1, plus particles share 3-momentum imbalance 
        compensation according to their longitudinal mass with respect to 
        the imbalance direction.
\item[ MSTV(8)] : (D=0) type of particles to be included in energy 
       and momentum conservation options under MSTV(7).\\
    {\bf = 0 :} all living  particles are included.\\
    {\bf = 1 :} only partons.\\
    {\bf = 2 :} only hadrons.
\item[ MSTV(9)] :  - not used -
\item[ MSTV(10)] : (D=111111) initial seed for random number generation.
\item[ MSTV(11)] : (D=250000) maximum number of collision events $A+B$, 
    after which program is forcibly terminated.
\item[ MSTV(12)] : (D=2500) maximum number of time steps per collision 
    event, after which a new collision event is generated. When
    increased, the dimension of the arrays in commonblock $VNITIM$ 
    need to be altered accordingly.
\item[ MSTV(13)] : (D=0) choice of time grid $TIME(I)$ and time increase
    $TINC(I)$. Generically, $t(i)=t(i-1)+[f(i)-f(i-1)]$, with 
    $i=0,\ldots,MSTV(12)$, and $t(0)=t_0=$PARV(12), $t_f=$PARV(13). \\
    {\bf = 0 :} power law increase $f(i)=a\cdot(i)^b$, where $a=$PARV(14) and 
    $b=$PARV(15). \\
    {\bf = 1 :} logarithmic increase $f(i)=a\cdot \ln(i)^b$, where $a=$PARV(14), 
    $b=$PARV(15).
\item[ MSTV(14)] : (D=0) Option to select or switch on/off certain physics
    processes $ISUB$ in the simulation, by specifying the array $MSUB$ 
    (c.f. Sec. 3.8 below), where $MSUB(ISUB) = 0\; (1)$ switches a 
    subprocess off (on).\\ 
    {\bf = 0 :} the relevant processes are  initialized  automatically.\\ 
    {\bf = 1 :} user selection of included processes is required.
\medskip

\noindent
\sline

\item[ PARV(12)] : (D=0. fm) choice of initial point of time $t_i =$ PARV(12) 
    when evolution of collision event begins.
\item[ PARV(13)] : (D=0. fm) choice of final point of time $t_f =$ PARV(13).
    If = 0., then it is set automatically to $TIME$(MSTW(2)), where MSTW(2)
    is the selected number of time steps. Otherwise, if $> 0.$, it specifies
    $t_f$ such that the minimum  $\min(t_f , t_{max})$ is taken as the point when the
    evolution of a collision event stops, where $t_{max} = TIME$(MSTV(12)).
\item[ PARV(14)] : (D=0.05) prefactor $a$ of the time-increment function $f$, see
    MSTV(13).
\item[ PARV(15)] : (D=1.) exponent $b$ of the time-increment function $f$, see
    MSTV(13). For longer collision time scales as  in hadron-nucleus
    or nucleus nucleus collisions, it is automatically scaled by
    PARV(15) to 1.5 $\times$ PARV(15) to extend the time range.

\noindent
\sline

\end{description}
\medskip

\noindent
\lline

\noindent {\bf 2. Initial state of collision system:}

\noindent
\lline

\begin{description}
\item[ MSTV(20)] : (D=0) choice of model for initial parton distributions.  \\
    {\bf = 0 :} standard structure functions are used.\\
    {\bf = 1 :} other parametrization, e.g. a la McLerran {\it et al.}
    (not implemented yet).
\item[ MSTV(21)] : (D=1) choice of nucleon structure-function 
      set (c.f. MSTV(22)). \\
    {\bf =  1 :} GRV94LO (Lowest order fit).\\
    {\bf =  2 :} GRV94HO (Higher order: MSbar fit).\\
    {\bf =  3 :} GRV94DI (Higher order: DIS fit).\\
    {\bf =  4 :}  - not used - \\
    {\bf =  5 :} CTEQ2M (best higher order MSbar fit).\\
    {\bf =  6 :} CTEQ2MS (singular at small $x$).\\
    {\bf =  7 :} CTEQ2MF (flat at small $x$).\\
    {\bf =  8 :} CTEQ2ML (large $\Lambda$).\\
    {\bf =  9 :} CTEQ2L (best lowest order fit).\\
    {\bf = 10 :} CTEQ2D (best higher DIS order fit).
\item[ MSTV(22)] : (D=1) choice of nucleon structure-function library.  \\
    {\bf = 1 :} the internal one with parton distributions chosen according 
    to MSTV(21).\\
    {\bf = 2 :} the PDFLIB one with MSTV(21)=$1000\cdot NGROUP+NSET$
    (requires PDFLIB to be linked). 
\item[ MSTV(23)] : (D=1) choice of pion structure-function set (c.f. MSTV(24)).\\
    {\bf = 1 :} Owens set 1.\\
    {\bf = 2 :} Owens set 2.\\
\item[ MSTV(24)] : (D=1) choice of pion structure-function library.  \\
    {\bf = 1 :} the internal one with parton distributions chosen according 
    to MSTV(23).\\
    {\bf = 2 :} the PDFLIB one with $MSTV(23)=1000\cdot NGROUP+NSET$
    (requires PDFLIB to be linked). 
\item[ MSTV(25)] : (D=2) choice of minimum $Q$-value $Q_{0}$ used 
in parton distributions of initial hadronic or nuclear state.\\
    {\bf = 0 :} fixed at $Q_{0}=$PARV(23).\\
    {\bf = 1 :} determined by the average momentum tranfer of primary parton 
    collisions $Q_0=\langle Q_{prim}\rangle$, 
    with initial value $Q_{0}=$PARV(23) before the first event.\\
    {\bf = 2 :} as =1, but in addition varying with total energy $\sqrt{s}$ of 
    the overall collision system, with 
    $\widetilde{Q}_{0} = \max[Q_0, \frac{a}{4} \cdot \left(E_h/GeV\right)^b]$, 
    where $Q_0=$PARV(23), $a=$PARV(24), $b=$PARV(25), and 
    $E_h=\sqrt{s}/\mbox{hadron}$ the average energy per hadron (nucleon).
\item[ MSTV(26)] : (D=1) option to choose $x$-dependent $Q_{in}$ 
in parton densities.  \\
    {\bf = 0 :} switched off, $Q_{0}$ according to selection of MSTV(25) 
    is used.\\
    {\bf = 1 :} varying with Bjorken-$x$ of struck initial state partons, as 
    well as with total energy $\sqrt{s}$ of collision system, with  
    $Q_{0}=Q(x)$ where $1/Q^2(x) = 1/Q_{0}^2+Q_{0}^2/(x E_{had})^2$, and 
    $E_h=\sqrt{s}/\mbox{hadron}$ the average energy per hadron (nucleon).
\item[ MSTV(31)] : (D=0) initial separation along the collision ($z$-)axis of 
    beam and target particle in the overall CM frame.\\
    {\bf =  0 :} fixed at $\Delta z=$PARV(31).\\
    {\bf$>$ 0 :} shifted towards minimal separation, such that beam and 
    target particles almost touch with $\Delta z =$MSTV(31)$\cdot$PARV(31).
\item[ MSTV(32)] : (D=2) selection of impact parameter $b$ transverse to the 
    collision ($z$-) axis of beam and target particle.\\
    {\bf = 0 :} fixed at $b =$ Max[PARV(32),PARV(33)].\\
    {\bf = 1 :} randomly sampled in between $b_{min}\le b\le b_{max}$, with 
    $b_{min} = $PARV(32) and $b_{max} = $PARV(33).\\
    {\bf = 2 :} as 1, but with $b_{min} =$ PARV(32)$\cdot\sqrt{\sigma_{nd}/\pi}$, 
    and $b_{max} =$ PARV(33)$\cdot\sqrt{\sigma_{nd}/\pi}$, 
    where $\sigma_{nd}$ is the non-diffractive crossection for given 
    beam/target and beam energy.
\item[ MSTV(33)] : (D=2) Choice of nuclear matter distribution for collisions
    involving one or two nuclei, i.e. of nucleons' positions within nucleus.\\
    {\bf = 0 :} uniform distribution with `sharp edge'.\\
    {\bf = 1 :} Gaussian distribution for light nuclei $A\le 12$, 
    and `smeared edge' distribution for intermediate and heavy nuclei 
    $A > 12$ (c.f. PARV(27)).\\
    {\bf = 2 :} Gaussian distribution for $A\le 12$, and Fermi distribution for
    $A > 12$ (c.f. PARV(28)).
\item[ MSTV(34)] : (D=1) spatial distribution of initial partons within their
    parent hadrons of beam and target particles, in the hadron restframe
    within a sphere of the hadron radius.\\
    {\bf = 0 :} uniform distribution.\\
    {\bf = 1 :} exponential distribution with width given by PARV(34).\\
    {\bf = 2 :} Gaussian distribution with width given by PARV(35).
\item[ MSTV(35)] : (D=2) Lorentz-boosted spatial distribution of initial partons.\\
    {\bf = 0 :} spatial positions of all partons (valence, sea, glue) inside 
    parent hadron are Lorentz-boosted with the parent hadron's (nucleon's) 
    $\gamma$-factor.\\ 
    {\bf = 1 :} only valence quarks are boosted with the parent  hadron's 
    (nucleon's) 
    $\gamma$-factor; seaquarks and gluons are smeared out symmetrically(!) 
    in hadron (nucleon) direction of motion around the valence quark disc.  \\ 
    {\bf = 2 :} as = 1, but now seaquarks and gluons are 
    distributed all behind(!) the valence quark disc.
\item[ MSTV(36)] : (D=1) choice of primordial $k_\perp$-distribution of 
    initial partons.\\
    {\bf = 0 :} no primordial $k_\perp$.\\
    {\bf = 1 :} exponential $k_\perp$-distribution with width 
    specified by PARV(37).\\
    {\bf = 2 :} Gaussian $k_\perp$-distribution with width specified by PARV(38).
\item[ MSTV(37)] : (D=1) choice of masses of initial partons.\\
    {\bf = 0 :} on mass shell with quark current masses.\\
    {\bf = 1 :} off mass shell with space-like virtuality as determined by
    the excess of the partons' momentum over their energies.
\item[ MSTV(38)] : (D=0) switch for including effects of parton shadowing in
    nuclear parton structure functions by multiplication with effective
    shadowing factor $R_A(x)=f_A(x)/f_N(x)$ as a simple parametrization
    of EMC data, where $f_A$ and $f_N$ are the measured nuclear and nucleon 
    structure functions, respectively.\\
    {\bf = 0 :} parton shadowing switched off.\\
    {\bf = 1 :} switched on.
\medskip

\noindent
\sline

\item[ PARV(21)] : (D=1. GeV$^{-1}$) the minimum resolvable partonic 
    Bjorken-$x$, denoted $x_c$,
    is set by the `confinement radius' $R_c = $PARV(21), which defines the 
    maximum possible longitudinal spread of a confined parton within 
    $\Delta z = 1/xP < 1/(x_c P) \equiv R_c$, where $P$ is the longitudinal 
    momentum of the mother hadron.
\item[ PARV(22)] : (D=0.95)  the minimum 
    summed $x$-value $\left(\sum_i x_i\right)_{min}$, 
    when `filling' a hadron with partons of energy(momentum) fractions
    $x_i$ of the hadron. The value depends, slightly on the value of $Q_0$ 
    (c.f. PARV(23)-PARV(25)), or if cuts on $x$ are made. It should be 
    adjusted such that when adding one more parton, the average 
    $\langle \sum_i x_i \rangle$ 
    should be = 1. If the sum is smaller than PARV(22), another parton 
    is added to the hadron, i.e. PARV(22) = $1 - \langle \sum_i x_i \rangle$. 
\item[ PARV(23)] : (D=1. GeV) minimum value $Q_0$ that is used in initial parton
    distributions (structure functions or alternative distributions).
    I.e., if $Q < Q_0$, then the value PARV(23) is used.
\item[ PARV(24)], {\bf PARV(25)} : (D=2.50,0.25) 
    possibility of using effective, energy-dependent
    $Q_{in}$ 
    value in structure functions if MSTV(25)=2, according to the function
    $\widetilde{Q}_{0} = \max[Q_0,\frac{a}{4}\cdot \left(E_h/GeV\right)^b]$, 
    where  $Q_0 = $PARV(23), $a = $PARV(24), 
    $b= $ PARV(25), and
    $E_h=\sqrt{s}/\mbox{hadron}$ the average energy per hadron (nucleon).
\item[ PARV(26)] : (D=1.18 fm) nuclear radius parameter $r_0$, with 
    $R(A) = r_0 A^1/3$. 
\item[ PARV(27)] : (D=3.0 fm) parameter $R_d$ in `smeared-edge' distribution of 
    nucleons' positions in nucleus with $A > 12$ (for MSTV(33) = 1).
\item[ PARV(28)] : (D=0.545 fm) parameter $R_a$ in Fermi distribution of 
    nucleons' positions in nucleus with $A > 12$ (for MSTV(33) = 2).
\item[ PARV(31)] : (D=0.25 fm) initial separation along the collision ($z$-)axis 
    of beam and target particle in the overall CM frame for MSTV(31)$\ge$ 1.
\item[ PARV(32)] : (D=0. fm) determines 
    value of {\it minimum} impact parameter $b_{min}$, 
    transverse to the collision ($z$-)axis, such that
    $b_{min} = $PARV(32) for MSTV(32) = 0 or 1, 
    or $b_{min} = $PARV(32)$\cdot\sqrt{\sigma_{nd}/\pi}$ for MSTV(32)=2.
\item[ PARV(33)] : (D=1. fm) determines 
    value of {\it maximum} impact parameter $b_{max}$, 
    transverse to the collision ($z$-)axis, such that
    $b_{max} = $PARV(33) for MSTV(32) = 0 or 1, 
    or $b_{max} = $PARV(33)$\cdot\sqrt{\sigma_{nd}/\pi}$ for MSTV(32)=2.
\item[ PARV(34)] : (D=1.19 GeV$^{-1}$) width $w$ in exponential distribution 
    $\exp[-r/w]$ of partons' positions around its parent-hadron center-of-mass 
    (c.f. MSTV(33)).
    (Note: $w=1/\nu$, where $\nu=0.84$ GeV is the measured nucleon form factor.)
\item[ PARV(35)] : (D=2.38 GeV$^{-1}$) sigma $s$ in Gaussian distribution 
    $\exp[-r^2/(2\,s^2)]$ of 
    partons' positions around its parent-hadron center-of-mass (c.f. MSTV(33)).
    (Note: $s=2/\nu$, where $\nu=0.84$ GeV is measured nucleon form factor.)
\item[ PARV(36)] : (D=2.) cut-off for sampling initial partons' spatial 
    distributions 
    from exponential or Gaussian above, such that $r < a\,R_h$, where 
    $a=$PARV(36) and $R_h$ is the radius of the mother hadron.
\item[ PARV(37)] : (D=0.44 GeV) width $w$ in exponential distribution 
    $\exp[-k_\perp/w]$ of primordial $k_\perp$ distribution of initial partons 
    (c.f. MSTV(36)).
\item[ PARV(38)] : (D=0.44 GeV) sigma $s$ in Gaussian distribution 
    $\exp(-k_\perp^2/(2\,s^2)]$ of primordial $k_\perp$ distribution of 
    initial partons (c.f. MSTV(36)).
\item[ PARV(39)] : (D=2. GeV) upper cut-off for sampling primordial 
    $k_\perp$ from above exponential or Gaussian distribution.

\noindent
\sline

\end{description}
\medskip

\noindent
\lline

\noindent {\bf 3. Parton scatterings:}

\noindent
\lline

\begin{description}
\item[ MSTV(40)] : (D=3) master switch for parton collisions. \\
    {\bf = 0 :} generally switched off.
\\
    {\bf = 1 :} only primary-primary collisions, i.e. those involving 
    2 partons that both have not interacted before.
\\
    {\bf = 2 :} as = 1 plus primary-secondary collisions, i.e. those involving 
    2 partons, of which one has had already at least one previous interaction.
\\
    {\bf = 3 :} as = 2 plus secondary-secondary interactions, i.e.
    all multiple collisions included.
\item[ MSTV(41)] : (D=1) choice of (semi)hard QCD collision cross-sections.
\\
    {\bf = 1 :} standard $O(\alpha_s^2)$ perturbative cross-sections for 
    $2 \rightarrow 2$ and $2 \rightarrow 1$ processes.
\\
    {\bf = 2 :} alternative  of $2 \rightarrow n$ cross-sections 
    (not implemented yet!).
\item[ MSTV(42)] : (D=1) choice of $p_{0}$ 
    to separate {\it hard} and {\it soft} parton collisions, and to act
    as a regulator for the parton cross-sections $d\hat{\sigma}/dQ^2$,
    that are divergent 
    for vanishing momentum transfer (or invariant mass) $Q$.
\\
    {\bf = 0 :} fixed at 
    $p_0 = \max \left[ PARV(42), \max (VKIN(3),VKIN(5) \right]$.
\\
    {\bf = 1 :} energy and mass dependent, with  $p_0 = p_0(s,A)$, 
    according to the parametrization given by PARV(42), PARV(43).
\\
    {\bf = 2 :} initially, before the first event, set to 
    $p_0(s,A)$, c.f. =1, 
    but in subsequent events set equal to the average momentum transfer 
    of primary parton-parton collisions 
    $p_0 = \langle Q_{prim}\rangle$ accumulated by statistics.
\item[ MSTV(43)] : (D=2) $Q^2$-definition in $2 \rightarrow 2$ 
    parton collision processes;
    for $\hat s$-channel  or fusion processes, 
    $Q^2$ is always chosen to be $m^2=\hat s$.
    Note that the parton-parton collisions invariants are
    denoted as $\hat s = (p_1+p_2)^2$,
    $\hat t = (p_1-p_1')^2$, $\hat u = (p_1-p_2')^2$.
\\
    {\bf = 1 :} $Q^2 = 2 \hat{s} \hat{t} \hat{u}/(\hat{s}^2 + 
    \hat{t}^2 + \hat{u}^2)$.
\\
    {\bf = 2 :} $Q^2 = 0.5 \,(m_{\perp\,1}^2 + m_{\perp\,2}^2)$.
\\
    {\bf = 3 :} $Q^2 = \min (-\hat{t}, - \hat{u})$.
\\
    {\bf = 4 :} $Q^2 = \hat{s}$.
\item[ MSTV(44)] : (D=2) choice of scheme for parton collision-time estimate.
\\
    {\bf = 0 :} zero collision-time, i.e. instantaneous process. 
\\
    {\bf = 1 :} finite collision-time sampled from an exponential distribution
    $\exp(-x)$, where $x=t/\tau$, $t$ the time in the lab, and $\tau$
    the mean life-time of the parton in the lab, see PARV(44).
\\
    {\bf = 2 :} finite collision-time sampled from Gyulassy-Wang distribution
    $(2/\tau) \;x/(1+x^2)^2$, where $x=t/\tau$, $t$ the time in the lab, 
    and $\tau$ the mean life-time of the parton in the lab, see PARV(44).
\item[ MSTV(45)] : (D=1) selection of maximum allowed distance between two partons
    in their $cm$-frame in order to be able to collide (used for ruling
    out widely separated pairs to speed up simulation).
\\
    {\bf = 0 :} fixed at a value given by $r_{12\;max} =$ PARV(45).
\\
    {\bf = 1 :} variable at a value 
    $r_{12\;max} =$ PARV(45)/$(\sqrt{\hat{s}}/GeV)$.
\item[ MSTV(46)] : (D=0) choice of probability distribution W(x) from which 
    collision probability of 2 partons is sampled, where
    $x = r_{12}^{(sep)}/r_{12}^{(\hat{\sigma})}$, with $r_{12}^{(sep)}$ 
    the transverse  2-parton separation 
    at closest approach and $r_{12}^{(\hat{\sigma})}$ 
    the transverse radius of interaction as given by the parton-parton cross-section $\hat{\sigma}$.
\\
    {\bf = 0 :} flat distribution, $W(x) = \theta(a-x)$, with $a =$ PARV(46).
\\
    {\bf = 1 :} exponential distribution, $W(x) = \exp(-x/a)$, $a =$ PARV(46).
\item[ MSTV(47)] : (D=2) special treatment of `flavor excitation' by parton 
    collisions involving one or two primary hadronic or nuclear 
    (anti)quarks by requiring a minimum momentum transfer of the
    resolving parton with the heavy quark.
\\
    {\bf = 0 :} no special treatment of heavy quarks over light quarks.\\
    {\bf = 1 :} requirement of minimum momentum transfer for liberation of
    heavy quark out of initial parton distribution by scattering, where
    the minimum required momentum trasnfer 
    is given by PARV(47) times the mass of the heavy quark.
\\
    {\bf = 2 :} as = 1, but now both the struck heavy quark as well as its
    initial antiquark sibling are liberated (i.e. get on mass shell).
\item[ MSTV(48)] : (D=0) switch for including soft parton collisions according 
    to a phenomenological cross-section 
    $d\hat{\sigma}/dQ^2 \propto \alpha_s^2(p_0^2)/(Q^2 + \mu^2)$
    for soft parton collisions (c.f. eq. (\ref{sigs})), 
    where $p_0 =$ PARV(42), $\mu =$ PARV(48), and $Q^2 = p_\perp^2$
    for $\hat t, \hat u$ channel and
    $Q^2 = \hat s$ for $\hat s$ channel.
\\
    {\bf = 0 :} switched off.
\\
    {\bf = 1 :} soft collisions are treated complementary to hard collisions,
    i.e. occur only if $Q < p_{0}$.
\\
    {\bf = 2 :} soft collisions are treated supplementary to hard collisions,
    i.e. compete with the latter, whichever cross-section dominates.
\item[ MSTV(49)] : (D=1) switch to allow for $2 \rightarrow 1$
    parton fusion processes
    $a+b\rightarrow c^\ast$ in competition with $2\rightarrow 2$ 
    scattering processes $a+b\rightarrow c^\ast\rightarrow d+e$,
    such that for a given pair a,b either of those processes is selected
    according to the relative probability given by the respective 
    cross-sections and depending on the `life-time' of the particle $c^\ast$.
\\
    {\bf = 0 :} fusion processes switched off.
\\
    {\bf = 1 :} switched on, with relative weight determined by PARV(49).
\medskip

\noindent
\sline

\item[ PARV(41)] : (D=0.25 GeV) nominal $\Lambda$-value used in 
    running $\alpha_s$ for parton-parton collisions.
\item[ PARV(42)] : (D=2.0 GeV) cut-off for perturbative QCD cross-sections 
    that are divergent when the momentum transfer $Q\rightarrow 0$. Here the 
    2-body collsion scale $Q$ is taken equal to $p_\perp$, the
    transverse momentum 
    exchanged in a  parton scattering, or equal to $\hat s$ for
    annihilation or fusion processes. For MSTV(42)=0, the value 
    $p_{0} = a$ with $a =$ PARV(42) defines {\it hard} ($Q>p_{0}$) and 
    {\it soft} ($Q<p_{0})$ collisions. For MSTV(42)=1, see PARV(43). 
\item[ PARV(43)] : (D=0.27) parameter for effective $p_{0}$ value, 
    if MSTV(42)=1, 
    instead of $p_{0} = const$ for  MSTV(42)=0. The effective $p_{0}$ 
    assumes a dependence on the total collision energy $\sqrt{s}$ and 
    the mass of the (nuclear) collision system. It is parametrized as 
    $p_{0}(\sqrt{s},A,B) = \frac{a}{4} \cdot \left(E_h/GeV\right)^b$, 
    where $a =$ PARV(42), $b =$ PARV(43), and $E_h= 2 \sqrt{s}/(A+B)$ 
    with $A$ ($B$) the mass number of beam (target) particle, if 
    it is a hadron or nucleus.
\item[ PARV(44)] : (D=0.1) if MSTV(44)=1 or MSTV(44)=2, the 
    proportionality between 
    mean collision time $\tau$ of a 2-parton collision process, and the 
    associated transverse momentum generated (for $\hat{s}$-channel processes, the 
    invariant mass of the pair). It is parametrized as $\tau=\,c\,(1/Q)$,
    with $c =$ PARV(45) and $Q = p_\perp$ or $\hat{s}$.
\item[ PARV(45)] : (D=5. GeV$^{-1}$) maximum separation of two partons to be 
    able to
    interact through a 2-body collision. For MSTV(45)=0, it is fixed at
    $r_{12} =$ PARV(45), whereas for MSTV(45)=1, it is taken s-dependent as 
    $r_{12} =$ PARV(45)$/(\sqrt{\hat{s}}/GeV)$.
\item[ PARV(46)] : (D=1.0) parameter in the collision-probability distribution 
    $W(x)=\theta(a-x)$ (for MSTV(46)=0) and $W(x)=\exp(-x/a)$ for MSTV(46)=1,
    with $x = r_{12}^{(sep)}/r_{12}^{(\hat{\sigma})}$, $r_{12}^{(sep)}$
    the relative transverse parton 
    separation of closest approach and $r_{12}^{(\hat{\sigma})}$ 
    the transverse radius of interaction as given by the 
    parton-parton cross-section $\hat{\sigma}$.
\item[ PARV(47)] : (D=1.0) for `flavor excitation' processes where a primary
    hadronic or nuclear (anti)quark is struck out by a collision, an
    effective minimum resolution scale is required, depending on
    the flavor, such that the momentum transfer
      $Q^2 > a \cdot m_f$, with $a=$ PARV(47) and $m_f$ the flavor-dependent
     quark mass. This parameter acts in addition to the 
    constraint set by PARV(42) above.
\item[ PARV(48)] : (D=0.5 GeV) parameter in the phenomenological 
    cross-section
    $d\hat{\sigma}/dQ^2 \propto \alpha_s^2(p_0^2)/(Q^2 + \mu^2)$
    for soft parton collisions, 
    where $\mu =$ PARV(48), $p_0$ is given by PARV(42), and $Q^2 = p_\perp^2$
    for $\hat t, \hat u$ channel and
    $Q^2 = \hat s$ for $\hat s$ channel.
\item[ PARV(49)] : (D=10.) parametric strength of $2 \rightarrow 1$ 
    parton fusion processes $a+b\rightarrow c$ competing with
    $2\rightarrow 2$ scattering processes $a+b\rightarrow c+d$. It is  
    given by the ratio of the two processes which is proportional 
    to $1/(R^2 \hat{s})$, where $1/R^2 = $PARV(49).

\noindent
\sline

\end{description}
\medskip

\noindent
\lline

\noindent {\bf 4.  Space-like parton branchings:}

\noindent
\lline

\begin{description}
\item[ MSTV(50)] : (D=1) master switch for space-like branchings.
\\
    {\bf = 0 :} switched off
\\
    {\bf = 1 :} on.
\item[ MSTV(51)] : (D=3) level of coherence imposed on the space-like parton
    shower evolution.
\\
    {\bf = 1 :} none, i.e. neither $k^2$ values nor angles need be ordered.
\\
    {\bf = 2 :} $k^2$ values at branches are strictly ordered, increasing
        towards the hard interaction.
\\
    {\bf = 3 :} $k^2$ values and opening angles of emitted (on-mass-shell or
        time-like) partons are both strictly ordered, increasing
        towards the hard interaction.
\item[ MSTV(52)] : (D=2) structure of associated time-like parton evolution, 
    i.e.  showers initiated by emission off the incoming space-like partons.
\\
    {\bf = 0 :} no associated showers are allowed, i.e. emitted partons are
        put on mass-shell.
\\
    {\bf = 1 :} a shower may evolve, with maximum allowed time-like virtuality
        set by the phase space only.
\\
    {\bf = 2 :} a shower may evolve, with maximum allowed time-like virtuality
        set by phase space or by the scale $Q^2$, the virtuality of the 
        space-like parton created at the same vertex, whichever is the 
        stronger constraint.
\medskip

\noindent
\sline

\item[ PARV(51)] : (D=0.25 GeV)  $\Lambda$-value used in 
   $\alpha_s$ and in structure functions for space-like parton evolution.
\item[ PARV(52)] : (D=1.0 GeV) invariant mass cutoff $q_0$ of 
    space-like parton 
    showers, below which parton branchings re not evolved. For consistency
    it is taken as $q_{0} = \max(Q_0$,PARV(52)) where $Q_0 = $PARV(23) is
    the initial resolution scale in the hadron structure functions.
\item[ PARV(53)] : (D=4.) $Q^2$-scale of the hard scattering is multiplied by 
    PARV(53) to define the maximum parton virtuality allowed in 
    space-like branchings associated with the hard interaction. This does not 
    apply to $\hat{s}$-channel resonances, where the maximum virtuality is set 
    by $m^2=\hat s$.

\noindent
\sline
             
\end{description}
\medskip

\noindent
\lline

\noindent {\bf 5.  Time-like parton branchings:}

\noindent
\lline

\begin{description}
\item[ MSTV(60)] : (D=2) master switch for time-like branchings.
\\
    {\bf = 0 :} no branchings at all, i.e. switched off.
\\
    {\bf = 1 :} QCD type branchings of quarks and gluons.
\\
    {\bf = 2 :} also emission of photons off quarks; the photons are assumed
        on mass-shell.
\item[ MSTV(61)] : (D=2) branching mode for time-like parton evolution.
\\
    {\bf = 1 :} conventional branching, i.e. without angular ordering.
\\
    {\bf = 2 :} coherent branching, i.e. with angular ordering.
\item[ MSTV(62)] : not used.
\item[ MSTV(63)] : (D=1) choice of formation-time scheme for parton emission.
\\
    {\bf = 0 :} no formation time, i.e. instantanous emission. 
\\
    {\bf = 1 :} finite formation time sampled from an exponential distribution
    $\exp(-x)$, where $x=t/\tau$, $t$ 
    the time in the lab, and  $\tau$ the mean life-time 
    of the parton in the lab, see PARV(63).
\\
    {\bf = 2 :} finite formation time sampled from Gyulassy-Wang distribution
    $ (2/\tau) \;x/(1+x^2)^2$, where $x=t/\tau$, $t$ the time in the lab, 
    and $\tau$ the mean life-time of the parton in the lab, see PARV(63).
\item[ MSTV(64)] : not used.
\item[ MSTV(65)] : (D=0) selection of kinematics reconstruction for branchings
    initiated by a single off-shell parton.
\\
    {\bf = 0 :} conservation of energy and jet direction, but longitudinal 
    momentum is not conserved.
\\
    {\bf = 1 :} with full energy-momentum conservation at each branching, but
    the jet direction is not conserved due to recoil.
\medskip

\noindent
\sline

\item[ PARV(61)] : (D=0.29 GeV) $\Lambda$-value used in running $\alpha_s$ for
    time-like parton showers.
\item[ PARV(62)] : (D=1.0 GeV) invariant mass cutoff $\mu_0$ of timeloke 
    parton evolution,
    below which partons are not assumed to radiate. Exception: for
    the case MSTV(84)=3, for reasons of consistency,
    the value is automatically 
    scaled  to 1.5 $\times$ PARV(62). 
\item[ PARV(63)] : (D=0.01) if MSTV(63)=1 or MSTV(63)=2, the proportionality 
    between mean-life time $\tau$ of an off-shell parton and its virtuality 
    $k^2$, is parametrized as $\tau = c\,(1/\sqrt{k^2})$, with $c =$ PARV(63). 
    (Is inversely related to PARV(65)).
\item[ PARV(64)] : (D=4.) $Q^2$-scale of the hard scattering is multiplied by 
    PARV(64) to define the maximum parton virtuality allowed in 
    time-like branchings associated with the hard interaction. This does not 
    apply to $\hat{s}$-channel resonances, where the maximum virtuality is set 
    by $m^2=\hat s$.
\item[ PARV(65)] : (D=4.) invariant virtuality $k^2$ 
    of off-shell partons is multiplied
    PARV(65) to define the maximum parton virtuality allowed in time-like 
    branchings of single parton decays. (Is inversely related to 
    PARV(63)).

\noindent
\sline
\end{description}
\medskip

\noindent
\lline

\noindent {\bf 6. Parton-cluster formation and cluster-hadron decay: }

\noindent
\lline

\begin{description}
\item[ MSTV(80)] : (D=1) master switch for hadronization of parton clusters.
\\
    {\bf = 0 :} off.
\\
    {\bf = 1 :} on.
\item[ MSTV(81)] : (D=1) choice of definition of 2-parton spatial separation 
    $r_{12}= r_1-r_2$ between two partons 1, 2, for measuring
    probability of pre-hadronic cluster formation of the two partons.
\\
    {\bf = 0 :} distance $r_12$ measured in the global frame of the collision 
     event.
\\
    {\bf = 1 :} distance $r_12$ measured in the center-of-mass frame of 1 and 2.
\item[ MSTV(82)] : (D=4) choice of definition of 2-parton momentum separation 
    $p_{12}$ of two partons 1, 2, for pre-hadronic cluster search.
\\
    {\bf = 1 :} three-momentum, i.e. 
    $p_{12}=\sqrt{p_{x\;12}^2+p_{y\;12}^2+p_{z\;12}^2}$.
\\
    {\bf = 2 :} transverse momentum, i.e. $p_{12}=\sqrt{p_{x\;12}^2+p_{y\;12}^2}$.
\\
    {\bf = 3 :} `pseudo-mass', i.e. $p_{12}=\sqrt{2\,E_1E_2\,(1-p_1p_2)/(m_1m_2)}$.
\\
    {\bf = 4 :} invariant mass', i.e. 
    $p_{12}=\sqrt{E_{12}^2-p_{x\;12}^2+p_{y\;12}^2+p_{z\;12}^2}$.
\item[ MSTV(83)] : (D=1) choice of probability distribution from which cluster
    formation of 2 partons is sampled. Simulates the conversion process as
    a `tunneling' of the partons through a potential barrier set by PARV(81) 
    and PARV(82), separating perturbative and non-perturbative regimes.
\\
    {\bf = 0 :} flat distribution, i.e. $\theta$-function.
\\
    {\bf = 1 :} exponential distribution.
\item[ MSTV(84)] : (D=3) level of including color correlations among pairs of 
     partons in the process of cluster formation.
\\
    {\bf = 0 :} none, except that clustering of two partons originating from the 
     same mother is vetoed.
\\
    {\bf = 1 :} probability of two cluster candidates being in a color singlet 
     state is sampled uniformly with factors 
     $C_{gg}=1/9, \;C_{q\bar q}=1/3, \;C_{gq}=1/3$.
\\
    {\bf = 2 :} exact matching of colors and anticolors of two partons 
     is required to give color singlet(s).
\\
    {\bf = 3 :} general case, where arbitrary color configuration of two partons
     is allowed, and color singlet formation is balanced by additional gluon
     emission(s) to conserve color locally at each vertex.
\item[ MSTV(85)] : (D=0) option allowing for diquark clustering, i.e. sequential
     clustering of two  (anti)quarks plus a third (anti)quark (not yet working
     properly!).
\\
    {\bf = 0 :} off.
\\
    {\bf = 1 :} on.
\item[ MSTV(87)] : (D=0) choice of effective invariant masses 
     of partons to ensure that
     kinematical thresholds for cluster-hadron decay can be satisfied.
\\
    {\bf = 0 :} `curent' masses are used, i.e. $m_f =$ (0.05, 0.01, 0.01, 0.2, 1.5, 
     5.0, 150) GeV for $f = g, d, u, s, c, b, t$, respectively.
\\
    {\bf = 1 :} `constituent' masses are used, i.e. $m_f =$ (0.5, 0.32, 0.32, 0.5,
     1.8, 5.2, 150) GeV for $f= g, d, u, s, c, b, t$, respectively.
\item[ MSTV(88)] : (D=0) choice of measure for rough distinction between  `slow' 
     and `fast' clusters in statistics accumulation of exogamous cluster 
     production with numerical value of boundary given by PARV(88).
\\
    {\bf = 0 :} energy $E_{cl}$.
\\
    {\bf = 1 :} 3-momentum $P_{cl}$
\\
    {\bf = 2 :} cluster velocity $\beta = P_{cl}/E_{cl}$.
\\
    {\bf = 3 :} Lorentz factor $\gamma = E_{cl}/M_{cl}$. 
\item[ MSTV(89)] : (D=1) smearing of cluster rapidities within
     an interval $| y | \le y_c$, where $y_c$ is determined by PARV(89).
\item[ MSTV(90)] : (D=1) master switch for hadronization of beam/target remnants.
\\
    {\bf = 0 :} off.
\\
    {\bf = 1 :} on.
\item[ MSTV(91)] : (D=1) switch for the decay of unstable hadrons, emerging either 
     directly from cluster-decays, or from previous unstable hadron decays,
\\
    {\bf = 0 :} off.
\\
    {\bf {\boldmath $\ge 1$} :} 
    the decay chain is iterated MSTV(91) times, but at most 5 times.
\item[ MSTV(92)] : (D=1) switch for simulating baryon stopping effect in
     nucleus-nucleus collisions of protons and neutrons (and associated $\Delta$'s)  
     by redistributing them  in transverse momentum $p_\perp$ according to
     a $1/(1+b \,p_\perp^2)^4$ distribution (approximately an exponential for
     $p_\perp \le 2 $ GeV), and in longitudinal momentum $p_z$ (or rapidity) 
     according to a Gaussian distribution $\exp[-P_{beam}^2/(2 c^2)]$.
     The parameters $b$ and $c$ are given by PARV(97) and PARV(98),
     respectively.
\\

    {\bf = 0 :} off.
\\
    {\bf = 1 :} on.

\noindent
\sline

\item[ PARV(81)] : (D=3.6 GeV$^{-1}$) minimum separation of 2 partons, below which
     a cluster cannot be formed (denoted $R_\chi$ in Ref. \cite{ms37}).
\item[ PARV(82)] : (D=4. GeV$^{-1}$) critical separation of 2 partons that sets the
     space-time scale for cluster formation (denoted $R_{crt}$ 
     in Ref. \cite{ms37}).
     Is related to the average cluster size and typically about 10 \% 
     above $R_\chi$.
\item[ PARV(83)] : (D=1.0) `slope' of the exponential probability distribution 
     $W(R)=\theta(R-R_\chi)\exp(-$PARV(83)$R/R_{crt})$ for MSTV(83)$>$ 0, 
     describing
     the probability of 2 partons to form a pre-hadronic 
     cluster by tunneling through 
     the `potential barrier' marked by $R_\chi$ and $R_{crt}$, with average
     cluster size $R_{cl} =2 R_{crt}/$PARV(83).
\item[ PARV(84)] : (D=0.3 GeV) minimum invariant mass of 2 partons required to 
     form a cluster being necessarily above threshold for hadron decay.
\item[ PARV(85)] : (D=1000. GeV) maximum allowed invariant mass of 2 partons to form
     a cluster with mass $M_{cl}$, in order to provide option to suppress 
     production of  heavy clusters - in addition to the  requirement of PARV(81),PARV(82).
     Note: default corresponds to no suppression.
\item[ PARV(86)] : (D=5. GeV$^{-1}$) maximum allowed separation of 
     2 partons that are
     unambigous cluster candidates. Cluster formation is enforced above
     this value, in order to avoid unphysical separation. In this case
     the requirement $M_{cl}$ $<$ PARV(85) is overridden.
\item[ PARV(87)] : (D=1.) `slope' of exponential Hagedorn-type distribution for
     sampling the masses of clusters in their 2-body decays, given by
     $\exp(- a \cdot M_{cl}/M_0)$, where $a =$ PARV(87), 
     $M_{cl}$ is the cluster mass and 
     $M_0 = 0.2$ GeV is the `Hagedorn temperature'  (see Ref. \cite{ms37}).
\item[ PARV(88)] : (D=1. GeV) boundary for distinguishing `slow' and `fast' 
     clusters in statistics accumulation of exogamous cluster production
     (c.f. MSTV(88)).
\item[ PARV(89)] : (D=0.2) determines value of $y_c$ = PARV(89) $y_{beam}$
     for smearing of cluster 
     rapidities within an interval $| y | \le y_c$, if MSTV(89) $>$ 0.
\item[ PARV(91)] : (D=0.6) Parametric factor for charged multiplicity $N_{ch}$ 
     from soft 
     fragmentation of beam/target remnants with respect to $p\bar{p}$ 
     collisions at CM energy $\sqrt{s}$, such that $N_{ch}(\sqrt{s}) = 
     N_{ch}^{(p\bar p)}(a\cdot \sqrt{s})$, where $a=$ PARV(91)
      (if MSTV(90) = 1).
\item[ PARV(92)] : (D=1.0) Nuclear dependence  in $eA$ or $\gamma A$ collisions of 
     charged multiplicity $N_{ch}$, resulting from soft fragmentation of nuclear 
     remnant of the original nucleus with mass number $A$. It is parametrized 
     as: $N_{ch}^{(eA)}(\sqrt{s},A) = A^\alpha \; N_{ch}^{(p\bar p)}(\sqrt{s})$, 
     where $\alpha =$ PARV(92).
\item[ PARV(93)] : (D=1.0) Nuclear dependence  in $hA$ or $AB$ collisions o the 
     charged multiplicity $N_{ch}$, resulting from soft fragmentation of nuclear 
     beam or/and target remnants of the original nuclei with mass number 
     $A\; (B)$ (case of hadron $h$ corresponds to $A=1$ or $B=1$). 
     It is parametrized as: 
     $N_{ch}^{(AB)}(\sqrt{s},A,B) = \left[(A+B)/2\right]^\beta \, 
     N_{ch}^{(p\bar p)}(\sqrt{s})$, where $\beta =$ PARV(93).
\item[ PARV(94)] : (D=25.GeV$^{-1}$) Time in the global collision frame when soft
     fragmentation starts to become gradually active (default corresponds to 5 fm.)
\item[ PARV(95)] : (D=250.GeV$^{-1}$) The positions of particles produced 
     in fragmentation
     of both parton clusters and beam/target clusters, are assigned such 
     they become active only after an effective formation time 
     $\Delta t = PARV(95) \times \min(E/M^2,$PARV(95)) (default value corresponds 
     $\Delta t \le 50$ fm).
\item[ PARV(96)] : (D=10.) proportionality factor in the uncertainty relation 
     $\Delta t = PARV(96) \times E/M^2$, which determines the mean formation
     time of particles produced in soft beam/target fragmentation (c.f. PARV(95)).
\item[ PARV(97)] : (D=0.35 GeV) 
     parameter $b$ in  $1/(1+b \,p_\perp^2)^4$ distribution for simulating
     baryon stopping effect in nucleus-nucleus collisions of protons and neutrons 
    (and associated $\Delta$'s)  
     by redistributing them  in transverse momentum $p_\perp$ (c.f. MSTV(92)).
\item[ PARV(98)] : (D=0.14 GeV) 
     parameter $c$ in Gaussian distribution $\exp[-P_{beam}^2/(2 c^2)]$ for
     simulating baryon stopping effect in nucleus-nucleus collisions 
      of protons and neutrons (and associated $\Delta$'s)  
     by redistributing them  in longitudinal momentum $p_z$ (c.f. MSTV(92)).

\noindent
\sline
\end{description}
\medskip
                    
\noindent
\lline

\noindent  {\bf 7.  Other settings:}

\noindent
\lline

\begin{description}
\item[$\bullet$]
{\it Handling of errors and warnings:}
\end{description}

\begin{description}
\item[ MSTV(100)] : (D=2) check on possible errors during program execution.
\\
    {\bf = 0 :} errors do not cause any immediate action.
\\
    {\bf = 1 :} possible errors are checked and in case of problem, it is
        excited from the subprogram, but the simulation is continued 
        from there on. For the first MSTV(101) errors a message is 
        printed; after that no messages appear.
\\
    {\bf = 2 :} possible errors are checked and in case of problem, the
        the simulation is forcibly terminated. For the first MSTV(101) 
        errors a message is printed; after that no messages appear.
\item[ MSTV(101)] : (D=10) max number of errors that are printed.
\item[ MSTV(102)] : (D=1) printing of warning messages.
\\
    {\bf = 0 :} no warnings are written.
\\
    {\bf = 1 :} first MSTV(103) warnings are printed, thereafter no warnings
        appear.
\item[ MSTV(103)] : (D=10) max number of warnings that are printed.
\end{description}

\begin{description}
\item[$\bullet$]
{\it Handling of input/output:}
\end{description}

\begin{description}
\item[ MSTV(110)] : (D=1) direction of program output.
\\
    {\bf = 0 :} all output is directed to standard output unit 6.
\\
    {\bf = 1 :} output is directed to units MSTV(111)-MSTV(113).
\item[ MSTV(111)] : (D=10) unit number to which general output is directed.
\item[ MSTV(112)] : (D=6) unit number for writing `on-line' information.
\item[ MSTV(113)] : (D=20) unit number for listing warnings and error messages.
\item[ MSTV(114)] : (D=2) switch for general information on unit MSTV(111), 
    which is directed automatically to file VNIRUN.DAT. Allows to
    select thr amount of output provided on the global performance of a 
    simulation consisting of of a sample of collision events:
\\
    {\bf = 0 :} only minimal output concerning selected process and 
    main parameters is written out.
\\
    {\bf = 1 :} as =0, plus a listing the 1st event in momentum and 
    position space.
\\
    {\bf = 2 :} as =1, plus listing of number and properties of elementary
    subprocesses that occurred during the simulation.
\item[ MSTV(115)] : (D=2) switch for `on-line' information on unit MSTV(112),
    i.e. writing of initialazation and termination info, and  initial 
    and final entries for each event as the simulation goes on.
\\
    {\bf = 0 :} no `online output'.
\\
    {\bf = 1 :} only initialization and finalization info is printed.
\\
    {\bf = 2 :} continous listing of minimum event information to keep
        control of the program performance.
\end{description}

\begin{description}
\item[$\bullet$]
{\it Miscellanous control switches for subroutines:}
\end{description}

\begin{description}
\item[ MSTV(120)] : (D=0) flag to indicate initialization of certain HERWIG 
    common blocks and default values that are necessary for using
    parts of the HERWIG program for the hadronization of clusters.
    Is set equal to 1 after first intitialization call. 
\item[ MSTV(121)] : (D=0) is set to 1 before a VNIROBO call, the V vectors are 
    reset (in the particle range to be rotated/boosted)  are set to
    0 before the rotation/boost. MSTV(121) is inactive during a VNIROBO
    call and is set back to 0 upon return.
\item[ MSTV(122)] : (D=0) specifies in a VNIROBO call the type of rotation of the 
    P, R, and V vectors. The rotation is clockwise (active rotation) for 
    MSTV(122)=0 and anticlockwise (passive rotation) otherwise. The 
    value of MSTV(122) is set back to 0 upon return.
\item[ MSTV(123)] : (D=0) specifies in a VNIROBO call for performing a boost,
    whether the vectors R are boosted (MSTV(123)=0) or not (MSTV(123)=1).
    The vectors P are always boosted. The value of MSTV(123) is set back
    to 0 upon return.
\item[ MSTV(124)] : (D=1) pointer to lowest entry in particle record VNIREC to be 
     included in data analysis routines VNIANA1-VNIANA5.
\item[ MSTV(125)] : (D=100000) pointer to highest entry in particle record VNIREC 
     to be included in data analysis routines VNIANA1-VNIANA5.
\item[ MSTV(126)] : (D=20) number of lines in the beginning of the particle 
     record that are reserved for internal event history information. The 
     value should not be reduced, but can be increased if necessary.
\item[ MSTV(127)] : (D=1) in listing of the current state of the particle record
     by calling the subroutine VNILIST, it can be chosen between listing 
     either color/anticolor $C$ $A$ and the space time origin of a particle $I$
     according to the information contained  $V(I,5)$ explained above, 
     ( = 0 ), or, alternatively the mother $KMO$ and color flow information
     contained in $K(I,3)-K(I,5)$, ( = 1 ). All other listed quantities are 
     the same for either option.
\item[ MSTV(128)] : (D=0) specifies in calls to VNIEDIT and VNIROBO the classes
   of particles to be included.
   The default = 0 {\it ex}cludes all inactive or decayed particles with $K(I,1) \le 0$
   or $K(I,1) > 10$, whereas the setting = 1 {\it in}cludes any entry 
    listed in the particle record.  MSTV(128) is set back to 0 upon return.
\end{description}

\begin{description}
\item[$\bullet$]
{\it Version, date of last change:}
\end{description}

\begin{description}
\item[ MSTV(140)] : (D=1) Print-out of VNI logo on first occasion; MSTV(140) 
       is reset to 0 afterwards.
\item[ MSTV(141)] : VNI version number.
\item[ MSTV(142)] : VNI subversion number.
\item[ MSTV(143)] : year of last change of VNI version
\item[ MSTV(144)] : month of last change of VNI version.
\item[ MSTV(145)] : day of last change of VNI version.

\end{description}
\medskip

\newpage
    
\noindent
\lline
         
\noindent {\bf 8. Statistics, event study and data analysis:} 

\noindent
\lline

\begin{description}
\item[$\bullet$]
{\it  Subroutine} VNILIST: {\it Full listing of first event.}
\end{description}

\begin{description}
\item[ MSTV(150)] : (D=0) Gives an example listing of the evolution
    history of the first sucessful event in  momentum and 
    position space. Note: if the general output switch MSTV(114)
    is set $\ge$ 1, then MSTV(150) is automatically set to 1.
\end{description}

\begin{description}
\item[$\bullet$]
{\it Subroutine} VNIANA1: {\it General event statistics.}
\end{description}

\begin{description}
\item[ MSTV(151)] : (D=0) Statistics on initial parton state.
\item[ MSTV(152)] : (D=0) Number and momenta of produced particles.
\item[ MSTV(153)] : (D=0) Factorial moments.
\item[ MSTV(154)] : (D=0) Energy-energy correlations.
\item[ MSTV(155)] : (D=0) Decay channels.
\end{description}

\begin{description}
\item[$\bullet$]
{\it Subroutine} VNIANA2: {\it Statistics on hadronic observables.}
\end{description}

\begin{description}
\item[ MSTV(161)] : (D=0) Distributions of particles in $y=\ln(1/x)$.
\item[ MSTV(162)] : (D=0) Bose-Einstein correlation analysis.
\end{description}

\begin{description}
\item[$\bullet$]
{\it Subroutine} VNIANA3: {\it Statistics on pre-hadronic clusters.}
\end{description}

\begin{description}
\item[ MSTV(171)] : (D=0) Rapidity spectra of clusters $dN/dy$.
\item[ MSTV(172)] : (D=0) Longitudinal space-time spectra  $dN/dz$.
\item[ MSTV(173)] : (D=0) Transverse momentum spectra $(1/p_\perp)\;dN/dp_\perp$
\item[ MSTV(174)] : (D=0) Transverse space-time spectra $(1/r_\perp)\;dN/dr_\perp$.
\item[ MSTV(175)] : (D=0) Distributions of cluster sizes and masses.
\item[ MSTV(176)] : (D=0) Polarization profile of cluster density.
\end{description}

\begin{description}
\item[$\bullet$]
{\it Subroutine} VNIANA4: {\it Statistics on partons and produced hadrons.}
\end{description}

\begin{description}
\item[ MSTV(181)] : (D=0) Rapidity spectra of partons and hadrons $dN/dy$.
\item[ MSTV(182)] : (D=0) Longitudinal space-time spectra  $dN/dz$.
\item[ MSTV(183)] : (D=0) Transverse momentum spectra $(1/p_\perp)\;dN/dp_\perp$.
\item[ MSTV(184)] : (D=0) Transverse space-time spectra $(1/r_\perp)\;dN/dr_\perp$.
\end{description}

\newpage
\begin{description}
\item[$\bullet$]
{\it Subroutine} VNIANA5: {\it Global thermodynamic multiparticle properties.}
\end{description}

\begin{description}
\item[ MSTV(191)] : (D=0) Flavor, energy and momentum composition.
\item[ MSTV(192)] : (D=0) Flow velocity profiles.
\item[ MSTV(193)] : (D=0) Particle densities, energy densities, pressures.
                  Note: in order to use this option, the {\it flow-velocity
                  profiles must! be obtained beforehand}
                  in a seperate run, because they are assumed to be
                  read in as input (see MSTV(192)).
\end{description}

\noindent
\lline

\bigskip
\bigskip

\begin{center}
{\bf MSTW(200), PARW(200): generated quantities and statistics}
\end{center}
\bigskip   

\noindent
\lline

\noindent{\bf 1. General:}

\noindent
\lline

\begin{description}

\item[ MSTW(1)]  : total number of collision events to be generated.
\item[ MSTW(2)]  : total number of time steps per collision event.
\item[ MSTW(3)]  : physics process $A+B$ that is simulated. 
     The current available
     beam ($A$) and target ($B$) particles and the collision processes are the
     ones listed before in Table 1, with MSTW(3)=IPRO.
\item[ MSTW(4)]  : Overall Lorentz frame of reference for event siumulation.
\\
    = 1 : center-of-momentum frame (CMS) of beam and target particle.
\\
    = 2 : fixed-target frame (FIXT) with target particle at rest.
\\
    = 3 : user-defined frame (USER) with given {\it 3-momentum} of beam
          and target. Particles are assumed on the mass shell.
\\
    = 4 : user-defined frame (FOUR) with given {\it 4-momentum} 
         ( i.e., 3-momentum and energy) of beam
          and target particle. The particles need not to be on the mass shell.
\\
    = 5 : user-defined frame (FIVE) with given {\it 5-momentum} 
         ( i.e., 3-momentum, energy and mass) of beam
          and target particle. The particles need not to be on the mass shell,
          but 4-momentum and mass information must(!)  match. 
\item[ MSTW(5)] : KF flavour code for beam  particle $A$.
\item[ MSTW(6)] : KF flavour code for target  particle $B$.
\item[ MSTW(7)] : type of incoming beam particle $A$: 
      1 for lepton, 2 for hadron, and 3 for nucleus.
\item[ MSTW(8)] : type of incoming target particle $B$: 
     1 for lepton, 2 for hadron, and 3 for nucleus.
\item[ MSTW(9)] : combination of incoming beam and target particles.
\\
    {\bf = 1 :} lepton  on  lepton
\\
    {\bf = 2 :} lepton  on  hadron
\\
    {\bf = 3 :} lepton  on nucleus
\\
    {\bf = 4 :} hadron  on  lepton
\\
    {\bf = 5 :} hadron  on  hadron
\\
    {\bf = 6 :} hadron  on nucleus
\\
    {\bf = 7 :} nucleus on  lepton
\\
    {\bf = 8 :} nucleus on  hadron
\\
    {\bf = 9 :} nucleus on  nucleus
\item[ MSTW(10)] : performance flag for current event. Is  =0 at beginning, and 
    set =1 if during the evolution the event turns out to be rejectable,
    either due to unphysical kinematics, particle combinations, etc.,
    or due to numerical errors.
\item[ MSTW(11)] : current collision event.
\item[ MSTW(12)] : current time step in this collision event.
\item[ MSTW(13)] : current number of successful collision events,
     i.e. those that completed gracefully with MSTW(10)=0.
\item[ MSTW(14)] : counter for "non-diffractive" collisions events,
     i.e. those that involved at least one parton 
    collision (in  hadron or nucleus collisions).
\item[ MSTW(15)] :
Status of Lorentz transformation between different global Lorentz frames
in which the collision event is simulated  (which may be different from
the initially specified frame, c.f. MSTW(4)). The value of MSTW(15)
saves the last performed transformation:
\\
{\bf = 1 :} from global $cm$-frame to fixed-target or
user-specified frame;
\\
{\bf = 2 :} from global $cm$-frame to hadronic center-of-mass in
DIS (photon-hadron $cm$-frame);
\\
{\bf = -1 :} from fixed-target or user-specified frame
to global $cm$-frame;
\\
{\bf = -2 :} from  hadronic center-of-mass in
DIS (photon-hadron $cm$-frame) to global $cm$-frame.
\medskip

\noindent
\sline

\item[ PARW(1)] : time increment $TINC(I)$ in current time step $I$.
\item[ PARW(2)] : time $TIME(I)$ in current time step $I$.
\item[ PARW(3)] : local machine time (in CPU) passed when simulation started.
\item[ PARW(4)] : local machine time (in CPU) when simulation ended.
\item[ PARW(5)] : conversion factor CPU $\rightarrow$ seconds 
     (machine-dependent: for most processors, it is equal to 0.01, i.e. 
      1 sec = 100 CPU.
\item[ PARW(11)] : Total invariant $\sqrt{s} = E_{CM}$, i.e. the total CM energy
     of the collision system.
\item[ PARW(12)] : Invariant $s=E_{CM}^2$ mass-square of complete system.
\item[ PARW(13)] : CM energy per nucleon $E_{CM}/(A+B)$ (where 
     $A$ ($B$) is the mass number of beam (target), and $A+B$ is the total 
     number of nucleons in the system) in collisions involving nuclei. 
     Is equal to $E_{CM}$ for elementary particle- and hadronic collisions.
\item[ PARW(14)] : Invariant $s_{eff} =E_{CM}^2/(A+B)$ mass-square per nucleon 
     in collisions involving nuclei. 
     Is equal to $s$ for elementary particle- and hadronic collisions.
\item[ PARW(15)] : mass $M_A$ of beam particle.
\item[ PARW(16)] : mass $M_B$ of target particle.
\item[ PARW(17)] : longitudinal momentum $P_{z\,A}$
      of beam  particle in specified global frame.
\item[ PARW(18)] : longitudinal momentum $P_{z\,B}$
      of target  particle in specified global frame.
\item[ PARW(19)] : angle $\theta$ of rotation from CM frame to user-defined 
      frame (also, in $e^+e^-$ via $W^+W^-$, the rotation of the 
      $W$-pair along the $z$-axis).
\item[ PARW(20)] : azimuthal angle $\phi$ of rotation from CM frame to 
      user-defined frame,  corresponding to PARW(18).

\noindent
\sline

\end{description}
\medskip

\noindent
\lline

\noindent {\bf 2. Initial state of collision system:}

\noindent
\lline

\begin{description}
\item[ MSTW(21)] : number of neutrons of beam particle $A$.
\item[ MSTW(22)] : number of neutrons of target particle $B$.
\item[ MSTW(23)] : number of protons of beam particle $A$.
\item[ MSTW(24)] : number of protons of target particle $B$.
\item[ MSTW(25)] : number of $d$-valence quarks of beam particle $A$.
\item[ MSTW(26)] : number of $d$-valence quarks of target particle $B$.
\item[ MSTW(27)] : number of $u$-valence quarks of beam particle $A$.
\item[ MSTW(28)] : number of $u$-valence quarks of target particle $B$.
\item[ MSTW(29)] : number of $s$-valence quarks of beam particle $A$.
\item[ MSTW(30)] : number of $s$-valence quarks of target particle $B$.
\item[ MSTW(31)] : total number of initial state partons in collisions 
           involving one or more hadron or nucleus.
\item[ MSTW(32)] : number of initial gluons.
\item[ MSTW(33)] : $10^5\times$ number of $d$ quarks + number of $\bar{d}$ antiquarks.
\item[ MSTW(34)] : $10^5\times$ number of $u$ quarks + number of $\bar{u}$ antiquarks.
\item[ MSTW(35)] : $10^5\times$ number of $s$ quarks + number of $\bar{s}$ antiquarks.
\item[ MSTW(36)] : $10^5\times$ number of $c$ quarks + number of $\bar{c}$ antiquarks.
\item[ MSTW(37)] : $10^5\times$ number of $b$ quarks + number of $\bar{b}$ antiquarks.
\item[ MSTW(38)] : $10^5\times$ number of $t$ quarks + number of $\bar{t}$ antiquarks.
\medskip

\noindent
\sline

\item[ PARW(21)] : absolute value of velocity $\vec{\beta}_A$ of beam particle $A$.
\item[ PARW(22)] : absolute value of velocity $\vec{\beta}_B$ of target particle$B$.
\item[ PARW(23)] : Lorentz factor $\gamma_A$ of beam particle $A$.
\item[ PARW(24)] : Lorentz factor $\gamma_B$ of target particle $B$.
\item[ PARW(25)] : rapidity $Y_A$ of beam particle $A$.
\item[ PARW(26)] : rapidity $Y_B$ of target particle $B$.
\item[ PARW(27)] : radius $R_A$  (in 1/GeV) of beam particle in its restframe.
\item[ PARW(28)] : radius $R_B$ (in 1/GeV) of target particle in its restframe.
\item[ PARW(29)] : radius $R_n$ (in 1/GeV) of neutron within in nucleus.
\item[ PARW(30)] : radius $R_p$ (in 1/GeV) of proton within in nucleus.
\item[ PARW(31)] : accumulated average $Q_0$ of initial hadron (nucleus) parton 
    distributions (equals the average momentum scale of primary parton 
    collisions $\langle Q_{prim} \rangle$).
\item[ PARW(32)] : accumulated average momentum fraction $x_A$ of beam-side
    initial hadron (nucleus) parton distributions.
\item[ PARW(33)] : accumulated average momentum fraction $x_B$ of target-side
    initial hadron (nucleus) parton distributions.
\item[ PARW(34)] : 3 times the total charge of beam-side particle.
\item[ PARW(35)] : 3 times the total charge of target-side particle.
\item[ PARW(36)] : ratio $b/b_{max}$ of actual impact parameter
    $b_{min} \le b \le b_{max}$ to maximum allowed impact parameter
    of beam and target particles, if variable impact parameter is
    chosen  (c.f. MSTV(32) and PARV(32), PARV(33)).
\item[ PARW(37)] : azimuthal angle of colliding beam and target particles
    around the beam axis, if variable impact parameter is chosen.
\item[ PARW(38)] : sign of $r_x$-coordinate of beam-particle $A$ in global frame
    for variable impact parameter selection (= $-$ sign of target-particle $B$).
\item[ PARW(39)] : sign of $r_y$-coordinate of beam-particle $A$ in global frame
    for variable impact parameter selection (= $-$ sign of target-particle $B$).

\noindent
\sline

\end{description}
\medskip

\noindent
\lline

\noindent {\bf 3. Parton scatterings:}

\noindent
\lline

\begin{description}

\item[ MSTW(40)] : Total number of 2-parton collisions $a+b \rightarrow N$ 
                   ($N = 1,2$).
\item[ MSTW(41)] : Number of hard $2 \rightarrow 2$ collisions of $q+q, q+\bar{q}, 
    \bar{q}+\bar{q}$.
\item[ MSTW(42)] : Number of hard $2 \rightarrow 2$ collisions of $q+g, \bar{q}+g$.
\item[ MSTW(43)] : Number of hard $2 \rightarrow 2$ collisions of $g+g$.
\item[ MSTW(44)] : Number of soft $2 \rightarrow 2$ collisions of $q+q, q+\bar{q}, 
    \bar{q}+\bar{q}$.
\item[ MSTW(45)] : Number of soft $2 \rightarrow 2$ collisions of $q+g, \bar{q}+g$.
\item[ MSTW(46)] : Number of soft $2 \rightarrow 2$ collisions of $g+g$.
\item[ MSTW(47)] : Number of $2 \rightarrow 1$ fusions of $q+\bar{q}$.
\item[ MSTW(48)] : Number of $2 \rightarrow 1$ fusions of $q+g, \bar{q}+g$.
\item[ MSTW(49)] : Number of $2 \rightarrow 1$ fusions of $g+g$.
\medskip

\noindent
\sline

\item[ PARW(40)] : total number of `primary' (i.e., first) 
     collisions among primary partons.
\item[ PARW(41)] : accumulated average $Q^2$ of the `primary' collisions.
\item[ PARW(42)] : accumulated average $\sqrt{\hat{s}}$ of hard $2 \rightarrow 2$ 
     collisions.
\item[ PARW(43)] : accumulated average rapidity $y^\ast=y_1^\ast-y_2^\ast$ of hard 
     $2\rightarrow 2$ collisions.
\item[ PARW(44)] : accumulated average $p_\perp$ of hard 
     $2\rightarrow 2$ collisions.
\item[ PARW(45)] : accumulated average $\sqrt{\hat{s}}$ of soft
     $2\rightarrow 2$ collisions.
\item[ PARW(46)] : accumulated average rapidity $y^\ast=y_1^\ast - y_2^\ast$ of soft
     $2\rightarrow 2$ collisions.
\item[ PARW(47)] : accumulated average $p_\perp$ of soft 
     $2\rightarrow 2$ collisions.
\item[ PARW(48)] : accumulated average $\sqrt{\hat{s}}$ of hard
     $2\rightarrow 1$ collisions.
\item[ PARW(49)] : accumulated average rapidity $y^\ast=y_1^\ast - y_2^\ast$ of hard 
     $2\rightarrow 1$ collisions.
\item[ PARW(50)] : integrated  2-parton cross-section $\hat{\sigma}(\hat{s})
    =\int dp_\perp^2 (d\hat{\sigma}(\hat{s},p_\perp^2)/dp_\perp^2)$.

\noindent
\sline

\end{description}
\medskip

\noindent
\lline

\noindent {\bf 4.  Space-like parton branchings:}

\noindent
\lline

\begin{description}

\item[ MSTW(50)] : Total number of space-like branchings $a \rightarrow b+c$.
\item[ MSTW(51)] : Number of space-like processes  $CMshower \rightarrow b+c$.
\item[ MSTW(52)] : Number of space-like processes  
     $q (\bar{q}) \rightarrow q (\bar{q})+g$.
\item[ MSTW(53)] : Number of space-like processes  $ g  \rightarrow   g+g$.
\item[ MSTW(54)] : Number of space-like processes  $ g   \rightarrow  q+\bar{q}$.
\item[ MSTW(58)] : Number of space-like processes  
     $q/q~ \rightarrow q (\bar{q}) +\gamma$.

\medskip

\noindent
\sline

\item[ PARW(50)] : average $\sqrt{Q^2}$ of starting scale for 
   space-like branching evolution.
\item[ PARW(51)] : average $\sqrt{-q^2}$ of virtuality of branching particle.
\item[ PARW(52)] : average value of $x$ in $x \rightarrow x'x''$ branching.
\item[ PARW(53)] : average fraction $z = x'/x$ of space-like branchings.
\item[ PARW(54)] : average relative transverse momentum $q_\perp$ of 
     space-like branchings.
\item[ PARW(55)] : average longitudinal momentum $q_z$ of branching particle.
\item[ PARW(56)] : average energy of $q^0$ branching particle.
\item[ PARW(57)] : average invariant mass $\sqrt{|q^2|}$ of branching particle.

\noindent
\sline
             
\end{description}
\medskip

\newpage
             
\noindent
\lline

\noindent {\bf 5.  Time-like parton branchings:}

\noindent
\lline

\begin{description}

\item[ MSTW(60)] : Total number of time-like branchings $a \rightarrow b+c$.
\item[ MSTW(61)] : Number of time-like processes  $CMshower \rightarrow b+c$.
\item[ MSTW(62)] : Number of time-like processes  
    $q (\bar{q}) \rightarrow q (\bar{q})+g$.
\item[ MSTW(63)] : Number of time-like processes   $g  \rightarrow   g+g$.
\item[ MSTW(64)] : Number of time-like processes   $g  \rightarrow  q+\bar{q}$.
\item[ MSTW(68)] : Number of time-like processes  
    $q (\bar{q}) \rightarrow q (\bar{q}) + \gamma$.
\medskip

\noindent
\sline

\item[ PARW(60)] : average $\sqrt{Q^2}$ of starting scale for 
time-like branching evolution.
\item[ PARW(61)] : average $\sqrt{k^2}$ of virtuality of branching particle.
\item[ PARW(62)] : average value of $x$ in $x \rightarrow x'x''$ branching.
\item[ PARW(63)] : average energy fraction $z = x'/x$ of time-like branchings.
\item[ PARW(64)] : average relative  transverse momentum $k_\perp$ 
    of time-like branchings.
\item[ PARW(65)] : average longitudinal momentum $k_z$ of branching particle.
\item[ PARW(66)] : average energy of branching particle.
\item[ PARW(67)] : average invariant mass $\sqrt{m^2}$ of branching particle.

\noindent
\sline
\end{description}
\medskip

\noindent
\lline

\noindent {\bf 6. Parton-cluster formation and cluster-hadron decay: }

\noindent
\lline

\begin{description}

\item[ MSTW(80)] : Total number of 2-parton cluster formations.
\item[ MSTW(81)] : Total number of $gg$ clusterings, $gg \rightarrow CC + X$, 
\     where $X$ is either `nothing', or $g$, or $gg$.
\item[ MSTW(82)] : Number of processes, $gg \rightarrow CC + g$.
\item[ MSTW(83)] : Number of processes, $gg \rightarrow CC + gg$.
\item[ MSTW(84)] : Total number of $q\bar{q}$ clusterings, 
     $q\bar{q} \rightarrow CC + X$, where $X$ is either `nothing', 
     or $g$, or $gg$.
\item[ MSTW(85)] : Number of processes, $q\bar{q} \rightarrow CC + g$.
\item[ MSTW(86)] : Number of processes, $q\bar{q} \rightarrow CC + gg$.
\item[ MSTW(87)] : Total number of $gq\;(g\bar{q})$ clusterings, 
     $gq \rightarrow C + X$, where $X$ is either $q$ or $gq$.
\item[ MSTW(88)] : Number of processes, $gq \rightarrow CC + gq$.
\item[ MSTW(89)] : Total number of $(qq)q \rightarrow CC$, 
     $(\bar{q}\bar{q})\bar{q} \rightarrow CC$ clusterings.
\item[ MSTW(90)]{\bf - MSTW(99)} : Same as MSTW(80) - MSTW(89), but now for the 
      corresponding numbers of `exogamously' produced clusters, with 
      `exogamy index' $e_{cl} = (e_i+e_j)/2$ unequal to 0 or 1 
       (see Ref. \cite{ms40}).
\item[ MSTW(100)] : Total number of `fast' clusters, classified according to
      the choices of MSTV(88)] and PARV(88).
\item[ MSTW(101)] : Total number of `slow' clusters, the fraction complimentary
      to the `fast' one.
\item[ MSTW(110)] : Total number of primary "neutral" particles per event, 
      produced directly by cluster-hadron decays.
\item[ MSTW(111)] : Number of primary leptons and gauge bosons.
\item[ MSTW(112)] : Number of primary light mesons.
\item[ MSTW(113)] : Number of primary strange mesons.
\item[ MSTW(114)] : Number of primary charm and bottom mesons.
\item[ MSTW(115)] : Number of primary tensor mesons.
\item[ MSTW(116)] : Number of primary light baryons.
\item[ MSTW(117)] : Number of primary strange baryons.
\item[ MSTW(118)] : Number of primary charm and bottom baryons.
\item[ MSTW(119)] : Number of other particles.
\item[ MSTW(120)]{\bf - MSTW(129)} : Same as MSTW(110) - MSTW(119), but now for 
      charged particles only.
\item[ MSTW(130)]{\bf - MSTW(139)} : Same as MSTW(110) - MSTW(119), but now for 
      the numbers of secondary neutral particles per event (i.e. 
      those produced by decays of unstable particles).
\item[ MSTW(140)]{\bf - MSTW(149)} : Same as MSTW(130) - MSTW(139), but now for 
      charged particles only.
\medskip

\noindent
\sline

\item[ PARW(81)] : accumulated average space-time separation of clustered 
     partons.
\item[ PARW(82)] : accumulated average energy-momentum separation of clustered 
     partons.
\item[ PARW(83)] : minimum  encountered separation of any two clustered
     partons with respect to chosen space-time measure (c.f. MSTV(81)).
\item[ PARW(84)] : maximum encountered separation of any two clustered
     partons with respect to chosen space-time measure (c.f. MSTV(81)).
\item[ PARW(85)] : minimum encountered separation of any two clustered
     partons with respect to chosen energy-momentum measure (c.f. MSTV(82)).
\item[ PARW(86)] : maximum encountered separation of any two clustered
     partons with respect to chosen energy-momentum measure (c.f. MSTV(82)).
\item[ PARW(91)]{\bf - PARW(99)} : Gives the `exo(endo)-gamy distribution' 
     (see Ref. \cite{ms40}) of clusters 
     formed from partons with `exogamy' index $e_i, e_j$, such that 
     $e_{cl} = (e_i+e_j)/2$. Inital beam/target particles have $e = 0 (1)$, so that 
     $e_{cl}$ lies in the interval [0,1], which is binned as 
     $0, 0.05, 0.2, 0.3, 0.45,0.55, 0.7, 0.8, 0.95, 1$, with 
     PARW(90)-PARW(99) giving the numbers of clusters 
     in the bins. Note: only those clusters are counted that are classified as
     `fast' (c.f. MSTW(88)).
\item[ PARW(101)]{\bf - PARW(109)} : Same as PARW(91)-PARW(99), but for `slow' clusters
     only. The sums PARW(91)+PARW(101), etc., hence give the total numbers.
\item[ PARW(110)]{\bf - PARW(149)} :  not used.

\noindent
\sline
\end{description}
\medskip

\newpage
                    
\noindent
\lline
         
\noindent {\bf 8. Statistics, event study and data analysis:} 

\noindent
\lline

\begin{description}

\item[ MSTW(151)]{\bf - MSTW(160)} : 
      (D=1,21,41,61,91,121,151,181,221,301) Sequential
      numbers of time steps, at which a snapshot analysis during the
      space-time evolution of the collision system may be performed
      in routines VNIANA4 - VNIANA5.
\item[ MSTW(161)] : (D=40) Number of bins for energy distribution in
      $\ln(1/x)$ (wher $x$ is the energy fraction) in routine VNIANA2.
\item[ MSTW(162)] : not used.
\item[ MSTW(163)] : (D=40) Number of bins for pair mass distribution of 
      Bose-Einstein correlations in routine VNIANA2.
\item[ MSTW(164)-MSTW(170)] : not used.
\item[ MSTW(171)] : (D=20) Number of bins for cluster $y$-distribution in 
      routine VNIANA3. 
\item[ MSTW(172)] : not used.
\item[ MSTW(173)] : (D=20) Number of bins for cluster $r_z$-distribution in 
      routine VNIANA3. 
\item[ MSTW(174)] : not used.
\item[ MSTW(175)] : (D=20) Number of bins for cluster $p_\perp$-distribution in 
      routine VNIANA3. 
\item[ MSTW(176)] : not used.
\item[ MSTW(177)] : (D=20) Number of bins for cluster $r_\perp$-distribution in 
      routine VNIANA3. 
\item[ MSTW(178)]{\bf -MSTW(180)} : not used.
\item[ MSTW(181)] : (D=20) Number of bins for parton/hadron $y$-distribution 
      in routine VNIANA4. 
\item[ MSTW(182)] : not used.
\item[ MSTW(183)] : (D=20) Number of bins for parton/hadron $r_z$-distribution 
      in routine VNIANA4. 
\item[ MSTW(184)] : not used.
\item[ MSTW(185)] : (D=20) Number of bins for parton/hadron $p_\perp$-distribution
      in routine VNIANA4. 
\item[ MSTW(186)] : not used.
\item[ MSTW(187)] : (D=20) Number of bins for parton/hadron $r_\perp$-distribution 
      in routine VNIANA4. 
\item[ MSTW(188)]{\bf - MSTW(200)} : not used.
\medskip

\noindent
\sline

\item[ PARW(150)]{\bf - PARW(160)} : not used.

\item[ PARW(161)],{\bf PARW(162)} : 
      (D=0.,6.) Lower and upper bound for energy distribution in $\ln(1/x)$ 
      (where $x$ is the energy fraction of particles), in routine VNIANA2.
\item[ PARW(163)],{\bf PARW(164)} : (D=0.,3. GeV) Lower and upper bound 
     for pair mass 
     distribution of Bose-Einstein correlations in routine VNIANA2.
\item[ PARW(165)]{\bf - PARW(170)} : not used.

\item[ PARW(171)],{\bf PARW(172)} : (D=-8.,8.) Lower and upper bound for cluster 
      $y$-distribution in routine VNIANA3.
\item[ PARW(173)],{\bf PARW(174)} : (D=-40.,40. GeV$^{-1}$) 
      Lower and upper bound for cluster 
      $r_z$-distribution in routine VNIANA3.
\item[ PARW(175)],{\bf PARW(176)} : (D=0.,10. GeV) 
      Lower and upper bound for cluster 
      $p_\perp$-distribution in routine VNIANA3.
\item[ PARW(177)],{\bf PARW(178)} : (D=0.,40. GeV$^{-1}$) 
      Lower and upper bound for cluster 
      $r_\perp$-distribution in routine VNIANA3.
\item[ PARW(179)]{\bf - PARW(180)} : not used.

\item[ PARW(181)],{\bf PARW(182)} : (D=-8.,8.) 
      Lower and upper bound for parton/hadron
      $y$-distribution in routine VNIANA4.
\item[ PARW(183)],{\bf PARW(184)} : (D=-100.,100. GeV$^{-1}$) 
      Lower and upper bound for 
      parton/hadron $r_z$-distribution in routine VNIANA4.
\item[ PARW(185)],{\bf PARW(186)} : (D=0.,10. GeV) 
      Lower and upper bound for parton/hadron 
       $p_\perp$-distribution in routine VNIANA4.
\item[ PARW(187)],{\bf PARW(188)} : (D=0.,100. GeV$^{-1}$) 
      Lower and upper bound for parton/hadron $r_\perp$-distribution 
      in routine VNIANA4.
\item[ PARW(189)]{\bf - PARW(190)} : not used.

\item[ PARW(191)],{\bf PARW(192)} : (D=-20.,20. GeV$^{-1}$) 
      Lower and upper bound for 
      $r_z$ in ($r_z\times r_\perp$) grid for density profiles in routine VNIANA5.
      $y$-distribution in routine VNIANA4.
\item[ PARW(193)],{\bf PARW(194)} : (D=0.,10. GeV$^{-1}$) 
      Lower and upper bound for 
      $r_\perp$ in ($r_z\times r_\perp$) grid for density profiles in 
      routine VNIANA5.
\item[ PARW(195)]{\bf - PARW(200)} : not used.

\noindent
\sline

\end{description}

\noindent
\lline
\bigskip
\bigskip

\subsection{Kinematics cuts and selection of subprocesses}  
\bigskip

  The commonblock VNIKIN contains the arrays VKIN and MSUB, which may be used 
to impose specific kinematics cuts and to switch on/off certain
 subprocesses, respectively. The notation used in the following 
is that 'hat' denotes internal variables in a partonic interaction, 
while '*' is for variables in the global CM frame of the system as-a-whole. 
All dimensionful quantities are in units of GeV or GeV$^{-1}$.

  Kinematics cuts can be set by the user before the VNIXRUN call. Most of 
the cuts are directly related to the kinematical variables used in the 
evolution of the system in each event, and in the event data analysis. 
Except for the entries VKIN(7) - VKIN(12), the assignments follow the
convention of JETSET/PYTHIA, in order to allow an easy interfacing.

  Note, that in the current version of VNI, only the values 
VKIN(7)-VKIN(12) are actively used. All other quantities are related to 
parton collision processes involving one or more partons, which are only
of relevance in reactions with hadrons (nuclei) in the initial state.
As stated before, the latter are presently worked out, but not yet
included in the package.

\newpage
    
\begin{verbatim}
            COMMON/VNIKIN/VKIN(200),MSUB(200),ISET(200)
\end{verbatim}

\noindent
{\it Purpose}: To impose kinematics cuts on the particles' interactions
         and their space-time evolution.  Also, to allow the user to 
         run the program with a desired subset of processes.

\begin{description}
\item[ VKIN(1),]{\bf VKIN(2) :} (D=2.,-1.) range of allowed 
     $\hat{m} = \sqrt{\hat{s}}$ values. 
     If VKIN(2) $<$ 0, the upper limit is inactive.
\item[ VKIN(3),]{\bf VKIN(4) :} (D=2.,-1.) range of allowed 
     (D=0.,-1.) range of allowed $\hat{p}_\perp$ values for
    parton collisions, with transverse momentum 
    $\hat{p}_\perp$ is defined
    in the rest-frame of the 2-body interaction. If VKIN(4) $<$ 0,
    the upper limit is inactive. For processes which are singular
    in the limit $\hat{p}_\perp\rightarrow 0$ (see VKIN(6)), VKIN(5) provides
    an additional constraint. The VKIN(3) and VKIN(4) limits can
    also be used in $2 \rightarrow 1 \rightarrow 2$ processes. 
    Here, however, the product
    masses are not known and hence assumed vanishing in the event
    selection. The actual $\hat{p}_\perp$ range for massive products is thus
    shifted downwards with respect to the nominal one.

\item[ VKIN(5) :] 
 (D=1.) lower cutoff on $\hat{p}_\perp$ values, in addition to the
    VKIN(3) cut above, for processes which are singular in the
    limit $\hat{p}_\perp \rightarrow 0$ (see VKIN(6)).
\item[ VKIN(6) :] 
 (D=1.)  2-body parton collision  processes, which do not proceed only
    via an intermediate resonance (i.e. are $2 \rightarrow 1 \rightarrow 2$ 
    processes),
    are classified as singular in the limit $\hat{p}_\perp\rightarrow 0$,
    if either
    or both of the two final state products has a mass $m$ $<$ VKIN(6).
\\
\item[ VKIN(7),]{\bf VKIN(8) :}
   (D=-100.,100.) range of allowed particle rapidities
    $y^\ast$ in the CM frame of the overall system.
\item[ VKIN(9),]{\bf VKIN(10) :} 
    (D=0.,1000.) range (in GeV) of allowed particle
    transverse momenta $k_\perp^\ast$ perpendicular to the 
    $z$-axis in the global 
    CM frame of the event.
\item[ VKIN(11),]{\bf VKIN(12) :}
    (D=-1000.,1000) range (in GeV$^{-1}$) of allowed 
    longitudinal positions $z^\ast$ of particles from the CM of the global
    collision system.
\item[ VKIN(13),]{\bf VKIN(14) :} 
    (D=0.,1000.) range (in GeV$^{-1}$) of allowed 
    transverse distances $r_\perp^\ast$ perpendicular to the $z$-axis in the CM 
    frame of the global collision system.
\\
\item[ VKIN(21),]{\bf VKIN(22) :}
    (D=0.,1.) range of allowed $x_A$ (Bjorken-$x$) values for the parton
    of beam-side $A$ that enters  a parton collision.
\item[ VKIN(23),]{\bf VKIN(24) :}
     (D=0.,1.) range of allowed $x_B$ (Bjorken-$x$) values for the parton
    of target-side $B$ that enters  a parton collision.
\item[ VKIN(25),]{\bf VKIN(26) :} 
    (D=-1.,1.) range of allowed Feynman $x_F$ values, 
    where $x_F = x_A-x_B$.
\item[ VKIN(27),]{\bf VKIN(28) :}
    (D=-1.,1.) range of allowed $\cos(\hat{\theta})$ values
    in a  $2\rightarrow 2$ parton collision, where $\hat{\theta}$ 
    is the scattering
    angle in the rest-frame of the 2-body collision.
\\
\item[ VKIN(31),]{\bf VKIN(32) :}
    (D=2.,-1.) range of allowed $\hat{m}'$ values, where
    $\hat{m}'$ is the mass of the complete three- or four-body final state
    in $2 \rightarrow 3$ or $2 \rightarrow 4$ processes 
    (while $\hat{m}$ (without a prime), constrained in VKIN(1)
    and VKIN(2), here corresponds to the one- or two-body central
    system). If VKIN(32) $<$ 0, the upper limit is inactive.
\item[ VKIN(35),]{\bf VKIN(36) :}
    (D=0.,-1.) range of allowed $\mbox{abs}(\hat{t}) = -\hat{t}$
    values in $2 \rightarrow 2$ processes. 
    Note that for deep inelastic scattering
    this is nothing but the $Q^2$ scale of the photon vertex, 
    in the limit that initial and
    final state radiation is neglected. If VKIN(36) $<$ 0, the upper
    limit is inactive.
\item[ VKIN(37),]{\bf VKIN(38) :}
    (D=0.,-1.) range of allowed $\mbox{abs}(\hat{u}) = - \hat{u}$
    values in $2 \rightarrow 2$ processes. If VKIN(38) $<$ 0, 
    the upper limit is inactive.
\\
\item[ VKIN(39) -]{\bf VKIN(100) :}
currently not used.
\\
\item[ VKIN(101) -]{\bf VKIN(200) :} 
     reserved for internal use of storing kinematical
    variables of an event.
\\
\\
\item[ MSUB(ISUB) :]  array to be filled 
    when MSTV(14) = 1 (for MSTV(14) = 0, the
    array MSUB is initialized in VNIXRIN automatically) to choose which 
    subset of subprocesses to include in the generation. The ordering 
    follows the ISUB code given before in Table 3.
\\
{ = 0 :} the subprocess ISUB is {\it excluded}.
\\
{ = 1 :} the subprocess ISUB is {\it included}.
\\
\item[ ISET(ISUB) :] gives the type of kinematical variable selection scheme
    used for subprocess ISUB.
\\
{\bf =  1 :} 2 $\rightarrow$ 1 processes (parton fusion processes).
\\
{\bf =  2 :} 2 $\rightarrow$ 2 processes (hard and soft parton scattering processes).
\\
{\bf =  3 :} 1 $\rightarrow$ 2 processes (parton branching processes).
\\
{\bf = -1 :} process  is not yet implemented, but space is reserved.
\\
{\bf = -2 :} process  is not defined.
\end{description}
\bigskip
\bigskip

\subsection{Instructions on how to use the program}

  Two example programs are included in the package VNI-3.1:
  {\tt vnixple1.f}  and {\tt vnixple2.f}.
  The first  one, {\it  vnixple1.f}  is disguised 
as a generic skeleton program that calls the steering routines VNIXRIN, 
VNIXRUN and VNIXFIN (c.f. Sec. 3.4) on a black-box-level.
The second example {\tt vnixpl2.f} illustrates more specifically how
to extract data from a simulation by using histograms.
Below the simple example {\tt vnixple1.f} is printed.
This example 
generates 1000 $p\bar{p}$ events at $\sqrt{s} = 546$ GeV.
Each event is evolved for  a time range 
of 0 - 30 fm in the $p\bar{p}$ center-of-mass frame (the global CM frame). 


\begin{verbatim}
      PROGRAM VNIXPL1

C...Purpose: example for p+p~ collisions at CERN collider with 546 GeV.
      INCLUDE 'vni1.inc'
      CHARACTER*(8) FRAME,BEAM,TARGET


C...Required user input:
C.....number of events and final time (in fm).
      NEVT  = 1000
      TFIN  =   30.
C.....colliding particles, global Lorentz frame, and total energy.
      BEAM  = 'p+      '
      TARGET= 'p-      '
      FRAME = 'cms '
      WIN   = 546.

C.....Specify your change of default parameters here:
C     ..........


C...Initialize simulation procedure.
      CALL VNIXRIN(NEVT,TFIN,FRAME,BEAM,TARGET,WIN)


C...Begin loop over of events.
      DO 500 IEVT=1,NEVT

C.....Generate collision event.
      CALL VNIXRUN(IEVT,TFIN)

C.....Analysis of event as-a whole.
C     ..........

C...End loop over events.
  500 CONTINUE


C...Finish up simulation procedure.
      CALL VNIXFIN()


      END
\end{verbatim}

This example is rather self-explanatory, and will become
further clear with the following subsections. The "$\ldots$"
indicate where one can interface on a black-box-level with the program 
as-a-whole
without needing to dig deeper into its subroutine structure. In particular, one
can insert here histograming (either using the included
portable package "vnibook.f", or any other package such  as 
the HBOOK or PAW of the CERN library).
Note also the pre-programmed histogram options 
discussed in Sec. 3.7 (switches MSTV(150)-MSTV(200)) and Appendix B (subroutines VNIANA1-VNIANA5).

In the above example only the final state at the end of each event
is collected. If the user is interested in the actual
space-time development of the collision events, he/she can access this by
inserting in the loop over events the   following  extension:

\begin{verbatim}

C...Generate collision event; analyze evolution every 5 fm.
      TSTEP=5.
      NTIM=20

C.....Begin loop over time intervals.
      DO 400 ITIM=1,NTIM
      TRUN=ITIM*TSTEP
      IF(TRUN.GT.TFIN.OR.ITIM.EQ.NTIM) TRUN=TFIN
      CALL VNIXRUN(IEVT,TRUN)

C.....Analysis of time development.
C     ..........

C.....End loop over time intervals.
  400 IF(TRUN.EQ.TFIN) GOTO 450
  450 CONTINUE      

\end{verbatim}

\noindent
{\bf Remarks:}
\begin{description}
\item[(i)]
      In case one desires to interface with the JETSET/PYTHIA programs,
       one should write a little program with common block

\begin{verbatim}
       COMMON/LUJETS/N,K(4000,5),P(4000,5),V(4000,5) 
\end{verbatim}

       and which copies the variables $N, K, P$, and $V$ of the particle 
       record VNIREC directly.  Be aware however, that VNIREC is dimensioned 
       for 100000 entries, whereas LUJETS has space for only 4000 
       entries.
\\
\item[(ii)]
       In case one desires to interface with the HERWIG program, one 
       should write a little program with common block

\begin{verbatim}
       PARAMETER (NMXHEP=2000) 
      COMMON/HEPEVT/NEVHEP,NHEP,ISTHEP(NMXHEP),IDHEP(NMXHEP), 
     &JMOHEP(2,NMXHEP),JDAHEP(2,NMXHEP),PHEP(5,NMXHEP),VHEP(4,NMXHEP) 
\end{verbatim}

       and which calls the routine HWHEPC($MCONV,IP,NP$) with $MCONV = 1$,
       $IP = 1$,and $NP = N$. Again, watch the dimensions, 100000 versus 2000.
\end{description}
\bigskip
\bigskip
\bigskip

\subsection{Example for a typical collision event}

  As a result of running the above example program {\it vnixple1.f},
a number of files are created in the directory, all of which are called 
VNI???.DAT, where "???" are either 3 letters, or 3 digits.
The latter are data files that contain event statistics and other information
on the performance of a run. 
(see Sec. 3.7, switches MSTV(150)-MSTV(200) and Appendix B  for further options).
By default only  a few files are created:
\begin{description}
\item
VNIRUN.DAT containing
the main summary of a simulation with print-out of results,
\item
VNICOL.DAT containing a detailed  summary of all parton collisions that occurred,
\item
VNITIM.DAT containing the generated time-step grid and its real-time
translation, 
\item
VNIRLU.DAT containing the status of the random number generator,
\item
VNIERR.DAT containing error messages in case problems occur.
\end{description}

Most important is the file VNIRUN.DAT, which
gives a summary of what
occurred during a simulation on the parton cascade level, the cluster 
formation level and the hadron decay level.
One can obtain a listing of an  event by simply calling VNILIST(1)
at the end or during an event,
which gives a first impression
(for a complete listing see Appendix D): 
\begin{description}
\item[(i)]
The event record begins with the beam/target particles, here $p$ and $\bar{p}$,
which are then transformed to the center-of-mass frame (here it coincides
with the global frame of reference). 
\item[(ii)]
Then follow the
three valence quarks of $p$, and after that its intrinsic  gluons and
seaquark-pairs. After that the initial-sate valence quarks and gluons and
seaquarks of $\bar{p}$ are listed.
\item[(iii)]
Then comes the history of the cascade evolution, including
hard scatterings with shower (bremsstrahlung), subsequent
cluster-formation of the materialized partons
and hadron decay of these parton clusters.
\item[(iv)]
Finally, all non-materialized (non-interacted) partons
are recombined to a beam- and a target cluster that
undergo a soft fragmentation, again via cluster-hadron decay.
\item[(v)]
At the very end the sum of charge, three-momentum, energy, and
the total invariant mass of the collision system is printed.
\end{description}
\bigskip
\bigskip

\newpage
\begin{center}
Example of the particle record for a $p\bar{p}$ collision event
at $\sqrt{s} = 546$ GeV.
\end{center}

\begin{verbatim}

                  Particle listing (summary: momentum space P in GeV)
                  event no.    1  time step  200  at time  34.75 fm

    I particle     KS     KF  C A     P_x      P_y      P_z       E        M

    1 (p+)         15   2212  0 0      .000     .000  272.998  273.000     .938
    2 (p~-)        15  -2212  0 0      .000     .000 -272.998  273.000     .938
 ==============================================================================
    3 (p+)         15   2212  0 0      .000     .000  272.998  273.000     .938
 ==============================================================================
    4 (p~-)        15  -2212  0 0      .000     .000 -272.998  273.000     .938
 ==============================================================================
    5 (u)          13      2  1 0      .650     .048    1.786    2.319     .006
    6 (d)          13      1  1 0     -.593    -.165    2.092    2.587     .010
    7 (d~)         13     -1  0 3     -.135    -.452    6.724    6.886    -.490
    8 (d)          13      1  3 0      .413    -.211     .408     .935    -.484
    9 (g)          14     21  1 1     1.751     .556    2.232    3.398    -.682
      .
      .
 ==============================================================================
   48 (d~)         13     -1  0 2      .526    -.150  -16.659   15.157     .010
   49 (u~)         13     -2  0 3     -.121    -.114   -4.941    4.106    -.174
   50 (u)          13      2  3 0     -.252    -.044   -1.576     .935    -.249
   51 (g)          13     21  3 2     -.110     .229  -12.733   11.449    -.234
      .
      .
 ==============================================================================
      .
      .
   83 (g)          14     21  2 1     2.052    2.858    1.999    1.045   -3.909
   84 (d)          14      1  1 0      .000     .000   -8.008    8.008     .000
   85 (d)          14      1  2 0      .000     .000    8.008    8.008     .000
   86 (CMshower)   11     94  0 0     2.661    3.728   -3.710    8.110    5.571
   87 (d)          14      1  2 0     1.128     .073   -1.364    3.504    3.023
   88 (cluster)    11     91  0 0     1.116    1.615   -2.687    3.412     .750
   89 (cluster)    11     91  0 0     -.450    -.605    -.228    1.213     .922
   90 (cluster)    11     91  0 0    -1.528    -.662    1.381    2.181     .280
   91 (cluster)    11     91  0 0     1.252     .689   -1.272    3.136    2.484
   92 (cluster)    11     91  0 0      .110    -.650    5.219    5.796    2.433
   93 pi0           6    111  0 0      .660     .523    -.880    1.171     .135
   94 pi0           6    111  0 0      .461    1.090   -2.094    2.377     .135
   95 rho-          6   -213  0 0     -.334    -.562    -.332    1.068     .769
   96 pi0           6    111  0 0     -.110    -.046    -.182     .186     .135
   97 pi-           6   -211  0 0     -.678    -.312     .548    1.066     .140
   98 pi+           6    211  0 0     -.845    -.352     .547    1.203     .140
    .
    .
 ==============================================================================
  109 (Beam-REM)   11     92  0 0     -.315     .737  274.428  281.871   70.164
  110 (Targ-REM)   11     93  0 0     -.907   -1.587 -276.508  247.044 -116.181
  111 (CMF)        15    100  0 0    -1.222    -.850   -2.080  528.915  528.909
 ==============================================================================
  112 (cluster)    11     91  0 0     -.315     .737  274.428  281.871   70.164
  113 (cluster)    11     91  0 0     -.907   -1.587 -276.508  247.044 -116.181
  114 (Beam-REM)   11     92  0 0      .411     .648  127.485  127.543    3.762
  115 (a_1-)       17 -20213  0 0      .493     .228   19.200   19.250    1.275
  116 (Delta+)     17   2214  0 0     -.081     .420  108.286  108.293    1.232
  117 pi0           7    111  0 0      .490     .124    4.473    4.831     .135
  118 (rho-)       17   -213  0 0      .005     .102   14.584   14.605     .769
  119 pi0           7    111  0 0     -.160     .015   35.475   37.042     .135
  120 p+            7   2212  0 0      .085     .402   72.525   75.579     .938
  121 pi0           7    111  0 0     -.122    -.268    8.519    9.014     .135
  122 pi-           7   -211  0 0      .133     .367    5.779    6.174     .140
  123 (Targ-REM)   11     93  0 0    -1.252    -.510  -63.766   63.825    2.384
  124 (Delta~-)    17  -2214  0 0     -.779    -.303  -21.201   21.253    1.232
  125 (eta)        17    221  0 0     -.473    -.208  -42.565   42.572     .549
  126 pi0           7    111  0 0     -.072    -.254   -3.205    3.199     .135
  127 p~-           7  -2212  0 0     -.701    -.052  -18.281   18.903     .938
  128 pi0           7    111  0 0     -.179     .069  -12.277   12.622     .135
  129 pi0           7    111  0 0     -.060    -.060  -12.901   13.269     .135
  130 pi0           7    111  0 0     -.225    -.220  -17.816   18.383     .135
    .
    .
    .
  336 (rho+)       17    213  0 0     -.263     .404   -5.474    5.549     .769
  337 (omega)      17    223  0 0      .067    -.237  -31.485   31.496     .783
  338 pi0           7    111  0 0     -.342     .522   -5.018    5.113     .135
  339 pi+           7    211  0 0      .085    -.120    -.742     .657     .140
  340 pi0           7    111  0 0     -.309     .029  -10.513   10.790     .135
 ==============================================================================
                  sum:   .00           .000     .000     .000  546.000  546.000

\end{verbatim}

More information on 
pre-programmed 
print-out of results concerning for particle distributions, time evolution, 
observables, etc., can be switched on with switches MSTW(151) - MSTW(200), as 
explained in the previous Sec. 3.7.
\bigskip
\bigskip
\bigskip
\bigskip

\noindent {\bf ACKNOWLEDGEMENTS}
\medskip

The present program is, with respect to both its
physics and computational aspects, the product of several
years of gradual experience and development,
during which I profited from many colleagues, most of all
from John Ellis, Miklos Gyulassy, and Berndt M\"uller.
Thank you!

This work was supported in part by the D.O.E under contract no.
DE-AC02-76H00016.
\bigskip

\newpage

\newpage
\appendix

\section{Further physics routines}
\label{sec:appa}

Aside from the three main routines VNIXRIN, VNIXRUN, and VNIXFIN,
the physics routines described 
below form the heart of the program. In general,  
for each of the included collision processes of Table 1,
there is an initialization 
routine VNI??IN that sets up the initial state, and an evolution 
routine VNIEV?? that carries out the time evolution of the particle 
distributions in phase-space, starting from the initial state. 
The routines VNICOLL, VNISBRA, VNITBRA, etc., therein perform the 
perturbative parton evolution in terms of 
parton collisions, space-like branchings, 
and time-like branchings, respectively. 
The routines VNICHAD, VNIBHAD, etc., deal with the cluster hadronization. 
These routines are universal and independent of the process under consideration.

\begin{verbatim}
SUBROUTINE VNICOL2(NCOL)
\end{verbatim}

\noindent
{\it Purpose}: to test for all possible pairings of `alive' partons, whether a 
    collision is allowed, and, if it is allowed, to evaluate the quantities 
    characterizing the 2-body collision at the parton level according to the 
    cross-section and to choose one of the possible subprocesses (channels).

\begin{description}
\item[{\boldmath $NCOL$} :] 
      total number of found collisions.
\end{description}
\medskip

\begin{verbatim}
SUBROUTINE VNICOLL(ISUB,Q2,IP1,IP2,IP3,IP4)
\end{verbatim}

{\it Purpose}: to perform a 2-body parton collision, find outgoing 
flavors	and colors and calculate  
kinematics and  parton-parton collision.

\begin{description}
\item[{\boldmath $ISUB$} :] 
      Subprocess of 2-body collision (c.f. Table 3).
\item[{\boldmath $Q2$} :] 
      Momentum transfer scale of subprocess (c.f. MSTV(43), Sec. 3.7).
\item[{\boldmath $IP1$,}]{\boldmath $IP2$} : 
      Line numbers of paerons entering the interaction vertex.
\item[{\boldmath $IP3$,}]{\boldmath $IP4$} : 
      Line numbers of particles emerging from the interaction vertex.
      For $2\rightarrow 1$ processes $IP4$ is equal to zero.
\end{description}
\medskip

\begin{verbatim}
SUBROUTINE VNISBRA(IP1,IP2,QMAX)
\end{verbatim}

\noindent
{\it Purpose}: to generate space-like parton evolution, by backward tracing of
    branching processes down to some initial resolution scale associated
    with a non-perturbative cut-off. The performance is regulated by 
    the switches MSTV(50)-MSTV(52) and parameters PARV(51)-PARV(53) 
    (c.f. Sec. 3.7). 

\begin{description}
\item[{\boldmath $IP1, $}]{\bf {\boldmath $IP2$} :} 
    partons in lines $IP1$ and $IP2$ in the commonblock VNIREC,
    are evolved by space-like parton branchings from initial mass scale 
    QMAX. If one of the particles is a non-radiating one, then only
    one branch is evolved, and the spectator particle role is then only
    to take up recoil due to emitted particles by the radiating one.
\item[{\boldmath $QMAX$} :]
    initial mass scale, from which the two particles are evolved
    from $Q = QMAX$ 
    backwards to lower $Q < QMAX$ until the cut-off $Q=q_0 \equiv $PARV(52)
    is reached.
\end{description}
\medskip

\begin{verbatim}
SUBROUTINE VNITBRA(IP1,IP2,QMAX)
\end{verbatim}

\noindent
{\it Purpose}:
    to generate time-like parton evolution, by coherent branching
    processes within a finite time interval, $q \rightarrow qg$, 
    $q \rightarrow \gamma$, $g \rightarrow gg$, $g \rightarrow q\bar q$.
    The performance is regulated by the switches 
    MSTV(61)-MSTV(65) and parameters PARV(61)-PARV(65) (c.f. Sec. 3.7). 

\begin{description}
\item[{\boldmath $IP1 > 0$}]{\bf {\boldmath $IP2 = 0$} :} 
    generate a time-like parton branching for the parton
    in line $IP1$ in commonblock VNIREC, with maximum allowed mass
    $QMAX$. 
\item[{\boldmath $IP1 > 0$}]{\bf {\boldmath $IP2 > 0$} :} 
    generate time-like parton branchings for the two
    partons in lines $IP1$ and $IP2$ in the commonblock VNIREC,
    with maximum allowed mass for each parton $QMAX$.
\item[{\boldmath $QMAX$} :]
    the maximum allowed mass of a radiating parton, i.e. the
    starting value for the subsequent evolution. (In addition, the
    mass of a single parton may not exceed its energy, the mass of a
    parton in a system may not exceed the invariant mass of the
    system.)
\end{description}
\medskip

\begin{verbatim}
SUBROUTINE VNICHAD(I1,I2)
\end{verbatim}

\noindent
{\it Purpose}: to search at a given point of time the 
particle record for possible clustering of materialized
partons,  which are those which have been struck out from the initial state
or have been produced as secondaries by 
parton collisions or radiative processes.
    It consists of checking for each possible pair of two partons, the 
    invariant space-time distance between them, their flavor and origin, 
    the matching of color and anticolor, their common invariant mass, 
    etc.. If a pair satisfies all necessary conditions for the formation
    of one or two color singlet clusters, then they are combined and
    converted into a composite cluster with internal flavor content and
    invariant mass determined by the partons, which then decays either
    into a pair of color singlet clusters, or into a singlet cluster plus
    an additional gluon or quark.

\begin{description}
\item[{\boldmath $I1, \, I2$} :]
    first and last entry of particle record VNIREC in between the
    cluster formation is carried out by
    looping over all particles $I$ with $I1 \le I \le I2$.
\end{description}
\medskip

\begin{verbatim}
SUBROUTINE VNICFRA(I1,I2)
\end{verbatim}

\noindent
{\it Purpose}: to fragment color singlet clusters with given 
    $q\bar q$ substructure
    into final state hadrons. For each cluster its invariant mass and
    flavor content are checked, and it is determined whether it decays
    isotropically into a pair of hadrons, or converts directly into a
    single hadron of corresponding flavor. The hadron decay is described
    by a phase-space model, in which the probability for forming a 
    certain hadron state is given by the density of hadronic states with 
    mass below the cluster mass. Once a hadron state is chosen, it either
    propagates on undisturbed if it is a stable particle, or it decays 
    further according to the particle data tables, if it is a resonance.

\begin{description}
\item[{\boldmath $I1 > 0$}  :] 
    entry of particle record VNIREC from which analysis starts.
\item[{\boldmath $I2 > I2$} :] 
    entry of particle record up to which analysis extends.
\end{description}
\medskip

\begin{verbatim}
SUBROUTINE VNIBHAD(I1,I2)
\end{verbatim}

\noindent
{\it Purpose}:
    to perform beam cluster formation and hadronization. The
    remnant partons of beam/target particles, those that originate from
    the partonic coherent initial state in collisions involving hadrons 
    or nuclei and have not interacted with decohereing effect. These
    primary partons are reassembled to clusters along the beam axis,
    which are then fragmented based on the minimum bias model of the
    UA5 collaboration, such that color, flavor and energy-momentum of 
    the collision system as-a-whole are conserved.

\begin{description}
\item[{\boldmath $I1, \, I2$} :] 
    first and last entry of particle record VNIREC in between the
    soft beam/target fragmentation is carried out by
    looping over all particles $I$ with $I1 \le I \le I2$.
\end{description}
\medskip

\newpage 

\begin{verbatim}
SUBROUTINE VNIBFRA(I1,I2)
\end{verbatim}

\noindent
{\it Purpose}: to hadronize beam/target clusters consisting of remnant
    beam partons,  using parts of the HERWIG 5.7 program. Beam remnants
    are those that emerge from the coherent partonic initial state 
    without having undergone any decohering interaction, and carry the
    remaining energy, flavor and color of the system. The beam remnants
    are lumped to beam clusters which are fragmented similarly as the
    clusters arising from the parton cascade evolution.
\begin{description}
\item[{\boldmath $I1 > 0$}  :] 
    entry of particle record VNIREC from which analysis starts.
\item[{\boldmath $I2 > I1$} :] 
    entry of particle record up to which analysis extends.
\end{description}
\medskip

\begin{verbatim}
SUBROUTINE VNIPADC(I1,I2,NUST) 
\end{verbatim}

\noindent
{\it Purpose}: to decay  unstable particles resulting from cluster decays
    to hadrons. Out of these primary hadrons, those that are unstable 
    resonances are decayed according to the Particle Data, which
    gives lower mass hadrons plus possibly gauge bosons and leptons.
    These secondary decay products may decay further.
	
\begin{description}
\item[{\boldmath $I1 > 0$}  :] 
     entry of particle record VNIREC from which analysis starts.
\item[{\boldmath $I2 > I1$} :] 
     entry of particle record up to which analysis extends.
\item[{\boldmath $NUST$}    :] 
      returns number of decay products that are still unstable.
\end{description}
\bigskip
\bigskip

\newpage

\section{Event analysis routines} 
\label{sec:appb}

  The following routines collect, analyze and print out information
on the result of a simulation, and give the user immediate access
to calculated particle data, spectra, and other observables. 
The following routines are adopted (with some extensions and modifications)
from the JETSET/PYTHIA program.
\begin{itemize}
\item
The routine VNILIST provides a useful tool to obtain a
quick overall view of the state of the particle record at any time during the 
simulation of one or many events, and (in parts) the space-time history 
of the system. 
\item
 The routine VNIEDIT allows to edit the particle record, e.g. to
compress, or to exclude certain particle species from the record. It is
used frequently during the simulation procedure to clean up used but 
further on unecessary particle information.
\item
 Via the functions KVNI and PVNI, values of some frequently appearing
variables may be obtained more easily. 
\item
 The routines VNIANAL and VNIANA1 - VNIANA5 carry out various data 
accumulation duties, and print out tables at the end of the simulation, 
which are automatically directed to files named VNI???.DAT where ??? is 
a 3-digit number, specified below.
\item
 Finally, the six routines VNISPHE, VNITHRU, VNICLUS, VNICELL, 
VNIJMAS and VNIFOWO 
allow the possibility to find some global event shape 
properties as discussed in detail in the JETSET/PYTHIA documentation \cite{jetset}.
\end{itemize}
\medskip

\begin{verbatim}
SUBROUTINE VNILIST(MSEL)
\end{verbatim}

\noindent
{\it Purpose}: to list the current state of the particle record during the
     evolution of an event, or certain specified excerpts of it. Also,
     to list parton or particle data, or current parameter values.

\begin{description}
\item[{\boldmath $MSEL$}  :] determines what is to be listed.
\\
{\bf =  0 :} writes a logo with program version number and last date of 
        change; is mostly for internal use.
\\
{\bf =  1 :} gives a simple list of current event record, in an 80 column
        format suitable for viewing directly on the computer terminal.
        For each entry, the following information is given: the entry
        number I, the parton/particle name, the status code $KS$ (K(I,1)), 
        the flavour code $KF$ (K(I,2)), the color and anticolor, $C$ (L(I,1)),
        and $A$ (L(I,2)), and the three-momentum, energy and mass 
        (P(I,1)-P(I,5)). A final line contains information on total charge, 
        momentum, energy and invariant mass.
\\
{\bf =  2 :} gives a more extensive list of the current event record, in a
        132 column format, suitable for printers or workstations.
        For each entry, the following information is given: the entry
        number I, the parton/particle name (with padding as described
        for $MSEL = 1$), the status code $KS$ (K(I,1)), the flavour code
        $KF$ (K(I,2)), the color and anticolor, $C$ (L(I,1)) and $A$ (L(I,2)), 
        the origin of the particle in a 3 digit $N1 N2 N3$, where $N1$ is
        the event and $N2$ the time step in this event, when the particle
        was first produced, and $N3$ is the number of the entry in the 
        particle record that the particle occupied when it first appeared.
        This labeling allows a unique specification of its `historical'
        origin. Finally, it follows the three-momentum, energy and mass 
        (P(I,1) - P(I,5)).  A final line contains information on total
        charge, momentum, energy and invariant mass. 
\\
{\bf =  3 :} gives the same basic listing as = 2, but with an additional
        line for each entry containing information on production vertex
        position and time (V(I,1) - V(I,4)).
\\
{\bf = -1 :} gives the format of listing as = 1, but instead of momentum, 
        energy and mass, it lists now the particle's position in the 
        overall reference frame, (R(I,1) - R(I,3)), its momentum-rapidity 
        (Y(P)) and its transverse momentum (PT) with respect to the 
        $z$-axis, which is usually the jet-axis, or beam axis. 
\\
{\bf = -2 :} gives the format of listing as = 2, but as for = -1, instead
        of momentum, energy and mass, now the particle's positions,
        rapidities and transverse momenta in the overall Lorentz frame
        are listed. As for = 2, additional information on the `historical'
        origin of the particle is listed.
\\
{\bf = -3 :} gives the format of listing as = 3, again, instead of momentum, 
        energy and mass, now the particle's positions, rapidities and 
        transverse momenta are listed.
\\
{\bf = 11 :} provides a simple list of all parton/particle codes defined
        in the program, with $KF$ code and corresponding particle name.
        The list is grouped by particle kind, and only within each group
        in ascending order.
\\
{\bf = 12 :} provides a list of all parton/particle and decay data used in
        the program. Each parton/particle code is represented by one
        line containing $KF$ flavour code, $KC$ compressed code, particle
        name, antiparticle name (where appropriate), electrical and
        color charge, mass, resonance width and maximum broadening, 
        average invariant lifetime and whether the particle is considered 
        stable or not.
\\
{\bf = 13 :} gives a list of current parameter values for MSTV, PARV, and for
        MSTW and PARW. This is useful to keep check of which default values 
        were changed in a given run.
\\
{\bf = 14 :} gives a list of time grid and size of time slizes used in the 
        current run.
\end{description}
\medskip

\begin{verbatim}
SUBROUTINE VNIEDIT(MEDIT)
\end{verbatim}

\noindent
{\it Purpose}: to exclude unstable or undetectable partons/particles from the
    event record. One may also use VNIEDIT to store spare copies of
    events (specifically initial parton configuration) that can be
    recalled to allow e.g. different fragmentation schemes to be run
    through with one and the same parton configuration. Finally, an
    event which has been analysed with VNISPHE, VNITHRU or VNICLUS
    (see below) may be rotated to align the event axis with
    the $z$-direction.

\begin{description}
\item[{\boldmath $MEDIT$}  :] tells which action is to be taken.
\\
{\bf = 0 :} empty ($KS = 0$) and documentation ($KS > 20$) lines are removed.
        The partons/particles remaining are compressed in the beginning of
        the VNIREC commonblock and the $N$ value is updated accordingly.
        The event history is lost, so that information stored in K(I,3),
        K(I,4) and K(I,5) is no longer relevant.
\\
{\bf = 1 :} as = 0, but in addition all partons/particles that have
        fragmented/decayed ($KS > 10$) are removed.
\\
{\bf = 2 :} as = 1, but also all neutrinos and unknown particles (i.e.
        compressed code $KC = 0$) are removed.
\\
{\bf = 3 :} as = 2, but also all uncharged, color neutral particles are
        removed, leaving only charged, stable particles (and
        unfragmented partons, if fragmentation has not been performed).
\\
{\bf = 5 :} as = 0, but also all partons which have branched or been
        rearranged in a parton shower and all particles which have
        decayed are removed, leaving only the fragmenting parton
        configuration and the final state particles.
\\
{\bf = 11 :} remove lines with K(I,1) $<$ 0. Update event history
        information (in K(I,3) - K(I,5)) to refer to remaining entries.
\\
{\bf = 12 :} remove lines with K(I,1) = 0. Update event history
        information (in K(I,3) - K(I,5)) to refer to remaining entries.
\\
{\bf = 13 :} remove lines with K(I,1) = 11, 12 or 15, except for any line
        with K(I,2) = 94. Update event history information (in K(I,3) -
        K(I,5)) to refer to remaining entries. In particular, try to
        trace origin of daughters, for which the mother is decayed,
        back to entries not deleted.
\\
{\bf = 14 :} remove lines with K(I,1) = 13 or 14, and also any line with
        K(I,2) = 94. Update event history information (in K(I,3) -
        K(I,5)) to refer to remaining entries. In particular, try to
        trace origin of rearranged partons back through the parton shower
        history to the shower initiator.
\\
{\bf = 15 :} remove lines with K(I,1) $>$ 20. Update event history
        information (in K(I,3) - K(I,5)) to refer to remaining entries.
\\
{\bf = 21 :} all partons/particles in current event record are stored
        (as a spare copy) in bottom of commonblock VNIREC.
\\
{\bf = 22 :} partons/particles stored in bottom of event record with = 21
        are placed in beginning of record again, overwriting previous
        information there (so that e.g. a different fragmentation scheme
        can be used on the same partons). Since the copy at bottom is
        unaffected, repeated calls with = 22 can be made.
\\
{\bf = 23 :} primary partons/particles in the beginning of event record
        are marked as not fragmented or decayed, and number of entries
        N is updated accordingly. Is simple substitute for = 21 plus = 22
        when no fragmentation/decay products precede any of the
        original partons/particles.
\\
{\bf = 31 :} rotate largest axis, determined by VNISPHE, VNITHRU or VNICLUS,
        to sit along the $z$-direction, and the second largest axis into
        the $xz$ plane. For VNICLUS it can be further specified to $+z$ axis
        and $xz$ plane with $x > 0$, respectively. Requires that one of
        these routines has been called before.
\\
{\bf = 32 :} mainly intended for VNISPHE and VNITHRU, this gives a further
        alignment of the event, in addition to the one implied by = 31.
        The "slim" jet, defined as the side ($z > 0$ or $z < 0$) with the
        smallest summed $p_\perp$ over square root of number of particles,
        is rotated into the $+z$ hemisphere. In the opposite hemisphere
        (now $z < 0$), the side of $x > 0$ and $x < 0$ which has the largest
        summed $p_z$ absolute is rotated into the $z < 0$, $x > 0$ quadrant.
        Requires that VNISPHE or VNITHRU has been called before.
\end{description}
\medskip

\begin{verbatim}
FUNCTION KVNI(I,J)
\end{verbatim}

\noindent
{\it Purpose:} to provide various integer-valued event data. Note that many
    of the options available (in particular $I > 0, J \ge 14$) which
    refer to event history will not work after a VNIEDIT call.

\begin{description}
\item[{\boldmath $I = 0, J = $} :] properties referring to the complete event.
\\
{\bf = 1 :} N, total number of lines in event record.
\\
{\bf = 2 :} total number of partons/particles remaining after
        fragmentation and decay.
\\
{\bf = 6 :} three times the total charge of remaining (stable) partons
        and particles.
\\
\item[{\boldmath $I > 0, J = $} :] 
properties referring to the entry in line no. $I$ of the event record.
\\
{\bf = 1 - 5 :} K(I,1) - K(I,5), i.e. parton/particle status KS, flavour
        code KF and origin/decay product/color-flow information.
\\
{\bf = 6 :} three times parton/particle charge.
\\
{\bf = 7 :} 1 for a remaining entry, 0 for a decayed/fragmented/
        documentation entry.
\\
{\bf = 8 :} $KF$ code (K(I,2)) for a remaining entry, 0 for a
        decayed/fragmented/documentation entry.
\\
{\bf = 9 :} $KF$ code (K(I,2)) for a parton (i.e. not color-neutral entry),
        0 for a particle.
\\
{\bf = 10 :} $KF$ code (K(I,2)) for a particle (i.e. color-neutral entry),
        0 for a parton.
\\
{\bf = 11 :} compressed flavour code $KC$.
\\
{\bf = 12 :} color information code, i.e. 0 for color neutral, 1 for
        color triplet, -1 for antitriplet and 2 for octet.
\\
{\bf = 13 :} flavour of "heaviest" quark or antiquark (i.e. with largest
        code) in hadron or diquark (including sign for antiquark),
        0 else.
\\
{\bf = 14 :} generation number. Beam particles or virtual exchange
        particles are generation 0, original partons/particles generation
        1 and then 1 is added for each step in the fragmentation/decay
        chain.
\\
{\bf = 15 :} line number of ancestor, i.e. predecessor in first generation
        (generation 0 entries are disregarded).
\\
{\bf = 16 :} rank of a hadron in a jet system it belongs to. 
        Rank denotes the
        ordering in flavour space, with hadrons containing the original
        flavour of the jet having rank 1, increasing by 1 for each step
        away in flavour ordering. All decay products inherit the rank
        of their parent. Whereas the meaning of a first-rank hadron
        in a quark jet is always well-defined, the definition of higher
        ranks is only meaningful for independently fragmenting quark
        jets. In other cases, rank refers to the ordering in the actual
        simulation, which may be of little interest.
\\
{\bf = 17 :} generation number after a collapse of a jet system into one
        particle, with 0 for an entry not coming from a collapse, and
        -1 for entry with unknown history. A particle formed in a
        collapse is generation 1, and then one is added in each decay
        step.
\\
{\bf = 18 :} number of decay/fragmentation products (only defined in a
        collective sense for fragmentation).
\\
{\bf = 19 :} origin of color for showering parton, 0 else.
\\
{\bf = 20 :} origin of anticolor for showering parton, 0 else.
\\
{\bf = 21 :} position of color daughter for showering parton, 0 else.
\\
{\bf = 22 :} position of anticolor daughter for showering parton, 0 else.
\\
{\bf = 23 :} number of 2-body collisions of parton/particle.
\\
{\bf = 24 :} number of space-like decays of parton/particle.
\\
{\bf = 25 :} number of time-like decays of parton/particle.
\\
{\bf = 27 :} entry of direct mother of parton/particle.
\\
{\bf = 28 :} timestep of production of parton/particle.
\\
{\bf = 29 :} index of generation in the cascade tree of parton/particle.
\end{description}
\medskip

\begin{verbatim}
FUNCTION PVNI(I,J)
\end{verbatim}

\noindent
{\it Purpose}: to provide various real-valued event data. Note that some
    of the options available ($I > 0, J = 20-25$) are primarily intended
    for studies of systems in their respective CM frame.

\begin{description}
\item[{\boldmath $I = 0, J = $} :] properties referring to the complete event.
\\
{\bf = 1 - 4 :} sum of $p_x, p_y, p_z$ and $E$, respectively, for the stable
        remaining entries.
\\
{\bf = 5 :} invariant mass of the stable remaining entries.
\\
{\bf = 6 :} sum of electric charge of the stable remaining entries.

\item[{\boldmath $I > 0, J = $} :]
    properties referring to the entry in line no. $I$ of the event
    record.
\\
{\bf = 1 - 5 :} P(I,1) - P(I,5), i.e. normally $p_x, p_y, p_z, E$ and $m$ for
        parton/particle.
\\
{\bf = 6 :} electric charge $q$.
\\
{\bf = 7 :} momentum squared $\vec{p}^2 = p_x^2 + p_y^2 + p_z^2$.
\\
{\bf = 8 :} momentum $p = \sqrt{\vec{p}^2}$.
\\
{\bf = 9 :} transverse momentum squared $p_\perp^2 = p_x^2 + p_y^2$.
\\
{\bf = 10 :} transverse momentum $p_\perp$.
\\
{\bf = 11 :} transverse mass squared $m_\perp^2 = m^2 + p_x^2 + p_y^2$.
\\
{\bf = 12 :} transverse mass $m_\perp$.
\\
{\bf = 13 - 14 :} polar angle $\theta$ in radians (between 0 and $\pi$) or
        degrees, respectively.
\\
{\bf = 15 - 16 :} azimuthal angle $\phi$ in radians (between $-\pi$ and $\pi$)
         or degrees, respectively.
\\
{\bf = 17 :} true rapidity $y = \frac{1}{2} \ln((E+p_z)/(E-p_z))$.
\\
{\bf = 18 :} rapidity $y_\pi$ obtained by assuming that the particle is a
        pion when calculating the energy $E$, to be used in the
        formula above, from the (assumed known) momentum $p$.
\\
{\bf = 19 :} pseudorapidity $\eta = \frac{1}{2} \ln((p+p_z)/(p-p_z))$.
\\
{\bf = 20 :} momentum fraction $x_p = 2p/W$, where $W$ is the total energy of
        initial parton/particle configuration.
\\
{\bf = 21 :} $x_F = 2p_z/W$ (Feynman-$x$ if system is studied in CM frame).
\\
{\bf = 22 :} $x_\perp = 2p_\perp/W$.
\\
{\bf = 23 :} $x_E = 2E/W$.
\\
{\bf = 24 :} $z_+ = (E+p_z)/W$.
\\
{\bf = 25 :} $z_- = (E-p_z)/W$.
\end{description}
\medskip

\begin{verbatim}
SUBROUTINE VNIANAL(MSEL)
\end{verbatim}

\noindent
{\it Purpose}: to administrate in each event, which of the available analysis 
     duties (described below) are carried out, and when during the time
     evolution of the particle system data samples are taken and accumula-
     ted to perform a statistical analysis at the end. It calls the 
     routines VNIANA1 - VNIANA5 to do this job, for the options that are 
     switched on by default or by the user. The switches and their effect
     are described in Sec. 3.7.  After the last event of a run, the 
     results are automatically printed out in tabular form to files 
     called VNI???.DAT, with numbers ???, corresponding to the selected 
     analysis options.

\begin{description}
\item[{\boldmath $MSEL$} :] gives the type of action to be taken.
\\
{\bf =  0 :} all arrays for data analysis in VNIANA1 - VNIANA5 are reset.
\\
{\bf =  1 :} data samples are accumulated, stored and analyzed in VNIANA1 - 
     VNIANA5.
\\
{\bf = 10 :} final results are written to output files called VNI.???.DAT,
     where `???' is a 3-digit number specified in the descriptions below.
\end{description}
\medskip

\begin{verbatim}
SUBROUTINE VNIANA1(MSEL)
\end{verbatim}

\noindent
{\it Purpose}: to provide a number of global analyses referring to time-
    independent quantities of events as-a-whole, as e.g. total charged 
    multiplicities, etc.. Is adopted from JETSET/PYTHIA 
    and only slightly altered.
    Accumulated statistics is written out at the end of a run. When errors 
    are quoted, these refer to the uncertainty in the average value for the 
    event sample as a whole, rather than to the spread of the individual 
    events, i.e.  errors decrease like one over the square root of the 
    number of events analyzed.

\begin{description}
\item[{\boldmath $MSEL$} :]
    determines which action is to be taken. Generally, a last digit
    equal to 0 indicates that the statistics counters for this option
    is to be reset. Last digit 1 leads to an analysis of current event 
    with respect to the desired properties. The statistics accumulated 
    is output in tabular form with last digit 2, and written to disc.
\\
\\
{\bf = 10 :} statistics on parton multiplicity is reset.
\\
{\bf = 11 :} the parton content of the current event is analyzed,
        classified according to the flavour content of the hard
        interaction and the total number of partons.
\\
{\bf = 12 :} gives a table on parton multiplicity distribution, written
        to file  VNI110.DAT.
\\
\\
{\bf = 20 :} statistics on particle content is reset.
\\
{\bf = 21 :} the particle/parton content of the current event is analyzed,
        also for particles which have subsequently decayed and partons
        which have fragmented. Particles are subdivided into primary
        and secondary ones, the main principle being that primary
        particles are those produced directly, while secondary come from 
        decay of other particles. 
\\
{\bf = 22 :} gives a table of particle content in events, written to file
        VNI120.DAT.
\\        
\\
{\bf = 30 :} statistics on factorial moments is reset.
\\
{\bf = 31 :} analyzes the factorial moments of the multiplicity
        distribution in different bins of rapidity and azimuth.
\\
{\bf = 32 :} prints a table of the first four factorial moments for various 
        bins of pseudorapidity and azimuth, written to file VNI130.DAT. 
        The moments are properly normalized so that they would be unity 
        (up to statistical fluctuations) for uniform and uncorrelated particle
        production according to Poissonian statistics, but increasing
        for decreasing bin size in case of "intermittent" behaviour.
\\        
\\
{\bf = 40 :} statistics on energy-energy correlation is reset.
\\
{\bf = 41 :} the energy-energy correlation (EEC) of the current event is
        analyzed. Events are assumed given in their CM frame. The weight 
        assigned to a pair $i$ and $j$ is $2E_iE_j/W^2$, where $W$ 
        is the sum of 
        energies of all analyzed particles in the event. Energies are 
        determined from the momenta of particles and masses. Statistics is 
        accumulated for the relative angle $\theta_{ij}$, 
        ranging between 0 and 180 
        degrees, subdivided into 50 bins.
\\
{\bf = 42 :} prints a table of the energy-energy correlation (EEC) and
        its asymmetry (EECA), with errors, written to file VNI140.DAT.  
        The definition of errors is not unique. Each event is viewed as 
        one observation, i.e. an EEC and EECA distribution is obtained by 
        summing over all particle pairs of an event, and then the average 
        and spread of this event-distribution is calculated in the standard 
        fashion, with the quoted error containing a factor of 
        $1/\sqrt{\mbox{no. events}}$. 
\\
\\
{\bf = 50 :} statistics on complete final states is reset.
\\
{\bf = 51 :} analyzes the particle content of the final state of the
        current event record. During the course of the run, statistics
        is thus accumulated on how often different final states appear.
        Only final states with up to 8 particles are analyzed, and there
        is only reserved space for up to 200 different final states.
        Most high energy events have multiplicities far above 8, so the
        main use for this tool is to study the effective branching
        ratios obtained with a given decay model for e.g. charm or
        bottom hadrons. 
\\
{\bf = 52 :} gives a list of the (at most 200) channels with up to 8
        particles in the final state, with their relative branching
        ratio, written to file VNI150.DAT. The ordering is according to 
        multiplicity, and within each multiplicity according to an ascending 
        order of $KF$ codes.  The $KF$ codes of the particles belonging to a 
        given channel are given in descending order.
\end{description}
\medskip

\begin{verbatim}
SUBROUTINE VNIANA2(MSEL)
\end{verbatim}

\noindent
{\it Purpose}: to analyze final state hadron distributions and calculate
    global distributions for the total (charged) ensemble of hadronic
    final states.

\begin{description}
\item[{\boldmath $MSEL$} :]
 determines which action is to be taken, with the convention noted before. 
\\
\\
{\bf = 10 :} statistics on $y=ln(1/x)$ distribution of final hadrons is reset,
        where $x = 2 E/\sqrt{s}$ is the energy fraction of a particle.
\\
{\bf = 11 :} hadron content at the end of each event is binned in $y=\ln(1/x)$
        individually for all, and for charged, hadrons.
        Covered range in $y$ is given by PARW(161) and PARW(162).
\\
{\bf = 12 :} gives a table of the obtained $y=ln(1/x)$ spectra, written to
        file VNI210.DAT.
\\
\\
{\bf = 20 :} statistics on Bose-Einstein correlations of final state pions
        and kaons is reset.
\\
{\bf = 21 :} Bose-Einstein effects are simulated according to a simple
        parametrization, with an algorithm adopted from JETSET/PYTHIA. The 
        distribution of pair masses is stored for both cases, without and 
        with Bose-Einstein correlations taken into account.
        Covered range in pair mass is given by PARW(163) and PARW(164).
\\
{\bf = 22 :} gives a table of the obtained pair mass spectrum of `like-sign'
        pions and kaons, without and with Bose-Einstein effects, written
        to file VNI220.DAT.
\end{description}
\medskip

\begin{verbatim}
SUBROUTINE VNIANA3(MSEL)
\end{verbatim}

\noindent
{\it Purpose}: to analyze the dynamics of pre-hadronic clusters formed from 
    partons during the parton-hadron conversion stage of the system.
    The cluster size- and mass-distributions, and momentum and spatial
    spectra are accumulated at various time steps during the time 
    evolution of the system, so that the time change of the spectra 
    gives a history of the cluster formation processes, locally in
    phase-space.

\begin{description}
\item[{\boldmath $MSEL$} :]
    determines which action is to be taken, with the convention noted
    before. The statistics accumulated is output in tabular form with
    bins, ideally covering the range of momenta and coordinates of relevant
    particles. Depending on the kinematics, however, the limits of this range
    may require an adjustment, and/or a finer resolution of bins may be
    necessary.
\\
\\
{\bf = 10 :} statistics on cluster sizes and their invariant mass spectrum is 
        reset.
\\
{\bf = 11 :}  
        the cluster content of the system is analyzed each single time step 
        during an an event, and cluster sizes and masses are stored in bins.
        Covered range of analysis is $0 \le R \le 1.5$ fm, and 
        $0 \le M \le 7$ GeV.
\\
{\bf = 12 :} gives a table on the obtained cluster size- and mass-spectra,
        written to file VNI310.DAT.
\\
\\
{\bf = 20 :} statistics on rapidity ($y$) spectra of clusters is reset.
\\
{\bf = 21 :} the cluster content at certain pre-selected time steps during an
        an event is analyzed. 
        Covered range in $y$ is given by PARW(171) and PARW(172).
\\
{\bf = 22 :} gives a table on the obtained $y$-spectra, written to file VNI320.DAT.
\\
\\
{\bf = 30 :} statistics on longitudinal ($r_z$) spatial distribitions of clusters
        is reset.
\\
{\bf = 31 :} the cluster positions at certain pre-selected time steps during an
        event are analyzed, with respect to the overall CM frame and to the 
        $z$-axis, usually the jet-axis, or beam-axis. 
        Covered range in $r_z$ is given by PARW(173) and PARW(174).
\\
{\bf = 32 :} gives a table on the obtained $r_z$-spectra, 
      written to file VNI330.DAT.
\\
\\
{\bf = 40 :} statistics on transverse momentum ($p_\perp$) 
      distribitions of clusters is reset.
\\
{\bf = 41 :} the cluster momenta $p_\perp$ 
       at certain pre-selected time steps during an
       an event are analyzed, transverse with respect to the z-axis. 
        Covered range in $p_\perp$ is given by PARW(175) and PARW(176).
\\
{\bf = 42 :} gives a table on the obtained $p_\perp$-spectra, 
       written to file VNI340.DAT.
\\
\\
{\bf = 50 :} statistics on transverse ($r_\perp$) 
       spatial distribitions of clusters is reset.
\\
{\bf = 51 :} the cluster content at certain pre-selected time steps during an
        an event is analyzed, with respect to the distance transverse to the 
        $z$-axis. 
        Covered range in $r_\perp$ is given by PARW(177) and PARW(178).
\\
{\bf = 52 :} gives a table on the obtained $r_\perp$-spectra, 
       written to file VNI350.DAT.
\end{description}
\medskip

\begin{verbatim}
SUBROUTINE VNIANA4(MSEL)
\end{verbatim}

\noindent
{\it Purpose}: to analyze momentum and spatial distributions of partons and
    hadrons at various time steps during the time evolution of the system, 
    so that the time change of the spectra gives a history of the dynamics 
    from the initial state all the way to the final state of hadrons.

\begin{description}
\item[{\boldmath $MSEL$} :]
    determines which action is to be taken, with the convention noted
    before. The statistics accumulated is output in tabular form with
    bins, ideally covering the range of momenta and coordinates of relevant
    particles. Depending on the kinematics, however, the limits of this range
    may require an extension. 
\\
\\
{\bf = 10 :} statistics on rapidity ($y$) spectra of partons and hadrons is reset.
\\
{\bf = 11 :} the particle content at certain pre-selected time steps during an
        an event is analyzed, for gluons, quarks, mesons and baryons seperately.
        Covered range in $y$ is given by PARW(181) and PARW(182).
\\
{\bf = 12 :} gives a table on the obtained $y$-spectra, 
        written to file VNI410.DAT.
\\
\\
{\bf = 20 :} statistics on longitudinal ($r_z$) spatial distribitions of 
        partons and hadrons is reset.
\\
{\bf = 21 :} the particle positions at certain pre-selected time steps during an
        event are analyzed, with respect to the overall CM frame and to the 
        $z$-axis, usually the jet-axis, or beam-axis. 
        Covered range in $r_z$ is given by PARW(183) and PARW(184).
\\
{\bf = 22 :} gives a table on the obtained $r_z$-spectra, 
       written to file VNI420.DAT.
\\
\\
{\bf = 30 :} statistics on transverse momentum ($p_\perp$) 
        distribitions of partons and hadrons is reset.
\\
{\bf = 31 :} the particle momenta $p_\perp$ at 
        certain pre-selected time steps during an
        an event are analyzed, transverse with respect to the $z$-axis.
        Covered range in $p_\perp$ is given by PARW(185) and PARW(186).
\\
{\bf = 32 :} gives a table on the obtained pT-spectra, 
        written to file VNI430.DAT.
\\
\\
{\bf = 40 :} statistics on transverse ($r_\perp$) spatial distribitions of 
        partons and 
        hadrons is reset.
\\
{\bf = 41 :} the particle content at certain pre-selected time steps 
        during an an event is analyzed, 
        with respect to the distance transverse to the 
        $z$-axis.
        Covered range in $r_\perp$ is given by PARW(187) and PARW(188).
\\
{\bf = 42 :} gives a table on the obtained $r_\perp$-spectra, 
        written to file VNI440.DAT.
\end{description}
\medskip

\begin{verbatim}
SUBROUTINE VNIANA5(MSEL)
\end{verbatim}

\noindent
{\it Purpose}: to analyze the time evolution of global quantities and 
    `thermodynamic' properties of particles at various time steps during 
    the time evolution of the system. 

\begin{description}
\item[{\boldmath $MSEL$} :]
    determines which action is to be taken, with the convention noted
    before. 
\\
\\
{\bf = 10 :} statistics on time evolution of particle numbers, energy 
        partitions, longitudinal momentum fractions and  average 
        transverse momentum is reset.
\\
{\bf = 11 :} the particle content is analyzed in each single time step 
        during an event, for the different particle species individually.
\\
{\bf = 12 :} gives a table on the obtained time evolution profiles, written 
        to file VNI510.DAT.
\\
\\
{\bf = 20 :} statistics on particle flow-velocity profiles is reset.
\\
{\bf = 21 :} at certain pre-assigned time steps, the phase-space densities 
        of the different particle species are analyzed, from this the 
        local flow-velocities are calculated in cellular space, and 
        stored in 2-dimensional bins in $z$ and $r_\perp$.
\\
{\bf = 22 :} gives a table on the obtained time evolution profiles, written 
        to file VNI520.DAT.
\\
\\
{\bf = 30 :} statistics on particle densities, energy densities, and 
        pressures of the various particle species is reset.
\\
{\bf = 31 :} at certain pre-assigned time steps, the phase-space densities 
        the different particle species are analyzed, convoluted with the 
        previously obtained flow-velocity profile, and Lorentz invariant
        particle densities, energy densities, diagonal and off-diagonal
        components of the pressures are stored in 2-dimensional bins 
        in $z$ and $r_\perp$.  IMPORTANT: In order to use this option, the flow 
        velocity profiles must! be obtained beforehand in a seperate run, 
        with the same parameters and conditions, because they are needed
        to calculate the densities, and are read in as input data.
\\
{\bf = 32 :} gives a table on the obtained density evolution profiles, 
        written to file VNI530.DAT.
\end{description}
\medskip

\begin{verbatim}
SUBROUTINE VNISPHE(SPH,APL)
\end{verbatim}

\noindent
{\it Purpose:} to diagonalize the momentum tensor, i.e. find the
    eigenvalues $\lambda_1 > \lambda_2 > \lambda_3$, with sum unity,
    and the corresponding eigenvectors.
    Momentum power dependence is given by PARU(41); default corresponds
    to sphericity, PARU(41)=1. gives measures linear in momenta.
    Which particles (or partons) are used in the analysis is determined
    by the MSTU(41) value.

\begin{description}
\item[{\boldmath $SPH$} :]
   $3(\lambda_2 + \lambda_3)/2$, i.e. sphericity (for PARU(41)=2.).
   Is returned  = -1, if analysis not performed because event contained less
        than two particles (or two exactly back-to-back particles, in
        which case the two transverse directions would be undefined).
\item[{\boldmath $APL$} :]
 $3/2 \lambda_3$, i.e. aplanarity (for PARU(41)=2.).
    Is returned = -1,  as $SPH = -1$
\item[Remark:]
    the lines $N+1$ through $N+3$ ($N-2$ through $N$ for MSTU(43) = 2) in
    VNIREC will, after a call, contain the following information:
\\
    K($N+i,1) = 31$;
\\
    K($N+i,2) = 95$;
\\
    K($N+i,3) : i$, the axis number, $i=1,2,3$;
\\
    K($N+i,4)$, K($N+i,5) = 0$;
\\
    P($N+i,1)$ - P($N+i,3)$ : the $i$-th eigenvector, $x,y$ and $z$ components;
\\
    P($N+i,4) : \lambda_i$, the $i$-th eigenvalue;
\\
    P($N+i,5) = 0$;
\\
    V($N+i,1)$ - V($N+i,5) = 0$.
\\
    Also, the number of particles used in the analysis is given in
    MSTU(62).
\end{description}
\medskip

\begin{verbatim}
SUBROUTINE VNITHRU(THR,OBL)
\end{verbatim}

\noindent
{\it Purpose}: to find the thrust, major and minor axes and corresponding
    projected momentum quantities, in particular thrust and oblateness.
    The performance of the program is affected by MSTU(44), MSTU(45),
    PARU(42) and PARU(48). In particular, PARU(42) gives the momentum
    dependence, with the default value 1. corresponding to linear
    dependence. Which particles (or partons) are used in the analysis
    is determined by the MSTU(41) value.

\begin{description}
\item[{\boldmath $THR$} :]
    thrust (for PARU(42)=1.).
    Is returned = -1, if analysis cannot be not performed,
        because event contained less than two particles.
    Is returned = -2, if remaining space in VNIREC 
        (partly used as working area)
        is not large enough to allow analysis.
\item[{\boldmath $OBL$} :]
   oblateness (for PARU(42)=1.).
   Is returned = -1, respectively = -2,  as for $THR$.
\item[Remark:]
     the lines $N+1$ through $N+3$ ($N-2$ through $N$ for MSTU(43) = 2) in
    VNIREC will, after a call, contain the following information:
\\
    K($N+i,1) = 31$;
\\
    K($N+i,2) = 96$;
\\
    K($N+i,3) : i$, the axis number, $i=1,2,3$;
\\
    K($N+i,4$), K($N+i,5) = 0$;
\\
    P($N+i,1$) - P($N+i,3)$ : the thrust, major and minor axis,
        respectively, for $i =$ 1, 2 and 3;
\\
    P($N+i,4)$ : corresponding thrust, major and minor value;
\\
    P($N+i,5) = 0$;
\\
    V($N+i,1$) - V($N+i,5) = 0$.
\\
    Also, the number of particles used in the analysis is given in
    MSTU(62).
\end{description}
\medskip

\begin{verbatim}
SUBROUTINE VNICLUS(NJET)
\end{verbatim}

\noindent
{\it Purpose}: to reconstruct an arbitrary number of jets using a cluster
    analysis method based on particle momenta. Two different distance
    measures are available, the traditional one roughly corresponding
    to relative transverse momentum, and  one based on the JADE method, 
    which roughly corresponds to invariant mass (in both cases with 
    some important modifications).  The choice is controlled by MSTU(46). 
    The distance scale $d_{join}$, above which two clusters may not be joined, 
    is normally given by PARU(44). In general, $d_{join}$ may be varied to 
     describe different "jet resolution powers". With the mass distance,
    PARU(44) can be used to set the absolute maximum cluster mass, or
    PARU(45) to set the scaled one, i.e. in $y = m^2/W^2$, where $W^2$ is
    the total invariant mass-squared of the particles being considered.
    It is possible to continue the cluster search from the configuration
    already found, with a new higher $d_{join}$ scale, by selecting MSTU(48)
    properly. In MSTU(47) one can also require a minimum number of jets
    to be reconstructed; combined with an artificially large $d_{join}$ this
    can be used to reconstruct a predetermined number of jets. Which
    particles (or partons) are used in the analysis is determined by the
    MSTU(41) value, whereas assumptions about particle masses are given
    by MSTU(42). The parameters PARU(43) and PARU(48) regulate more
    technical details (for events at high energies and large
    multiplicities, however, the choice of a larger PARU(43) may be
    necessary to obtain reasonable reconstruction times).

\begin{description}
\item[{\boldmath $NJET$} :]
    the number of clusters reconstructed.
    Is returned = -1, if analysis cannot performed, 
        because event contained less than
        MSTU(47) (normally 1) particles, or analysis failed to
        reconstruct the requested number of jets.
    Is returned = -2, if remaining space in VNIREC 
        (partly used as working area)
        is not large enough to allow analysis.
\item[Remark:] if the analysis does not fail, further information is found
    in MSTU(61) - MSTU(63) and PARU(61) - PARU(63). In particular,
    PARU(61) contains the invariant mass for the system analyzed, i.e.
    the number used in determining the denominator of $y = m^2/W^2$.
    PARU(62) gives the generalized thrust, i.e. the sum of (absolute
    values of) cluster momenta divided by the sum of particle momenta
    (roughly the same as multicity). PARU(63) gives the minimum
    distance d (in $p_\perp$ or $m$) between two clusters in the final cluster
    configuration, 0 in case of only one cluster. Further, the lines
    $N+1$ through $N+NJET$ ($N-NJET+1$ through $N$ for MSTU(43) = 2) in VNIREC
    will, after a call, contain the following information:
\\
    K($N+i,1) = 31$;
\\
    K($N+i,2) = 97$;
\\
    K($N+i,3) : i$, the jet number, with the jets arranged in falling
        order of absolute momentum;
\\
    K($N+i,4) :$ the number of particles assigned to jet i;
\\
    K($N+i,5) = 0$;
\\
    P($N+i,1$) - P($N+i,5) :$ momentum, energy and invariant mass of jet i;
\\
    V($N+i,1$) - V($N+i,5) = 0$.
\\
    Also, for a particle which was used in the analysis, K($I,4) = i$,
    where $I$ is the particle number and $i$ the number of the jet it has
    ben assigned to. Undecayed particles not used then have K($I,4) = 0$.
    An exception is made for lines with K($I,1) = 3$ (which anyhow are not
    normally interesting for cluster search), where the color flow
    information stored in K($I,4)$ is left intact.
\end{description}
\medskip

\begin{verbatim}
SUBROUTINE VNICELL(NJET)
\end{verbatim}

\noindent
{\it Purpose}: to provide a simpler cluster routine more in line with
    what is currently used in the study of high-$p_\perp$ collider events.
    A detector is assumed to stretch in pseudorapidity between -PARU(51)
    and +PARU(51) and be segmented in MSTU(51) equally large $\eta$
    (pseudorapidity) bins and MSTU(52) $\phi$ (azimuthal) bins. Transverse
    energy $E_\perp$ for undecayed entries are summed up in each bin. For
    MSTU(53) nonzero, the energy is smeared by calorimetric resolution
    effects, cell by cell. This is done according to a Gaussian
    distribution; if MSTU(53) = 1 the standard deviation for the $E_\perp$ is
    PARU(55)$ \sqrt{E_\perp}$, if MSTU(53) = 2 the standard deviation for the
    $E$ is PARU(55)$ \sqrt{E}$, $E_\perp$ and $E$ expressed in GeV. 
    The Gaussian is
    cut off at 0 and at a factor PARU(56) times the correct $E_\perp$ or $E$.
    All bins with $E_\perp > $PARU(52) are taken to be possible initiators
    of jets, and are tried in falling $E_\perp$ sequence to check whether
    the total $E_\perp$ summed over cells no more distant than PARU(54) in
    $\sqrt{(\Delta\eta)^2 + (\Delta\phi)^2}$ exceeds PARU(53). If so, these
    cells define one jet, and are removed from further consideration.
    Contrary to VNICLUS, not all particles need be assigned to jets.
    Which particles (or partons) are used in the analysis is determined
    by the MSTU(41) value.

\begin{description}
\item[{\boldmath $NJET$} :]
    the number of jets reconstructed (may be 0).
    Is returned = -2, if remaining space in VNIREC (partly used as working area)
        is not large enough to allow analysis.
\item[Remark:] the lines $N+1$ through $N+NJET$ ($N-NJET+1$ through $N$ for
    MSTU(43) = 2) in VNIREC will, after a call, contain the
    following information:
\\
    K($N+i,1) = 31$;
\\
    K($N+i,2) = 98$;
\\
    K($N+i,3) : i$, the jet number, with the jets arranged in falling
        order in $E_\perp$;
\\
    K($N+i,4) :$ the number of particles assigned to jet $i$;
\\
    K($N+i,5) = 0$;
\\
    V($N+i,1$) - V($N+i,5) = 0$;
\\
\\
    Further, for MSTU(54) = 1:
\\
    P($N+i,1$), P($N+i,2) =$ position in $\eta$ and $\phi$ of the center of the
        jet initiator cell, i.e. geometrical center of jet;
\\
    P($N+i,3), P($N+i,4) = position in $\eta$ and $\phi$ of the 
        $E_\perp$-weighted
        center of the jet, i.e. the center of gravity of the jet;
\\
    P($N+i,5) = \sum E_\perp$ of the jet;
\\
\\
    while for MSTU(54) = 2:
\\
    P($N+i,1$) - P($N+i,5) :$ the jet momentum vector, constructed from the
        summed $E_\perp$ and the $\eta$ and $\phi$ of the 
        $E_\perp$-weighted center of
        the jet as $(p_x, p_y, p_z, E, m) =
        E_\perp (\cos\phi, \sin\phi, \sinh\eta, \cosh\eta, 0)$;
\\
\\
    and for MSTU(54) = 3:
\\
    P($N+i,1$) - P($N+i,5) :$ the jet momentum vector, constructed by adding
        vectorially the momentum of each cell assigned to the jet,
        assuming that all the $E_\perp$ was deposited at the center of the
        cell, and with the jet mass in P($N+i,5)$ calculated from the
        summed $E$ and $p$ as $m^2 = E^2 - p_x^2 - p_y^2 - p_z^2$.
\\
\\
    Also, the number of particles used in the analysis is given in
    MSTU(62), and the number of cells hit in MSTU(63).
\end{description}
\medskip

\begin{verbatim}
SUBROUTINE VNIJMAS(PMH,PML)
\end{verbatim}

\noindent
{\it Purpose}: to reconstruct high and low jet mass of an event, by dividing
    the collision event into two hemispheres, transversely to the 
    sphericity axis. Then one particle at a time is reassigned to the 
    other hemisphere if that reduces the sum of the two jet masses-
    squared $m_H^2 + m_L^2$. The procedure is stopped when no further
    significant change (see PARU(48)) is obtained. Often, the original
    assignment is retained as it is. Which particles (or partons)
    are used in the analysis is determined by the MSTU(41) value, whereas
    assumptions about particle masses are given by MSTU(42).

\begin{description}
\item[{\boldmath $PMH$} :] heavy jet mass (in GeV).
    Is returned = -2m, if remaining space in VNIREC 
     (partly used as working area)
      is  not large enough to allow analysis.
\item[{\boldmath $PML$} :]  light jet mass (in GeV).
    Is returned 
    = -2, if remaining space in VNIREC (partly used as working area)
        is not large enough to allow analysis.
\item[Remark:]  light jet mass (in GeV).
    After a successful call, MSTU(62) contains the number of
    particles used in the analysis, and PARU(61) the invariant mass of
    the system analyzed. The latter number is helpful in constructing
    scaled jet masses.
\end{description}
\medskip

\begin{verbatim}
SUBROUTINE VNIFOWO(H10,H20,H30,H40)
\end{verbatim}

\noindent
{\it Purpose}: to do an event analysis in terms of the Fox-Wolfram moments
    The moments $H_i$ are normalized to the lowest one, $H_0$.  Which 
    particles (or partons) are used in the analysis is determined by 
    the MSTU(41) value.

\begin{description}
\item[{\boldmath $H10$} :]
 $H_1/H_0$. Is = 0 if momentum is balanced.
\item[{\boldmath $H20$} :]
 $H_2/H_0$.
\item[{\boldmath $H30$} :]
 $H_3/H_0$.
\item[{\boldmath $H40$} :]
 $H_4/H_0$.
\item[Remark:]
    the number of particles used in the analysis is given in
    MSTU(62).
\end{description}
\bigskip
\bigskip

\newpage

\section{Further commonblocks}
\label{sec:appc}
\bigskip

The subroutines and commonblocks that one normally comes in direct
contact with have already been described. For the sake of completeness,
two further commonblocks are here described, which might be of interest 
to a user that
desires a more detailed insight.
The common block VNIINT serves as a temporary storage of
quantities during the simulation of a specific parton collision plus its
associated branching evolution.
It is very similar to a corresponding array in the JETSET/PYTHIA program.

\begin{verbatim}
                COMMON/VNIINT/MINT(400),VINT(400)
\end{verbatim}

\noindent
{\it Purpose}: to provide an intermediate storage for integer and real valued 
    variables as work area, used in the program to carry out the evolution 
    of a single partonic subsystem at a time, consisting of a parton 
    collision and its associated space-like (initial state) and time-like 
    (final state) branchings. These variables must not be changed by the 
    user.

\begin{description}
\item[ MINT(1) :]
     specifies the general type of partonic subprocess that has 
    occured, according to the ISUB code given in Table 3.
\item[ MINT(2) :] whenever MINT(1) (together with MINT(15) and MINT(16)) are
    not enough to specify the type of a $2 \rightarrow 2$
    parton collision uniquely, 
    MINT(2) provides an ordering of the different possibilities, related
    to different color-flow topologies. The different possibilities
    are discussed in Refs. \cite{bengt88}. With $i$ = MINT(15), $j$ = MINT(16)
    and $k$ = MINT(2), they are:
\begin{description}
  \item[ ]
    ISUB = 1, $i = j, q_i q_i \rightarrow q_i q_i$;
\\
        $k = 1$ : color configuration A.
\\
        $k = 2$ : color configuration B.
  \item[ ]
    ISUB = 1, $i = j, q_i q_j \rightarrow q_i q_j$;
\\
        $k = 1$ : only possibility.
  \item[ ]
    ISUB = 2, $q_i \bar{q}_i \rightarrow q_l \bar{q}_l$;
\\
        $k = 1$ : only possibility.
  \item[ ]
    ISUB = 3, $q_i \bar{q}_i \rightarrow g g$;
\\
        $k = 1$ : color configuration A.
\\
        $k = 2$ : color configuration B.
  \item[ ]
    ISUB = 6, $q_i g \rightarrow q_i g$;
\\
        $k = 1$ : color configuration A.
\\
        $k = 2$ : color configuration B.
  \item[ ]
    ISUB = 8, $g g \rightarrow q_l \bar{q}_l$;
\\
        $k = 1$ : color configuration A.
\\
        $k = 2$ : color configuration B.
  \item[ ]
    ISUB = 9, $g g \rightarrow g g$;
\\
        $k = 1$ : color configuration A.
\\
        $k = 2$ : colour configuration B.
\\
        $k = 3$ : color configuration C.
\end{description}
\item[ MINT(3) :] number of partons produced in the hard interactions, i.e.
    the number $n$ of the $2 \rightarrow n$ matrix elements used, 
     where currently
     only $n = 1$ or $n = 2$ processes are included.
\item[ MINT(4) :] number of documentation lines in common block VNIREC that are
     reserved and used as work space for simulating a parton collision.
\item[ MINT(5) :] not used.
\item[ MINT(6) :] current frame of event as initialized, cf. MSTW(4).
\item[ MINT(7) :] line number for documentation of outgoing particle 1 from 
    a parton collision $2\rightarrow 2$ or $2\rightarrow 1$.
\item[ MINT(8) :] line number for documentation of outgoing particle 2 from 
    a parton collision $2 \rightarrow 2$. Is 0 for $2 \rightarrow 1$.
\item[ MINT(9) :] not used.
\item[ MINT(10) :] not used.
\item[ MINT(11) :] $KF$ flavour 
       code of initial beam (side 1) particle (c.f. MSTW(5)).
\item[ MINT(12) :] $KF$ flavour 
       code of initial target (side 1) particle (c.f. MSTW(6)).
\item[ MINT(13) :] $KF$ flavour 
    code of side 1 parton that initiates a space-like
    cascade (only for primary partons, is 0 else).
\item[ MINT(14) :] $KF$ flavour code of 
    side 2 parton that initiates a space-like
    cascade (only for primary partons, is 0 else).
\item[ MINT(15) :] $KF$ flavour code 
    for side 1 incoming parton to collision subprocess.
\item[ MINT(16) :] $KF$ flavour code 
     for side 2 incoming parton to collision subprocess.
\\
\item[ MINT(17) -]{\bf MINT(20) :} not used.
\\
\item[ MINT(21) :] $KF$ flavour code for 
     parton 3 outgoing from a parton collision.
\item[ MINT(22) :] $KF$ flavour code for 
     parton 4 outgoing from a parton collision (is 0
     for $2\rightarrow 1$ processes).
\item[ MINT(23) :] $KF$ flavour code for 
   parton 5 outgoing from a parton collision - not
   used currently (relevant only for $2 \rightarrow 3, 2\rightarrow 4$ 
   processes).
\item[ MINT(24) :] $KF$ flavour code for parton 6 
   outgoing from a parton collision - not
   used currently (relevant only for  $2 \rightarrow  4$ processes).
\\
\item[ MINT(25) -]{\bf MINT(30) :} not used.
\\
\item[ MINT(31) -]{\bf MINT(40) :} not used.
\\
\item[ MINT(41) :] type of initial beam particle; 1 for lepton and 2 for hadron
    and 3 for nucleus (c.f. MSTW(7)).
\item[ MINT(42) :] type of initial  target 
    particle; 1 for lepton and 2 for hadron
    and 3 for nucleus (c.f. MSTW(8)).
\item[ MINT(43) :] combination of 
    incoming beam and target particles (c.f. MSTW(9)).
\item[ MINT(44) :] as MINT(43), but a photon counts as a lepton.
\item[ MINT(45), ]{\bf MINT(46) :}
    structure of incoming beam and target particles.
\\
{\bf = 1 :} no internal structure, i.e. lepton carries full beam energy.
\\
{\bf = 2 :} defined with structure functions, i.e. hadrons and nuclei.
\item[ MINT(47) :] combination of incoming beam and target particle structure
    function types.
\\
{\bf = 1 :} no structure function either for beam or target.
\\
{\bf = 2 :} structure functions for target but not for beam.
\\
{\bf = 3 :} structure functions for beam but not for target.
\\
{\bf = 4 :} structure functions both for beam and target.
\item[ MINT(48) :] total number of subprocesses switched on.
\item[ MINT(49), ]{\bf MINT(50) :} not used.
\item[ MINT(51) :] 
    internal flag that the simulation of a 2-parton collision system,
    including space-like initial state and time-like final state cascades, 
    has failed cuts.
\\
{\bf = 0 :} no problem.
\\
{\bf = 1 :} failed and to be skipped.
\\
\item[ MINT(52) -]{\bf MINT(60) :} not used.
\item[ MINT(61) -]{\bf MINT(70) :} not used.
\item[ MINT(81), ]{\bf MINT(82) :} not used.
\item[ MINT(83) :] number of lines in the event record already filled by
    previously considered pileup events.
\item[ MINT(84) :] MINT(83) + MSTP(126), i.e. number of lines already filled
    by previously considered events plus number of lines to be kept
    free for event documentation.
\\
\\
\item[ VINT(1)  :] $E_{CM}$, CM energy.
\item[ VINT(2)  :] $s =E_{CM}^2$, mass-square of complete system.
\item[ VINT(3)  :] mass of beam particle.
\item[ VINT(4)  :] mass of target particle.
\item[ VINT(5)  :] momentum of beam (and target) particle in CM frame.
\item[ VINT(6)  :] angle $\Theta$ for rotation from CM frame 
    to fixed target frame.
\item[ VINT(7)  :] angle $\Phi$ for rotation from CM 
    frame to fixed target frame.
\item[ VINT(8)  :] $\beta_x$ velocity for boost from CM frame to fixed target frame.
\item[ VINT(9)  :] $\beta_y$ velocity for boost from CM frame to fixed target frame.
\item[ VINT(10) :] $\beta_z$ velocity for boost from CM frame to fixed target frame.
\item[ VINT(11) :] $\hat{\beta}_x$ 
     velocity for boost from overall CM frame to `microscopic'
     cm frame of a parton-parton collision subsystem.
\item[ VINT(12) :] $\hat{\beta}_y$ 
     velocity for boost from overall CM frame to parton-parton 
     cm frame.
\item[ VINT(13) :] $\hat{\beta}_z$ 
     velocity for boost from overall CM frame to parton-parton 
     cm frame.
\item[ VINT(14) :] angle $\hat{\theta}$ 
      for rotation from overall CM frame to `microscopic'
     cm frame of a parton-parton collision subsystem.
\item[ VINT(15) :] angle $\hat{\phi}$ 
     for rotation from overall CM frame to parton-parton
     cm frame.
\item[ VINT(16) :] Lorentz boost factor $\gamma$  
     to parton-parton collision cm frame.
\item[ VINT(17) :] Rapidity $\hat{y}_{cm}$ of 
    2-parton cm sytem with respect to global CM frame.
\item[ VINT(18) :] Transverse momentum $\hat{p}_{\perp\;cm}$ 
 of 2-parton cm sytem in global CM frame.
\item[ VINT(19) :] Longitudinal separation 
    $\Delta\hat{r}_z$  of 2 partons in their c.m. frame
     of collision, at the point of `closest approach'.
\item[ VINT(20) :] Transverse separation 
    $\Delta\hat{r}_\perp$  of 2 partons in their c.m. frame
     of collision, at the point of `closest approach'.
\item[ VINT(21) :] $\tau = x_1x_2 s$ with $x_{1,2}$ the 
     longitudinal momentum fractions (Bjorken-$x$) of 2 colliding partons.
\item[ VINT(22) :] $\hat{y}^\ast=1/2 \ln(x_1/x_2)$, 
     the `relative' rapidity of 2 colliding partons.
\item[ VINT(23) :] $\cos(\hat{\theta})$) (scattering angle) 
    in $2\rightarrow 2$ parton collision.
\item[ VINT(24) :] $\hat{\phi}$ (azimuthal angle) in 
    in $2\rightarrow 2$ parton collision.
\\
\item[ VINT(25) -]{\bf VINT(30) :} not used.
\\
\item[ VINT(31) :] $\hat{t}_l$, lower value of kinematic limit on 
     $\hat{t}$ in parton collision.
\item[ VINT(32) :] $\hat{t}_u$, upper  value of kinematic limit on 
     $\hat{t}$ in parton collision.
\item[ VINT(33) :] $|\hat{t}|_1$, upper absolute value of 
     limit of integration over $d\hat{t}$ of 2-parton cross-section.
\item[ VINT(34) :] $|\hat{t}|_2$, lower absolute value of 
     limit of integration over $d\hat{t}$ of 2-parton cross-section.
\item[ VINT(35) :] $\cos(\hat{\theta})_{max}$ for $\hat{\theta}\le 0$.
\item[ VINT(36) :] $\cos(\hat{\theta})_{max}$ for $\hat{\theta}\ge 0$.
\item[ VINT(37) :] integrated 2-parton cross-section 
     $\hat{\sigma}(\hat{s}) = \int d\hat{p}_\perp^2 
     (d\hat{\sigma}(\hat{s},\hat{p}_\perp^2)/d\hat{p}_\perp^2$.
\item[ VINT(38) :] differential 2-parton cross-section 
     $d\hat{\sigma}(\hat{s},\hat{p}_\perp^2)/d\hat{p}_\perp^2$.
\item[ VINT(39) :] characteristic proper collision time 
     $\tau_col(p_\perp^2) \propto 1/p_\perp$ in the 2-body $c.m.$-frame.
\item[ VINT(40), ] not used.
\item[ VINT(41) :] the momentum fraction 
     $x_1$ taken by parton 1 at the 2-parton vertex.
\item[ VINT(42) :] the momentum fraction 
     $x_2$ taken by parton 2 at the 2-parton vertex.
\item[ VINT(43) :] $\hat{m} = \sqrt{\hat{s}}$, 
     mass of 2-parton collision subsystem.
\item[ VINT(44) :] $\hat{s}$ of the collision subprocess 
     ($2 \rightarrow 2, 2 \rightarrow 1$).
\item[ VINT(45) :] $\hat{t}$ of the collision subprocess 
     ($2 \rightarrow 2, 2 \rightarrow 1$).
\item[ VINT(46) :] $\hat{u}$ of the collision subprocess 
     ($2 \rightarrow 2, 2 \rightarrow 1$).
\item[ VINT(47) :] $\hat{p}_\perp$ of the collision subprocess, 
    i.e., transverse momentum evaluated in the rest frame of the collision.
\item[ VINT(48) :] $\hat{p}_\perp^2$ of the collision subprocess; see VINT(47).
\item[ VINT(49) :] $\hat{p}_{\perp\, min}$ of process, 
    i.e. PARV(42), or max(VKIN(3),VKIN(5)), 
    depending on which is larger, and whether the process is singular in 
    $\hat{p}_\perp\rightarrow 0$, or not.
\item[ VINT(50) :] not used.
\item[ VINT(51) :] $Q$ of the hard subprocess. 
    The precise definition is process-dependent, 
    see MSTV(43).
\item[ VINT(52) :] $Q^2$ of the hard subprocess; see VINT(51).
\item[ VINT(61) ]{\bf - VINT(65) :} $\theta$, $\varphi$ and
    \mbox{{\boldmath $\beta$}} for rotation and boost
    from c.m. frame to hadronic c.m. frame of a lepton-hadron 
    or lepton-nucleus event.
\end{description}
\bigskip

\begin{verbatim}
              COMMON/VNICOL/MCOL(1000,0:10),VCOL(1000,0:10)
\end{verbatim}

\noindent
{\it Purpose}: 
    to provide an intermediate storage for a minimal set of integer and 
    real valued variables for all detected parton collisions that occur
    during a single timestep during an event. The first index $I$ in the arrays
    MCOL and VCOL labels the sequence number of a collision and the second
    index $J$ is a pointer to interface with the above 
    MINT and VINT arrays wherein
    the calculated quantities are written to.

\begin{description}
\item[ MCOL(I,1) :] 
line number in VNIREC of parton 1 entering the collision subprocess.
\item[ MCOL(I,2) :] 
line number in VNIREC of parton 2 entering the collision subprocess.
\item[ MCOL(I,3) :] = MINT(1)
\item[ MCOL(I,4) :] = MINT(2)
\item[ MCOL(I,5) :] = MINT(3)
\item[ MCOL(I,6) :] = MINT(15)
\item[ MCOL(I,7) :] = MINT(16)
\item[ MCOL(I,8) :] = MINT(21)
\item[ MCOL(I,9) :] = MINT(22)
\item[ MCOL(I,10):] not used.
\\
\\
\item[ VCOL(I,1) :] = VINT(23)
\item[ VCOL(I,2) :] = VINT(24)
\item[ VCOL(I,3) :] = VINT(26)
\item[ VCOL(I,4) :] = VINT(41)
\item[ VCOL(I,5) :] = VINT(42)
\item[ VCOL(I,6) :] = VINT(44)
\item[ VCOL(I,7) :] = VINT(45)
\item[ VCOL(I,8) :] = VINT(46)
\item[ VCOL(I,9) :] = VINT(48)
\item[ VCOL(I,10):] = VINT(52)
\end{description}
\bigskip
\bigskip

\newpage

\section{Sample output}
\label{sec:appd}

\begin{center}
{\bf 1. General information on the simulation as-a-whole}
\end{center}
\smallskip

The following output stems from the example program
{\it vnixpl1.f}, simulating 1000 $p\bar{p}$ collisions at
CERN collider energy $\sqrt{s} = 546$ GeV. 
Unless switched of by setting $MSTV(114)=0$, this type of
output information is
provided automatically and written to 
the file VNIRUN.DAT at the end of a simulation. The output
includes a summary of the initialization overall collision system,
the gluon and quark content of the initial state, 
the type and number of occured interaction processes during
the parton-cascade development, the number and type of
produced pre-hadronic clusters from partons, and
the numbers of resulting final-state hadrons.
\bigskip

\begin{center}
\begin{verbatim}
 ******************************************************************************
 ******************************************************************************
 **                                                                          **
 **                                                                          **
 **   V       V   NN    N   IIIIII                                           **
 **    V     V    N N   N     II            The Monte Carlo Program of the   **
 **     V   V     N  N  N     II                                             **
 **      V V      N   N N     II                 PARTON CASCADE MODEL        **
 **       V       N    NN   IIIIII                                           **
 **                                                                          **
 **                                                                          **
 **                                                                          **
 **                                                                          **
 **          Version 3.109                                                   **
 **                                                                          **
 **  Last date of change:  6 Jan 1997            Main references:            **
 **                                                                          **
 **                                        1. K. Geiger, hep-ph/9701226      **
 **         Klaus Kinder-Geiger               submitted to Comp. Phys. Com.  **
 **                                        2. K. Geiger,                     **
 **   Brookhaven National Laboratory          Phys. Rep. 258, 237 (1995)     **
 **   Physics Department, Bldg 510-A       3. K. Geiger and B. Mueller,      **
 **      Upton, N.Y. 11973, U.S.A.            Nucl. Phys. B 369, 600 (1992)  **
 **        phone: (516) 344-3791           4. K. Geiger,                     **
 **                                           Phys. Rev. D 47, 133 (1993)    **
 **        e-mail: klaus@bnl.gov           5. J. Ellis and K. Geiger,        **
 **  http:/penguin.phy.bnl.gov/~klaus         Phys. Rev. D52, 1500 (1995)    **
 **                                                                          **
 ******************************************************************************
 ******************************************************************************

 ******************* VNIXRIN: initialization of VNI routines ******************

 ==============================================================================
 I                                                                            I
 I                       VNI will be initialized for a                        I
 I                                                                            I
 I                             p+ on p- collider                              I
 I                                                                            I
 I                  at     546.00 GeV center-of-mass energy                   I
 I                                                                            I
 I           corresponding to     546.00 GeV cm-energy per hadron             I
 I                                                                            I
 I                      h + h  ->  jets + X  ->  hadrons                      I
 I                                                                            I
 I                    Number of collision events:     10000                   I
 I                    Number of time steps per event:   193                   I
 I                                                                            I
 I          Initial time:     .00 1/GeV =       .00 fm  =   .00E+00 sec       I
 I          Final time:    166.75 1/GeV =     32.90 fm  =   .22E-22 sec       I
 I                                                                            I
 I                                                                            I
 I                  Initial parton distributions from VNI:                    I
 I                  nucleon structure functions GRV set  1                    I
 I                     pion structure functions  DO set  1                    I
 I                                                                            I
 I          Minimum initial resolution scale Q_0:           3.02 GeV          I
 I          Minimum sqrt(s) in parton collisions:           2.00 GeV          I
 I          Minimum pT_0 in parton collisions:              2.07 GeV          I
 I          Minimum q_0 in space-like branchings:           3.02 GeV          I
 I          Minimum mu_0 in time-like branchings:           1.48 GeV          I
 I                                                                            I
 I                                                                            I
 I                         Initial Collision System:                          I
 I                                                                            I
 I                                          Beam           Target             I
 I          Particle:                       p+             p-                 I
 I                                                                            I
 I          Number of neutrons:                  0              0             I
 I          Number of protons:                   1             -1             I
 I          Number of d_v quarks:                1             -1             I
 I          Number of u_v quarks:                2             -2             I
 I          Number of s_v quarks:                0              0             I
 I          Radius R_ms (fm):                 .810           .810             I
 I                                                                            I
 I          Velocity:                      .999994        .999994             I
 I          Gamma factor:                  290.505        290.505             I
 I          Rapidity:                        6.365         -6.365             I
 I                                                                            I
 I          p_x (GeV):                        .000           .000             I
 I          p_y (GeV):                        .000           .000             I
 I          p_z (GeV):                     272.998       -272.998             I
 I          E (GeV):                       273.000        273.000             I
 I          M (GeV):                          .938           .938             I
 I                                                                            I
 I          charge   px_tot    py_tot    pz_tot     E_tot     M_tot           I
 I             .0        0.        0.        0.      546.      546.           I
 I                                                                            I
 I          Initial separation parallel to beam:  Delta Z =   .25 fm          I
 I          Min/max impact transverse to beam:    .00 < B <  2.29 fm          I
 I                                                                            I
 I                                                                            I
 ==============================================================================

 ********************** VNIXRIN: initialization completed *********************



 *********************** VNIXFIN: run finished normally ***********************

 ==============================================================================
 I                                                                            I
 I          Total CPU time needed:    7 hours    8 minutes   55 seconds       I
 I                                                                            I
 ==============================================================================


 ==============================================================================
          INITIAL STATE
 ==============================================================================


       -  Internal parton structure functions selected:

          set  1 ( 1) for nucleons (pions) at initial Q0 =     2.870 GeV


       -  Minimum resolvable x-value resulting from spread of confined
          partons with 1/xP < R_conf =  .20 fm:

            Beam   p+    :       x_min =  .366E-02
          Target   p~-   :       x_min =  .366E-02


       -  Average sum of momentum fractions per hadron (nucleon):

            Beam   p+    :      <sum x> =    .9940
          Target   p~-   :      <sum x> =    .9958


       -  Average number of initial state partons:

                 gluons :    46.7329
               u+u~+d+d~:    17.7952
               u-u~+d-d~:    16.0016
                   s+s~ :     3.4325
                   c+c~ :      .6430
                   b+b~ :      .0000
                   t+t~ :      .0000

                  total :    87.6017



 ==============================================================================
          PARTON CASCADE EVOLUTION
 ==============================================================================

          Total number of events:     10000
          Total number of time steps:   193
          Initial/final time:   .0 fm  /  32.9 fm
          Analysis based on  10000 event samples

          All numbers per non-diffractive event:  fraction N_nd/N_tot =  .3870

          Actual minimum momentum transfer for hard collisions:    2.0699 GeV

          hard 2 -> 2 collisions:        switched on 
          soft 2 -> 2 collisions:        switched off
          fusions 2 -> 1:                switched on 
          space-like branchings 1 - > 2: switched on 
          time-like branchings  1 - > 2: switched on 


       -  Average number of parton collisions a + b -> c ( + d ):

            hard 2 -> 2
                q+q, q+q~, q~+q~ :     .0399
                     q+g,  q~+g  :     .3146
                           g +g  :     .7230

            soft 2 -> 2
                q+q, q+q~, q~+q~ :     .0000
                     q+g,  q~+g  :     .0000
                           g +g  :     .0000

            fusion 2 -> 1
                            q+q~ :     .0004
                     q+g,  q~+g  :     .0000
                           g +g  :     .0343

            total a+b -> c(+d)   :     1.122


       -  Average number of space-like branchings a -> b + c:

              ( CMshower -> b+c  :    1.6852 )
                 q/q~ -> q/q~+g  :     .2068
                  g   ->   g+g   :     .5309
                  g   ->   q+q~  :     .0226
               q/q~ -> q/q~+gamma:     .0000

                 total a -> b+c  :     .7603


       -  Average number of time-like branchings a -> b + c:

              ( CMshower -> b+c  :    4.3426 )
                 q/q~ -> q/q~+g  :     .3128
                  g   ->   g+g   :    2.0833
                  g   ->   q+q~  :     .1996
               q/q~ -> q/q~+gamma:     .0007

                 total a -> b+c  :    2.5964



 ==============================================================================
          PARTON CLUSTER FORMATION
 ==============================================================================

          NOTE: numbers refer to clusters from interacted partons only;
                beam/target remnant clusters (if present) are not included.

          Total number of events:     10000
          Total number of time steps:   193
          Initial/final time:   .0 fm  /  32.9 fm
          Analysis based on  10000 event samples

          All numbers per non-diffractive event:  fraction N_nd/N_tot =  .3870


       -  Total numbers of produced clusters:
          all:           3.1235
          fast:          2.3974 ( >  PARV(88)=    1.000   MSTV(88)=  0 )
          slow:           .7261 ( <  PARV(88)=    1.000   MSTV(88)=  0 )
          exogamous:      .6583

       -  Number of elementary cluster formation processes
                     type                total   exogamous
               g + g    -> C + C         .2244       .0287
               g + g    -> C + g         .5929       .0757
               g + g    -> C + g + g     .9401       .1436
               q + q~   -> C + C         .0946       .0357
               q + q~   -> C + g         .1752       .0646
               q + q~   -> C + g + g     .0000       .0000
               g + q    -> C + q         .3629       .0964
               g + q    -> C + q + g     .7334       .2136
               qq + q   -> C + C         .0000       .0000
               sum:                     3.1235       .6583

       -  Distribution of exogamous cluster processes:
             origin              all        fast        slow
            0.00 - 0.05         .3509       .2902       .0607
            0.05 - 0.20         .0427       .0262       .0165
            0.20 - 0.30         .0587       .0430       .0157
            0.30 - 0.45         .0401       .0247       .0154
            0.45 - 0.55         .2006       .1666       .0340
            0.55 - 0.70         .0521       .0304       .0217
            0.70 - 0.80         .1275       .0906       .0369
            0.80 - 0.95         .1357       .0731       .0626
            0.95 - 1.00        2.1152      1.6526       .4626



 ==============================================================================
          CLUSTER HADRONIZATION AND UNSTABLE PARTICLE DECAYS
 ==============================================================================

          NOTE: numbers refer to both, clusters from interacted partons
                as well as beam/target remnant clusters (if present).

          Total number of events:     10000
          Total number of time steps:   193
          Initial/final time:   .0 fm  /  32.9 fm
          Analysis based on  10000 event samples

          parton-cluster fragmentation:        switched on 
          beam/target-cluster fragmentation:   switched on 


       -  Average number of primary particles from cluster decays:

                          type         charged + neutral  charged only

               leptons, gauge bosons:          .0000          .0000
                        light mesons:        15.2396         7.3508
                      strange mesons:         4.5977         1.4296
                charm, bottom mesons:          .0442          .0216
                       tensor mesons:         1.4084          .7034
                       light baryons:         2.2769         1.5013
                     strange baryons:         1.7426          .3953
               charm, bottom baryons:          .0004          .0004
                               other:          .0012          .0004

                               total:        40.1700        11.4028

       -  Average number of secondary particles from unstable particle decays:

                          type         charged + neutral  charged only

               leptons, gauge bosons:         1.1555          .0018
                        light mesons:        32.2698        20.2790
                      strange mesons:         3.1257         1.5487
                charm, bottom mesons:          .0000          .0000
                       tensor mesons:          .0000          .0000
                       light baryons:         4.6264         2.3977
                     strange baryons:          .0000          .0000
               charm, bottom baryons:          .0000          .0000
                               other:          .0000          .0000

                               total:        41.0902        24.1830


 ==============================================================================


 ******************************************************************************
\end{verbatim}
\end{center}
\bigskip
\bigskip

\newpage

\begin{center}
{\bf 2. A typical event listing}
\end{center}
\smallskip

The event listing below is part of the standard output
obtained with the example program {\it vnixpl1.f}
for 1000 $p\bar{p}$ collisions at
CERN collider energy $\sqrt{s} = 546$ GeV. 
The listing, which is  is written to the file VNIRUN.DAT,
embodies the complete history of the first event
from the initial beam particles $p$ and $\bar{p}$,
their original gluon and quark content, the newly 
produced partons by hard scattering and radiation during
the parton-cascade development, the formed pre-hadronic clusters 
and resulting final state particles. The latter
emerge from both the parton clusters that result from
parton-cascade products, and the  beam/target clusters
that consist of the remnant initial partons
which have not participated in the parton cascades.
As explained in Sec. 3.6,
each entry contains name, status, flavor, and origin of a particle,
as well as it's three-momentum, energy and mass. Particles in
brackets are inactive, decayed or fragmented particles. 
The final line of the listing shows the total charge, three-momentum,
anargy and invariant mass of the particle system, given by
the sums over all active, stable particles.
\bigskip

\begin{center}
{\small
\begin{verbatim}
                  Particle listing (summary: momentum space P in GeV)
                  event no.    1  time step  200  at time  34.75 fm

    I particle     KS     KF  C A     P_x      P_y      P_z       E        M

    1 (p+)         15   2212  0 0      .000     .000  272.998  273.000     .938
    2 (p~-)        15  -2212  0 0      .000     .000 -272.998  273.000     .938
 ==============================================================================
    3 (p+)         15   2212  0 0      .000     .000  272.998  273.000     .938
 ==============================================================================
    4 (p~-)        15  -2212  0 0      .000     .000 -272.998  273.000     .938
 ==============================================================================
    5 (u)          13      2  1 0      .650     .048    1.786    2.319     .006
    6 (d)          13      1  1 0     -.593    -.165    2.092    2.587     .010
    7 (d~)         13     -1  0 3     -.135    -.452    6.724    6.886    -.490
    8 (d)          13      1  3 0      .413    -.211     .408     .935    -.484
    9 (g)          14     21  1 1     1.751     .556    2.232    3.398    -.682
   10 (u)          13      2  2 0      .263    -.405     .751    1.258    -.507
   11 (u~)         13     -2  0 1     -.056    -.437     .408     .935    -.460
   12 (d)          13      1  1 0     -.034    -.126     .826    1.328    -.149
   13 (d~)         13     -1  0 3      .027     .378     .408     .935    -.359
   14 (d~)         13     -1  0 3     -.287     .276    4.047    4.364    -.376
   15 (d)          13      1  1 0      .017     .556     .408     .935    -.536
   16 (u~)         13     -2  0 3      .396     .020    7.817    7.916    -.409
   17 (u)          13      2  3 0      .076    -.052     .473     .996    -.114
   18 (g)          13     21  1 1     -.050     .419     .539    1.058    -.400
   19 (g)          13     21  1 2      .197    -.140    1.146    1.630    -.263
   20 (d)          13      1  2 0     -.585     .022     .751    1.258    -.574
   21 (d~)         13     -1  0 1     -.054     .096     .408     .935    -.087
   22 (g)          13     21  1 2     -.223    -.221     .903    1.401    -.322
   23 (u~)         13     -2  0 2     -.473    -.482     .408     .935    -.683
   24 (u)          13      2  1 0     -.260     .767     .408     .935    -.786
   25 (u)          13      2  1 0      .096     .052    2.199    2.622    -.112
   26 (u~)         13     -2  0 2      .178     .298     .408     .935    -.336
   27 (g)          13     21  1 3      .286     .066    1.231    1.710    -.301
   28 (g)          13     21  1 2      .434    -.103    1.886    2.328    -.463
   29 (u~)         13     -2  0 3      .131    -.005     .408     .935    -.145
   30 (u)          13      2  1 0     -.138    -.218     .408     .935    -.271
   31 (s)          13      3  2 0     -.510     .046    2.092    2.522    -.537
   32 (s~)         13     -3  0 2     -.105    -.467     .408     .935    -.535
   33 (g)          13     21  1 3     -.238     .440    2.645    3.043    -.477
   34 (u)          13      2  2 0     -.026     .071    8.787    8.831    -.062
   35 (u~)         13     -2  0 2      .274    -.108     .408     .935    -.313
   36 (g)          13     21  1 2     -.266    -.039   47.203   45.031    -.309
   37 (u~)         13     -2  0 1     -.189     .247     .826    1.328    -.288
   38 (u)          13      2  3 0     -.118     .433     .408     .935    -.426
   39 (u~)         13     -2  0 3      .409     .237    2.762    3.153    -.473
   40 (u)          13      2  3 0     -.384    -.308     .408     .935    -.497
   41 (u)          13      2  1 0     -.001     .148     .408     .935    -.128
   42 (u~)         13     -2  0 2      .040    -.094     .408     .935    -.126
   43 (g)          13     21  1 3      .451     .038    1.688    2.141    -.463
   44 (d~)         13     -1  0 3     -.236     .582    1.593    2.051    -.605
   45 (d)          13      1  2 0     -.026    -.061     .408     .935    -.084
   46 (g)          13     21  3 1      .233    -.474  163.962  155.058    -.792
   47 (u~)         13     -2  0 2     -.126     .025   -9.757    8.645     .006
 ==============================================================================
   48 (d~)         13     -1  0 2      .526    -.150  -16.659   15.157     .010
   49 (u~)         13     -2  0 3     -.121    -.114   -4.941    4.106    -.174
   50 (u)          13      2  3 0     -.252    -.044   -1.576     .935    -.249
   51 (g)          13     21  3 2     -.110     .229  -12.733   11.449    -.234
   52 (d~)         13     -1  0 3      .051    -.270   -3.475    2.724    -.298
   53 (d)          13      1  1 0      .175     .560   -1.576     .935    -.571
   54 (d~)         13     -1  0 2     -.357    -.255   -2.231    1.552    -.442
   55 (d)          13      1  1 0     -.142    -.121   -1.576     .935    -.193
   56 (d~)         13     -1  0 2      .527    -.078   -3.586    2.828    -.547
   57 (d)          13      1  2 0      .288    -.539   -1.576     .935    -.636
   58 (d)          13      1  1 0      .259    -.102   -1.994    1.328    -.297
   59 (d)          14      1  2 0    -1.532    -.540   -1.937    2.087    -.108
   60 (d~)         13     -1  0 2      .004    -.395   -1.576     .935    -.416
   61 (d)          13      1  2 0      .139    -.028   -2.314    1.630    -.158
   62 (d~)         13     -1  0 3      .020    -.413   -1.576     .935    -.435
   63 (u~)         13     -2  0 1      .071    -.222   -4.675    3.855    -.257
   64 (u)          13      2  3 0      .287    -.066   -1.576     .935    -.311
   65 (d~)         13     -1  0 3      .115    -.103   -2.071    1.401    -.178
   66 (d)          13      1  2 0      .173     .196   -1.576     .935    -.255
   67 (u)          13      2  3 0     -.800    -.123   -1.994    1.328    -.802
   68 (u~)         13     -2  0 3     -.434    -.218   -1.576     .935    -.486
   69 (g)          13     21  1 2      .092    -.434  -13.900   12.548    -.469
   70 (d~)         13     -1  0 3     -.200    -.122   -2.231    1.552    -.237
   71 (d)          13      1  3 0     -.153    -.233   -1.576     .935    -.291
   72 (g)          13     21  2 1     -.558     .600  -16.397   14.901    -.798
   73 (g)          14     21  1 1     -.646    -.141    4.334    4.734    -.682
   74 (d)          14      1  2 0     -.014     .151   -1.915    1.286    -.108
 ==============================================================================
   75 !g!          21     21  0 0     2.954    3.873    -.435    5.206    -.682
   76 !d!          21      1  0 0     -.354    -.234   -2.970    2.639    -.108
   77 !g!          21     21  2 1     2.181    5.310   -3.320    6.631     .000
   78 !d!          21      1  1 0      .521   -1.530    -.381    1.660     .010
   79 (g)          14     21  1 1     2.954    3.873    -.435    5.206    -.682
   80 (d)          14      1  2 0     -.354    -.234   -2.970    2.639    -.108
   81 (g)          14     21  2 1     2.160    5.284   -3.324    6.486     .000
   82 (d)          14      1  1 0      .500   -1.556    -.386    1.624     .010
   83 (g)          14     21  0 0     2.052    2.858    1.999    1.045   -3.909
   84 (d)          14      1  0 0      .000     .000   -8.008    8.008     .000
   85 (d)          14      1  2 1      .000     .000    8.008    8.008     .000
   86 (CMshower)   11     94  0 0     2.661    3.728   -3.710    8.110    5.571
   87 (d)          14      1  2 0     1.128     .073   -1.364    3.504    3.023
   88 (cluster)    11     91  0 0     1.116    1.615   -2.687    3.412     .750
   89 (cluster)    11     91  0 0     -.450    -.605    -.228    1.213     .922
   90 (cluster)    11     91  0 0    -1.528    -.662    1.381    2.181     .280
   91 (cluster)    11     91  0 0     1.252     .689   -1.272    3.136    2.484
   92 (cluster)    11     91  0 0      .110    -.650    5.219    5.796    2.433
   93 pi0           6    111  0 0      .660     .523    -.880    1.171     .135
   94 pi0           6    111  0 0      .461    1.090   -2.094    2.377     .135
   95 rho-          6   -213  0 0     -.334    -.562    -.332    1.068     .769
   96 pi0           6    111  0 0     -.110    -.046    -.182     .186     .135
   97 pi-           6   -211  0 0     -.678    -.312     .548    1.066     .140
   98 pi+           6    211  0 0     -.845    -.352     .547    1.203     .140
   99 a_2-          6   -215  0 0      .161     .747    -.904    1.772    1.318
  100 rho+          6    213  0 0     1.097    -.060    -.654    1.490     .769
  101 rho+          6    213  0 0     1.139    -.488    2.544    3.178     .769
  102 pi0           6    111  0 0    -1.023    -.165    2.389    2.849     .135
  103 (cluster)    11     91  0 0     -.054    -.150     .360     .686     .562
  104 pi0           6    111  0 0      .146    -.046     .232     .425     .135
  105 pi0           6    111  0 0     -.194    -.107    -.157     .261     .135
  106 (cluster)    11     91  0 0      .729     .554    -.798    1.246     .280
  107 pi0           6    111  0 0      .491     .351    -.622     .810     .135
  108 pi0           6    111  0 0      .244     .200    -.462     .467     .135
  109 (Beam-REM)   11     92  0 0     -.315     .737  274.428  281.871   70.164
  110 (Targ-REM)   11     93  0 0     -.907   -1.587 -276.508  247.044 -116.181
  111 (CMF)        15    100  0 0    -1.222    -.850   -2.080  528.915  528.909
 ==============================================================================
  112 (cluster)    11     91  0 0     -.315     .737  274.428  281.871   70.164
  113 (cluster)    11     91  0 0     -.907   -1.587 -276.508  247.044 -116.181
  114 (Beam-REM)   11     92  0 0      .411     .648  127.485  127.543    3.762
  115 (a_1-)       17 -20213  0 0      .493     .228   19.200   19.250    1.275
  116 (Delta+)     17   2214  0 0     -.081     .420  108.286  108.293    1.232
  117 pi0           7    111  0 0      .490     .124    4.473    4.831     .135
  118 (rho-)       17   -213  0 0      .005     .102   14.584   14.605     .769
  119 pi0           7    111  0 0     -.160     .015   35.475   37.042     .135
  120 p+            7   2212  0 0      .085     .402   72.525   75.579     .938
  121 pi0           7    111  0 0     -.122    -.268    8.519    9.014     .135
  122 pi-           7   -211  0 0      .133     .367    5.779    6.174     .140
  123 (Targ-REM)   11     93  0 0    -1.252    -.510  -63.766   63.825    2.384
  124 (Delta~-)    17  -2214  0 0     -.779    -.303  -21.201   21.253    1.232
  125 (eta)        17    221  0 0     -.473    -.208  -42.565   42.572     .549
  126 pi0           7    111  0 0     -.072    -.254   -3.205    3.199     .135
  127 p~-           7  -2212  0 0     -.701    -.052  -18.281   18.903     .938
  128 pi0           7    111  0 0     -.179     .069  -12.277   12.622     .135
  129 pi0           7    111  0 0     -.060    -.060  -12.901   13.269     .135
  130 pi0           7    111  0 0     -.225    -.220  -17.816   18.383     .135
  131 (Beam-REM)   11     92  0 0     -.538     .372   74.393   74.429    2.212
  132 (f_2)        17    225  0 0     -.685     .038   30.734   30.768    1.273
  133 pi0           7    111  0 0      .150     .332   43.516   45.406     .135
  134 pi-           7   -211  0 0     -.479     .511    9.266    9.813     .140
  135 pi+           7    211  0 0     -.201    -.476   21.182   22.184     .140
  136 (Beam-REM)   11     92  0 0     -.016     .852   11.680   11.879    1.995
  137 (K*_1+)      17  20323  0 0     -.298     .808    8.322    8.473    1.340
  138 (K~0)        17   -311  0 0      .282     .044    3.358    3.406     .498
  139 pi+           7    211  0 0     -.201     .595    2.569    2.899     .140
  140 (K*0)        17    313  0 0     -.094     .211    5.610    5.686     .892
  141 K_L0          7    130  0 0      .284     .043    3.215    3.542     .498
  142 pi-           7   -211  0 0     -.263    -.066    2.426    2.691     .140
  143 K+            7    321  0 0      .174     .274    2.899    3.222     .494
  144 (Beam-REM)   11     92  0 0     1.284    -.646   18.839   19.323    4.053
  145 (f_2)        17    225  0 0      .927   -1.937   15.847   16.042    1.273
  146 pi-           7   -211  0 0      .360    1.290    2.849    3.412     .140
  147 pi-           7   -211  0 0      .541   -1.029    4.275    4.752     .140
  148 pi+           7    211  0 0      .391    -.911   11.286   11.931     .140
  149 (Beam-REM)   11     92  0 0     -.328    -.501    4.561    4.836    1.491
  150 (eta)        17    221  0 0      .101    -.808    2.154    2.367     .549
  151 pi+           7    211  0 0     -.426     .306    2.265    2.568     .140
  152 pi0           7    111  0 0      .078    -.224     .225     .457     .135
  153 pi-           7   -211  0 0     -.073    -.350     .820    1.078     .140
  154 pi+           7    211  0 0      .105    -.238     .680     .909     .140
  155 (Beam-REM)   11     92  0 0     -.145     .122    2.373    2.824    1.518
  156 (K0)         17    311  0 0      .108     .164    1.292    1.399     .498
  157 (K*-)        17   -323  0 0     -.254    -.043    1.081    1.425     .892
  158 (K_S0)       17    310  0 0      .108     .164    1.292    1.399     .498
  159 pi-           7   -211  0 0     -.328     .128     .432     .718     .140
  160 (K~0)        17   -311  0 0      .077    -.172     .506     .734     .498
  161 pi-           7   -211  0 0      .018    -.010    -.064     .161     .140
  162 pi+           7    211  0 0      .096     .171    1.071    1.287     .140
  163 K_L0          7    130  0 0      .080    -.174     .363     .764     .498
  164 (Beam-REM)   11     92  0 0     -.335    -.537    2.570    3.545    2.358
  165 (rho+)       17    213  0 0     -.239    -.514     .111     .962     .769
  166 (omega)      17    223  0 0     -.097    -.023    2.460    2.583     .783
  167 pi0           7    111  0 0      .245    -.072    -.118     .287     .135
  168 pi+           7    211  0 0     -.478    -.445    -.057     .702     .140
  169 pi0           7    111  0 0      .120    -.032     .138     .334     .135
  170 pi-           7   -211  0 0     -.147    -.207     .861    1.087     .140
  171 pi+           7    211  0 0     -.061     .212    1.032    1.252     .140
  172 (Beam-REM)   11     92  0 0     -.022    1.276    2.371    3.452    2.159
  173 pi0           7    111  0 0     -.936     .564     .746    1.475     .135
  174 (rho-)       17   -213  0 0      .917     .711    1.482    2.033     .769
  175 pi0           7    111  0 0      .150    -.108    -.042     .247     .135
  176 pi-           7   -211  0 0      .773     .816    1.239    1.857     .140
  177 (Beam-REM)   11     92  0 0      .028    -.987    1.765    2.579    1.601
  178 (omega)      17    223  0 0      .159    -.572     .855    1.303     .783
  179 (omega)      17    223  0 0     -.131    -.414     .910    1.277     .783
  180 pi0           7    111  0 0      .142     .023     .159     .360     .135
  181 pi-           7   -211  0 0     -.177    -.323     .175     .527     .140
  182 pi+           7    211  0 0      .203    -.276     .092     .436     .140
  183 pi0           7    111  0 0     -.257    -.051    -.044     .313     .135
  184 pi-           7   -211  0 0     -.004    -.213     .396     .619     .140
  185 pi+           7    211  0 0      .139    -.155     .129     .368     .140
  186 (Beam-REM)   11     92  0 0     1.076     .165    1.776    2.647    1.633
  187 pi+           7    211  0 0      .429    -.147    -.255     .504     .140
  188 (rho0)       17    113  0 0      .650     .311    1.888    2.162     .769
  189 pi-           7   -211  0 0      .471    -.021     .292     .680     .140
  190 pi+           7    211  0 0      .185     .329    1.310    1.568     .140
  191 (Beam-REM)   11     92  0 0      .627    -.168     .968    1.814    1.390
  192 (rho0)       17    113  0 0      .136    -.396     .264     .915     .769
  193 pi-           7   -211  0 0      .494     .227     .561     .935     .140
  194 pi-           7   -211  0 0      .340    -.491     .007     .654     .140
  195 pi+           7    211  0 0     -.198     .092    -.029     .286     .140
  196 (Beam-REM)   11     92  0 0    -1.186    -.195    1.298    2.721    2.067
  197 pi0           7    111  0 0     -.368     .146    -.593     .641     .135
  198 (rho0)       17    113  0 0     -.815    -.343    1.748    2.105     .769
  199 pi-           7   -211  0 0     -.417    -.533    1.092    1.472     .140
  200 pi+           7    211  0 0     -.392     .188     .370     .717     .140
  201 (Beam-REM)   11     92  0 0     -.184    -.529     .592    1.588    1.362
  202 pi0           7    111  0 0      .011     .329    -.331     .403     .135
  203 pi+           7    211  0 0     -.188    -.861     .637    1.232     .140
  204 (Beam-REM)   11     92  0 0      .879     .300     .513    2.125    1.841
  205 (rho-)       17   -213  0 0      .937     .336     .136    1.265     .769
  206 (rho+)       17    213  0 0     -.058    -.036     .377     .859     .769
  207 pi0           7    111  0 0      .523     .541    -.130     .794     .135
  208 pi-           7   -211  0 0      .420    -.208    -.021     .521     .140
  209 pi0           7    111  0 0      .103    -.135    -.307     .270     .135
  210 pi+           7    211  0 0     -.155     .096     .399     .613     .140
  211 (Beam-REM)   11     92  0 0      .116   -1.151     .357    2.408    2.081
  212 (a_20)       17    115  0 0     -.476    -.521     .333    1.532    1.318
  213 pi-           7   -211  0 0      .595    -.632    -.119     .912     .140
  214 pi0           7    111  0 0     -.037     .290     .182     .459     .135
  215 (eta)        17    221  0 0     -.435    -.812     .008    1.073     .549
  216 pi0           7    111  0 0     -.308    -.509    -.196     .637     .135
  217 pi0           7    111  0 0     -.048    -.163    -.031     .244     .135
  218 pi0           7    111  0 0     -.070    -.145    -.194     .216     .135
  219 (Targ-REM)   11     93  0 0     -.558    -.387    -.074    1.754    1.615
  220 pi+           7    211  0 0     -.079     .095     .024     .252     .140
  221 (a_2-)       17   -215  0 0     -.476    -.483    -.241    1.502    1.318
  222 pi0           7    111  0 0      .011    -.471     .009     .532     .135
  223 (rho-)       17   -213  0 0     -.484    -.014    -.393     .990     .769
  224 pi0           7    111  0 0     -.074    -.305    -.459     .465     .135
  225 pi-           7   -211  0 0     -.405     .289    -.220     .546     .140
  226 (Targ-REM)   11     93  0 0     -.584     .922    -.910    2.344    1.863
  227 (omega)      17    223  0 0     -.237    -.080    -.504     .964     .783
  228 (omega)      17    223  0 0     -.347    1.002    -.406    1.379     .783
  229 pi0           7    111  0 0     -.037     .207    -.270     .282     .135
  230 pi-           7   -211  0 0     -.286    -.188    -.360     .430     .140
  231 pi+           7    211  0 0      .095    -.102    -.302     .252     .140
  232 pi0           7    111  0 0     -.286     .288    -.157     .431     .135
  233 pi-           7   -211  0 0     -.047     .065    -.114     .165     .140
  234 pi+           7    211  0 0     -.006     .645    -.563     .814     .140
  235 (Targ-REM)   11     93  0 0     -.472   -1.489   -1.060    2.438    1.543
  236 (rho0)       17    113  0 0     -.075   -1.404   -1.183    1.992     .769
  237 pi+           7    211  0 0     -.393    -.087    -.020     .446     .140
  238 pi-           7   -211  0 0     -.109    -.365    -.137     .405     .140
  239 pi+           7    211  0 0      .039   -1.041   -1.332    1.650     .140
  240 (Targ-REM)   11     93  0 0      .263     .697   -1.440    2.287    1.613
  241 (f_2)        17    225  0 0      .163     .602    -.823    1.639    1.273
  242 pi-           7   -211  0 0      .103     .094    -.760     .674     .140
  243 pi-           7   -211  0 0      .404    -.097    -.916     .923     .140
  244 pi+           7    211  0 0     -.235     .696    -.193     .782     .140
  245 (Targ-REM)   11     93  0 0     -.514    -.122   -1.450    2.061    1.366
  246 (omega)      17    223  0 0     -.621     .269   -1.015    1.450     .783
  247 pi+           7    211  0 0      .110    -.393    -.578     .636     .140
  248 pi0           7    111  0 0     -.181     .225    -.261     .342     .135
  249 pi-           7   -211  0 0     -.157    -.169    -.463     .419     .140
  250 pi+           7    211  0 0     -.274     .209    -.720     .716     .140
  251 (Targ-REM)   11     93  0 0      .521    1.161   -2.393    3.205    1.711
  252 pi0           7    111  0 0      .194     .629    -.845    1.011     .135
  253 (a_20)       17    115  0 0      .329     .530   -1.691    2.233    1.318
  254 pi+           7    211  0 0      .155     .570    -.573     .774     .140
  255 (rho-)       17   -213  0 0      .177    -.042   -1.261    1.488     .769
  256 pi0           7    111  0 0      .214    -.025   -1.454    1.388     .135
  257 pi-           7   -211  0 0     -.031    -.019    -.093     .153     .140
  258 (Targ-REM)   11     93  0 0      .406    -.313   -2.421    3.135    1.925
  259 (f_1)        17  20223  0 0     -.212    -.339   -1.807    2.252    1.283
  260 pi-           7   -211  0 0      .621     .025    -.756     .918     .140
  261 (eta)        17    221  0 0      .066    -.030    -.883    1.042     .549
  262 pi-           7   -211  0 0      .004    -.207    -.974     .902     .140
  263 pi+           7    211  0 0     -.277    -.105    -.237     .343     .140
  264 pi0           7    111  0 0      .143    -.029    -.607     .524     .135
  265 pi0           7    111  0 0      .032    -.032    -.434     .324     .135
  266 pi0           7    111  0 0     -.100     .027    -.270     .214     .135
  267 (Targ-REM)   11     93  0 0      .900    -.332   -6.866    7.556    3.006
  268 (Delta++)    17   2224  0 0     -.308    -.535   -3.638    3.890    1.232
  269 (Delta~-)    17  -2214  0 0     1.209     .203   -3.227    3.665    1.232
  270 pi+           7    211  0 0      .096    -.090   -1.275    1.194     .140
  271 p+            7   2212  0 0     -.399    -.448   -2.649    2.852     .938
  272 pi0           7    111  0 0      .571     .009   -1.308    1.355     .135
  273 p~-           7  -2212  0 0      .644     .191   -2.206    2.457     .938
  274 (Targ-REM)   11     93  0 0     -.423    2.411   -7.871    8.635    2.572
  275 (K*0)        17    313  0 0      .285    1.878   -6.966    7.276     .892
  276 K-            7   -321  0 0     -.705     .532   -1.047    1.414     .494
  277 pi-           7   -211  0 0      .385     .806   -3.537    3.652     .140
  278 K+            7    321  0 0     -.095    1.069   -3.716    3.914     .494
  279 (Targ-REM)   11     93  0 0      .454     .090   -6.177    6.678    2.496
  280 (a_2+)       17    215  0 0      .653     .059   -6.253    6.424    1.318
  281 pi0           7    111  0 0     -.196     .030    -.067     .254     .135
  282 pi0           7    111  0 0      .231     .367   -1.287    1.280     .135
  283 (rho+)       17    213  0 0      .425    -.310   -5.108    5.192     .769
  284 pi0           7    111  0 0      .205    -.443   -4.489    4.550     .135
  285 pi+           7    211  0 0      .227     .131    -.905     .850     .140
  286 (Targ-REM)   11     93  0 0      .279   -1.845   -8.534    9.228    2.973
  287 (eta)        17    221  0 0     -.507   -1.723   -5.714    6.014     .549
  288 (a_2-)       17   -215  0 0      .786    -.121   -2.821    3.214    1.318
  289 pi0           7    111  0 0     -.050    -.557   -1.864    1.887     .135
  290 pi-           7   -211  0 0     -.101    -.501   -2.002    2.010     .140
  291 pi+           7    211  0 0     -.347    -.669   -2.277    2.358     .140
  292 pi0           7    111  0 0     -.054    -.298   -1.183    1.135     .135
  293 (rho-)       17   -213  0 0      .843     .175   -1.780    2.122     .769
  294 pi0           7    111  0 0     -.048    -.113    -.299     .240     .135
  295 pi-           7   -211  0 0      .897     .286   -1.767    1.956     .140
  296 (Targ-REM)   11     93  0 0      .537    -.194   -4.665    4.841    1.163
  297 pi0           7    111  0 0      .379     .133   -2.137    2.119     .135
  298 (rho0)       17    113  0 0      .161    -.328   -2.671    2.803     .769
  299 pi-           7   -211  0 0      .140     .210   -1.211    1.150     .140
  300 pi+           7    211  0 0      .027    -.541   -1.746    1.765     .140
  301 (Targ-REM)   11     93  0 0    -1.516     .775   -9.672   10.040    2.086
  302 K+            7    321  0 0     -.822     .830   -2.936    3.191     .494
  303 (K*-)        17   -323  0 0     -.691    -.056   -6.879    6.972     .892
  304 pi-           7   -211  0 0     -.067     .030   -3.431    3.424     .140
  305 (K~0)        17   -311  0 0     -.622    -.087   -3.591    3.680     .498
  306 (K_S0)       17    310  0 0     -.622    -.087   -3.591    3.680     .498
  307 pi0           7    111  0 0     -.373    -.224   -2.763    2.766     .135
  308 pi0           7    111  0 0     -.243     .134   -1.114    1.061     .135
  309 (Targ-REM)   11     93  0 0     -.684    -.224   -7.561    7.791    1.736
  310 (K*+)        17    323  0 0     -.060    -.267   -5.535    5.613     .892
  311 (K~0)        17   -311  0 0     -.624     .043   -2.026    2.178     .498
  312 pi0           7    111  0 0      .214    -.170   -3.101    3.092     .135
  313 K+            7    321  0 0     -.268    -.100   -2.720    2.745     .494
  314 (K_S0)       17    310  0 0     -.624     .043   -2.026    2.178     .498
  315 pi0           7    111  0 0     -.503     .096   -1.122    1.159     .135
  316 pi0           7    111  0 0     -.115    -.055   -1.190    1.106     .135
  317 (Targ-REM)   11     93  0 0     1.053   -1.107   -7.787    8.016    1.137
  318 (rho0)       17    113  0 0      .693    -.993   -6.816    6.965     .769
  319 pi-           7   -211  0 0      .363    -.115   -1.114    1.093     .140
  320 pi-           7   -211  0 0      .194    -.534   -1.738    1.767     .140
  321 pi+           7    211  0 0      .504    -.462   -5.364    5.477     .140
  322 (Targ-REM)   11     93  0 0      .585     .227  -23.490   23.597    2.149
  323 K+            7    321  0 0      .796     .530   -7.564    7.799     .494
  324 (K*-)        17   -323  0 0     -.208    -.305  -16.069   16.098     .892
  325 pi-           7   -211  0 0     -.071     .040   -9.208    9.429     .140
  326 (K~0)        17   -311  0 0     -.135    -.346   -7.003    7.031     .498
  327 K_L0          7    130  0 0     -.132    -.348   -7.146    7.312     .498
  328 (Targ-REM)   11     93  0 0    -1.689     .202  -60.528   60.729    4.635
  329 (Delta++)    17   2224  0 0    -1.569   -1.491  -21.311   21.456    1.232
  330 (Delta~--)   17  -2224  0 0     -.120    1.693  -39.217   39.273    1.232
  331 pi+           7    211  0 0     -.301    -.495   -4.010    4.070     .140
  332 p+            7   2212  0 0    -1.262    -.999  -17.587   18.244     .938
  333 pi-           7   -211  0 0     -.226     .567  -12.176   12.531     .140
  334 p~-           7  -2212  0 0      .112    1.123  -27.327   28.312     .938
  335 (Targ-REM)   11     93  0 0     -.196     .168  -36.959   37.045    2.502
  336 (rho+)       17    213  0 0     -.263     .404   -5.474    5.549     .769
  337 (omega)      17    223  0 0      .067    -.237  -31.485   31.496     .783
  338 pi0           7    111  0 0     -.342     .522   -5.018    5.113     .135
  339 pi+           7    211  0 0      .085    -.120    -.742     .657     .140
  340 pi0           7    111  0 0     -.309     .029  -10.513   10.790     .135
 ==============================================================================
                  sum:   .00           .000     .000     .000  546.000  546.000


\end{verbatim}
}
\end{center}

\newpage

\section{Summary list of subroutines}
\label{sec:appe}
\bigskip
                                                                 
\noindent
  In the following list, the notation is 
{\bf S} = subroutine,
{\bf F} = function 
{\bf B} = block data.                 
\bigskip
\bigskip

\noindent
{\bf PART 1: Main steering routines and general utilities}          
\begin{description}
\item[ S   VNICLOC]  to measure and return the CPU time used                
\item[ S   VNIXRIN]  to initialize the overall run and simulation procedure 
\item[ S   VNIXRUN]  to steer the simulation of a single collision event    
\item[ S   VNIXFIN]  to finish up the overall run and simulation procedure  
\item[ S   VNICBIN]  to interface commonblocks of VNI and JESTSET/PYTHIA    
\item[ S   VNILOGO]  to write program header                                
\item[ S   VNILIST]  to list event record, particle, or parameter values    
\item[ S   VNIEDIT]  to perform global manipulations on the event record.   
\item[ F   KVNI   ]  to provide integer-valued event information            
\item[ F   PVNI   ]  to provide real-valued event information               
\item[ F   RLU    ]  to provide a random number generator                   
\item[ S   RLUGET ]  to save the state of the random number generator       
\item[ S   RLUSET ]  to set the state of the random number generator        
\item[ S   VNIERRM]  to write error messages and abort faulty run           
\item[ S   VNIFILE]  to open and close output files                         
\item[ S   VNIDUMP]  to dump(read) all commonblocks to(from) file
\item[ B   VNIDATA]  to give default values to particle and decay data       
\end{description}                                                                      
\bigskip

\noindent
{\bf 
PART 2: Initialization routines 
\\
\\
\noindent
{\it general as well as process-dependent ones}    
}
\medskip

\begin{description}                                                  
\item[ S   VNICSIN]  to initialize collision system and kinematics          
\item[ S   VNIABXS]  to give parametrized cross-sections for beam/target    
\item[ S   VNITMIN]  to initialize time slizes and form of time increment   
\item[ S   VNI2JIN]  to set up an initial 2-jet in 
   $l^+l^-\rightarrow \gamma /Z_0 \rightarrow  q\bar{q}$ 
\item[ S   VNI2JFL]  to select flavour for initially produced 
  $q\bar{q}$ 2-jet      
\item[ S   VNI4JIN]  to set up an initial 4-jet in $l^+l^- 
\rightarrow W^+W^- \rightarrow q\bar{q}\;q'\bar{q}'$
\item[ S   VNI4JWW]  to initialize $W^+$ and $W^-$ 
decay vertices and the 4 jets  
\item[ S   VNILHIN]  to set up initial system for 
  $l^- + h \rightarrow l^- + \mbox{jet(s)} + X$ 
\item[ S   VNILAIN]  to set up initial system for 
  $l^- + A \rightarrow l^- + \mbox{jet(s)} + X$ 
\item[ S   VNIXDIS]  to generate the variables of DIS photon-hadron(nucleus) 
\item[ F   VNIGAQX]  to give elementary photon-quark cross-sections         
\item[ S   VNIGAQQ]  to give $q\bar{q}$ fluctuation of photon from VDM
\item[ S   VNIHHIN]  to set up initial system for 
  $h  + h'\rightarrow  \mbox{jets} + X$       
\item[ S   VNIHAIN]  to set up initial system for 
  $h  + A\rightarrow  \mbox{jets} + X$       
\item[ S   VNIABIN]  to set up initial system for 
  $A  + A'\rightarrow  \mbox{jets} + X$       
\item[ S   VNIBTIN]  to set up initial state of hadronic or nuclear beams   
\item[ S   VNINPDT]  to tabulate nuclear neutron/proton distribution        
\item[ S   VNIRPIN]  to initialize single hadrons, or nucleons in nucleus   
\item[ S   VNIQGDT]  to tabulate distributions in $x$ and flavor of partons   
\item[ F   VNIEFFQ]  to give parametrization of eff. parton resolution $Q(x)$ 
\item[ F   VNIQGSH]  to give simple parametrization of parton shadowing     
\item[ S   VNIVQKF]  to extract the valence quark content of a hadron       
\item[ S   VNIQGXF]  to initialize $x$ and flavor of partons in parent hadron 
\item[ S   VNIQGCS]  to arrange a set of partons as color singlet hadron    
\item[ S   VNIQGIN]  to distribute initialized partons inside parent hadron 
\item[ S   VNIPRKT]  to give primordial kT to initial partons               
\item[ S   VNIXTIN]  to find spatial extension of initial particle system   
\end{description}
\bigskip
                                                                       
\noindent
{\bf 
PART 3: Particle structure function package               
}
\\
\\
{\it 
portable, self-contained program unit with interface to PDFLIB  
}
\medskip

\begin{description}                                                  
\item[ S   VNISTFU]  calls structure functions: photon, $\pi, n, p$, hyperon  
\item[ S   VNISTGA]  gives photon structure function                         
\item[ S   VNISTAG]  gives anomalous part of photon structure function      
\item[ S   VNISTGS]  parametrized parton distributions in a virtual photon  
\item[ S   VNISTPI]  gives $\pi$ structure function                            
\item[ S   VNISTFL]  gives proton structure function at small $x$ and/or $Q^2$ 
\item[ S   VNISTPR]  gives proton structure function                        
\item[ S   VNIFCTQ]  CTEQ2'94 parametrization of parton distributions       
\item[ S   VNIFGRV]  GRV'94 parametrizations of parton distributions        
\item[ F   FV     ]  Function for GRV parametrizations                      
\item[ F   FW     ]  Function for GRV parametrizations                      
\item[ F   FWS    ]  Function for GRV parametrizations                      
\item[ F   FHS    ]  Function for GRV parametrizations                      
\item[ F   VNIHFTH]  gives threshold factor for heavy flavour production     
\item[ F   VNIDILN]  gives Dilogarithm function                             
\item[ F   VNIGAMM]  gives ordinary Gamma function                          
\item[ S   PDFSET ]  dummy routine, to be removed when PDFLIB is linked      
\item[ S   STRUCTM]  dummy routine, to be removed when PDFLIB is linked     
\end{description}
\bigskip
                                                                       
\noindent
{\bf 
PART 4: Main routines for cascade evolution and hadronization     
}
\medskip

\begin{description}
\item[ S   VNIEV2J]  to perform evolution for 
  $l^+ + l^- \rightarrow 2 \mbox{jets} \rightarrow \mbox{hadrons}$
\item[ S   VNIEV4J]  to perform evolution for 
  $l^+ + l^- \rightarrow 4 \mbox{jets} \rightarrow \mbox{hadrons}$
\item[ S   VNIEVLH]  to perform evolution for
  $l^- + h \rightarrow \mbox{jets} \rightarrow \mbox{hadrons}$
\item[ S   VNIEVLA]  to perform evolution for 
  $l^- + A \rightarrow \mbox{jets} \rightarrow \mbox{hadrons}$
\item[ S   VNIEVAB]  to perform evolution for 
  $A + B \rightarrow \mbox{jets} \rightarrow \mbox{hadrons}$
\item[ S   VNICOL2]  to find 2-parton collisions and their kinematics       
\item[ S   VNICOLX]  to give cross-sections for $2\rightarrow 2$ 
  and $2 \rightarrow 1$ collisions
\item[ S   VNICOLI]  to set up a found 2-parton collision subsystem         
\item[ S   VNICOLL]  to find outgoing flavors and type of parton collision  
\item[ S   VNISBRA]  to generate space-like parton branchings               
\item[ S   VNITBRA]  to generate time-like parton branchings                
\item[ S   VNICHAD]  to organize parton cluster formation and hadronization 
\item[ S   VNICFOR]  to form one or two cluster(s) from a pair of partons    
\item[ S   VNICREC]  to reconnect cluster-partons according to color flow   
\item[ S   VNICFRA]  to fragment found parton clusters into hadrons          
\item[ S   VNIBHAD]  to do beam remnant cluster formation and hadronization 
\item[ S   VNIBFRA]  to fragment beam/target remnant clusters into hadrons  
\item[ S   VNIPADC]  to decay unstable particles according to particle data 
\item[ S   VNIPAKF]  to give a hadronic particle from a flavor pair         
\item[ S   VNIBSTP]  to simulate baryon stopping in nuclear A+B collisions 
\end{description}
\bigskip
                                                                       
\noindent
{\bf 
PART 5: Auxiliary routines I:                     
}
\\
\\
{\it parton cascades \& cluster formation}
\medskip

\begin{description}
\item[ F   VNIALEM]  to give the $\alpha_{em}$ value                
\item[ F   VNIALPS]  to give the $\alpha_s$ value                         
\item[ F   VNIANGL]  to give the angle from known $x$ and $y$ components        
\item[ F   INVCHGE]  to give three times the electric charge                
\item[ F   INVCOMP]  to compress standard $KF$ flavour code to internal $KC$    
\item[ S   VNICOLB]  to boost/rotate 2-body collision subsystem along $p_z$   
\item[ S   VNIDEC2]  to generate 2-body decay                               
\item[ S   VNIFRAM]  to perform boosts between different frames 
\item[ S   VNIKLIM]  to find kinematic limits in 2-parton collsions         
\item[ S   VNICCOL]  to connect color indices in 2-parton collisions        
\item[ S   VNIJCOL]  to connect  partons with color flow indices            
\item[ S   VNILCOL]  to label color indices according to color flow tracing 
\item[ F   VNIMASS]  to give the mass of a particle or parton               
\item[ F   VNIMTHR]  to give `current' or `constituent' masses for partons  
\item[ S   VNINAME]  to give the name of a particle or parton               
\item[ F   VNIGORI]  to give the origin of a particle in space-time history 
\item[ F   VNISORI]  to sey the origin of a particle in space-time history  
\item[ S   VNIPROP]  to propagate a given number of particles                
\item[ S   VNIPSEA]  to generate  a hadron pair from the Dirac `sea'        
\item[ S   VNIQSEA]  to generate a $q$ or $\bar{q}$, or a 
    $q\bar{q}$-pair from the `sea'    
\item[ S   VNIROBO]  to rotate and/or boost a given number of particles     
\item[ S   VNIRSCL]  to rescale particle momenta for numerical precision    
\item[ S   VNISORT]  to count and sort particles                             
\item[ S   VNIWIDT]  to calculate width of particle resonances              
\end{description}
\bigskip
                                                                       
\noindent
{\bf 
PART 6: Auxiliary routines II:                     
}
\\
\\
{\it  cluster-hadron decay and beam/target fragmentation}      
\medskip

\begin{description}
\item[ S   HWHEPC]  to convert VNI event record to or from standard HEP     
\item[ S   HWLIST]  to list event record in HEP format                      
\item[ S   HWIGIN]  to set parameters of HERWIG 5.7                         
\item[ S   HWUINC]  to compute constants and lookup tables                 
\item[ S   HWUINE]  to perform re-initialization of a new event             
\item[ S   HWURES]  to load common blocks with particle data                
\item[ S   HWUIDT]  to translate particle identifiers                       
\item[ S   HWCCUT]  to cut up heavy cluster into two clusters               
\item[ S   HWCFLA]  to to set up flavors for cluster decay                  
\item[ S   HWCHAD]  to hadronize prepared beam/target clusters              
\item[ S   HWDHAD]  to decay unstable hadrons from beam/target remnants     
\item[ S   HWDTWO]  to generate two-body decay $0\rightarrow 1+2$
\item[ S   HWDTHR]  to generate three-body decay $0\rightarrow 1+2+3$
\item[ F   HWDPWT]  to give matrix element squared for phase-space decay    
\item[ F   HWDWWT]  to give matrix element squared for $V_A$ weak decay       
\item[ F   HWMNBI]  to compute negative binomial probability                
\item[ S   HWMLPS]  to generate cylindrical phase-space                     
\item[ S   HWMULT]  to pick charged multiplicity of beam/target diffraction 
\item[ F   HWUPCM]  to compute c.m. momentum for masses in 2-body decay     
\item[ S   HWULOB]  to transform momenta from rest frame into lab frame     
\item[ S   HWULOF]  to transform momenta from lab frame into rest frame     
\item[ S   HWUMAS]  to put mass in 5th component of momentum vector         
\item[ S   HWUROB]  to rotate vectors by inverse of rotation matrix R       
\item[ S   HWUROT]  to perform rotation with rotation matrix R              
\item[ S   HWUSOR]  to sort n-tupel A(N) into ascending order               
\item[ F   HWUSQR]  to give square root with sign retention                 
\item[ S   HWUSTA]  to make a particle type stable                          
\item[ S   HWVDIF]  to give vector difference                               
\item[ F   HWVDOT]  to give vector dot product                              
\item[ S   HWVEQU]  to give vector equality                                
\item[ S   HWVSCA]  to give product of vector times scalar                  
\item[ S   HWVSUM]  to give vector sum                                      
\item[ S   HWVZRO]  to give zero vector                                     
\item[ F   HWRGEN]  to provide random number generator                      
\item[ S   HWRAZM]  to randomly rotate 2-vector $p_\perp$
\item[ F   HWREXP]  random number from 1-dim. exponential distribution      
\item[ F   HWREXT]  random number from 2-dim. exponential distribution      
\item[ F   HWRGAU]  random number from 2-dim Gaussian distribution          
\item[ F   HWRINT]  random integer in finite interval                       
\item[ F   HWRLOG]  random logical                                          
\item[ S   HWRPOW]  random number from power-law distribution               
\item[ F   HWRUNG]  random number from flat-top distribution                
\item[ F   HWRUNI]  uniform random in finite interval                       
\item[ S   HWUTIM]  dummy routine
\item[ S   HWWARN]  dummy routine                                           
\item[ B   HWUDAT]  to provide HERWIG common blocks with data  
\end{description}
\bigskip

{\bf
PART 7: Event study and data analysis routines              
}
\medskip

\begin{description}
\item[ S   VNIANAL]  to steer pre-programmed event analysis
\item[ S   VNIANA1]  statistics on global particle properties \& observables 
\item[ S   VNIANA2]  statistics on hadronic distributions                   
\item[ S   VNIANA3]  statistics on pre-hadronic clusters                    
\item[ S   VNIANA4]  statistics on partons and produced hadrons            
\item[ S   VNIANA5]  global thermodynamic multiparticle properties          
\item[ S   VNISPHE]  to perform sphericity analysis                          
\item[ S   VNITHRU]  to perform thrust analysis                              
\item[ S   VNICLUS]  to perform three-dimensional jet cluster analysis          
\item[ S   VNICELL]  to perform jet cluster analysis in ($\eta, \phi, E_\perp$)
\item[ S   VNIJMAS]  to give high and low jet mass of event                  
\item[ S   VNIFOWO]  to give Fox-Wolfram jet moments                             
\end{description}
\bigskip
\bigskip

\newpage

\section{Bibiliography}


\bigskip

\begin{thebibliography}{99}                                                                 
\bibitem{ms39}     K. Geiger,  Phys. Rev. {\bf D54}, 949 (1996);
\bibitem{ms42}     K. Geiger, preprint BNL-63632, hep-ph/9611400.
\bibitem{kogut73}  J. Kogut and L. Susskind, Phys. Rep. {\bf 8}, 75 (1973).
\bibitem{dok80}    Yu. L. Dokshitzer, D. I. Dyakonov, and S. I. Troyan,  
                   Phys. Rep. {\bf 58}, 269 (1980).
\bibitem{glr83}    L. V. Gribov, E. M. Levin, and M. G. Ryskin, 
                   Phys. Rep. {\bf 100}, 1 (1983).
\bibitem{ms0}      K. Geiger and B. M\"uller, 
                   Nucl. Phys. {\bf B369}, 600 (1992) 
\bibitem{ms3}      K. Geiger, Phys. Rev. {\bf D47}, 133 (1993).
\bibitem{msrep}    K. Geiger,  Phys. Rep. {\bf 258}, 238 (1995). 
\bibitem{ms37}     J. Ellis and K. Geiger, Phys. Rev. {\bf D54}, 949 (1996), 
\bibitem{ms40}     J. Ellis and K. Geiger, Phys. Rev. {\bf D54}, 1755 (1996).
\bibitem{ms41}     J. Ellis, K. Geiger, and H. Kowalski,
                   Phys. Rev. {\bf D54}, 5443 (1996). 
\bibitem{jetcalc}  K. Konishi, A. Ukawa, and G. Veneziano, 
                   Nucl. Phys. {\bf B157}, 45 (1979).
\bibitem{bassetto} A. Bassetto, M. Ciafaloni and G. Marchesini,
                   Phys. Rep. {\bf 100}, 203 (1983).  
\bibitem{dokbook}  {\it Basics of Perturbative QCD}, Advanced Series on 
                   Directions in High Energy Physics, Vol. 5 (Editions 
                   Frontieres, Gif-sur-Yvette Cedex, France, 1991).
\bibitem{miklos}   M. Gyulassy, {\it private communication}.
\bibitem{feyfld}   R. P. Feynman and R. D. Field,
                   Phys. Rev. {\bf D15}, 2590 (1977); 
                   Nucl. Phys. {\bf B136}, 1 (1978).
\bibitem{emc}      European Muon Collaboration: J. Ashman {\it et al.},
                   Phys. Lett. {\bf B202} 603 (1988);
                   M. Arneodo {\it et al.},  
                   Phys. Lett. {\bf B211}, 493 (1988).
\bibitem{nucshad4} L. L. Frankfurt and M. I. Strikman,
                   Phys. Rep. {\bf 160}, 235 (1988).
\bibitem{nucshad2} A. H. Mueller and J. Qui, Nucl. Phys. {\bf B268}, 
                   427 (1986);
                   J. Qui, Nucl. Phys. {\bf B291}, 746 (1987), 746;
\bibitem{nucshad3} F. E. Close, J. Qiu and R. G. Roberts,
                   Phys. Rev. {\bf D 40}, 2820 (1989);
                   A. H. Mueller, Nucl. Phys. {\bf B335}, 115 (1990).
\bibitem{hijing2}  X. N. Wang and M. Gyulassy, 
                   Phys. Rev. {\bf D44}, 3501 (1991).
\bibitem{pTprim}   V. M. Belyaev and B. L. Ioffe,
                   Nucl. Phys. {\bf B313}, 647 (1989);
                   D. Antreasyan {\it et al.}, 
                   Phys. Rev. Lett. {\bf 48}, 302 (1983).
\bibitem{bj76}     J. D. Bjorken in {\it Current Induced Reactions},
                   {\sl Proceedings of the International Summer Institute on
                   Theoretical Particle Physics}, Hamburg, 1975, edited by 
                   J. K\"orner et al.  ( Lecture Notes in Physics, Vol. 56, 
                   Springer, New York 1976).
\bibitem{hwa86}    R. C. Hwa and K. Kajantie, 
                   Phys. Rev. Lett. {\bf 56}, 696 (1986)
\bibitem{mueller89} A. H. Mueller, Nucl. Phys. {\bf A498}, 41c (1989).
\bibitem{ms1}      K. Geiger, Phys. Rev. {\bf D46}, 4965 (1992).
\bibitem{dok88}    Yu. L. Dokshitzer, V. A. Khoze, A. H. Mueller, 
                   and S. I. Troyan, Rev. Mod. Phys. {\bf 60}, 373 (1988).
\bibitem{DGLAP}    L. N. Lipatov, Sov. J. Nucl. Phys.  {\bf 20}, 94 (1975);
                   Yu. L. Dokshitzer, Sov. Phys. JETP {\bf 46}, 641  (1977);
                   G. Altarelli and G. Parisi, 
                   Nucl. Phys. {\bf B126}, 298 (1977).
\bibitem{sig1}     R. Cutler and D. Sivers, Phys. Rev. {\bf D17}, 196 (1978);
                   B. L. Combridge, J. Kripfganz and J. Ranft,  
                   Phys. Lett. {\bf 70B}, 234 (1977).
\bibitem{sig2}     B. L. Combridge, Nucl. Phys. {\bf B151}, 429 (1979);
                   P. Nason, S. Dawson and R. K. Ellis, 
                   Nucl. Phys. {\bf B303}, 607 (1988);  
                   {\bf B327}, 49(1989); {\bf B335}, 260 (1990).
\bibitem{gustaf82} G. Gustafson, Z. Phys. {\bf C15}, 155 (1982).
\bibitem{webber88} G. Marchesini and B. R. Webber, 
                   Nucl. Phys. {\bf 310}, 461 (1988); 
                   Nucl. Phys. {\bf B349}, 617 (1991).
\bibitem{backevol} T. D. Gottschalk, Nucl. Phys. {\bf B277}, 100 (1986);
                   M. Bengtsson, T. Sj\"ostrand and M. van Zijl, 
                   Z. Phys. {\bf C32}, 67 (1986);
                   M. Bengtsson and T. Sj\"ostrand,
                   Z. Phys. {\bf C37}, 465 (1988).
\bibitem{webber86} G. Marchesini and B. R. Webber, 
                   Nucl. Phys. {\bf B238}, 1 (1984); 
                   B. R. Webber,  
                   Ann. Rev. Nucl. Part. Sci. {\bf 36}, 253 (1986).
\bibitem{webber84} G. Marchesini and B. R. Webber,
                   Nucl. Phys. {\bf B238}, 1 (1984);
                   B. R. Webber, Nucl. Phys. {\bf B238}, 492 (1984).
\bibitem{string}   B. Andersson, G. Gustafson, G. Ingelman,
                   and T. Sj\"ostrand, Phys. Rep. {\bf 97}, 33  (1983);
                   B. Andersson, G. Gustafson and B. S\"oderberg,
                   Nucl. Phys. {\bf B264}, 29 (1986).
\bibitem{preconf} D. Amati, A. Bassetto, M. Ciafaloni, G. Marchesini,
                  and G. Veneziano, Nucl. Phys. {\bf B173}, 429 (1980);
                  G. Marchesini, L. Trentadue and G. Veneziano, 
                  Nucl. Phys. {\bf B181}, 335 (1981).
\bibitem{hagedorn} R. Hagedorn, {\it Thermodynamics of Strong Interactions},
                   Carg\`ese Lectures in Physics, Vol. 6 (1973).
\bibitem{UA5}      UA5 Collaboration, G. J. Alner {\it et al.},
                   Nucl. Phys. {\bf B291}, 445 (1987);
                   Phys. Rep. {\bf 154}, 247 (1987).
\bibitem{Mdiff}    K. Goulianos, Phys. Rep. {\bf 101}, 169 (1983).
\bibitem{jetset}   T. Sj\"ostrand, Comp. Phys. Com. {\bf 82}, 74 (1994).
\bibitem{herwig}   G. Marchesini, B. R. Webber, G. Abbiendi, I. G. Knowles,
                   M. H. Seymor and L. Stanco,
                   Comp. Phys. Com.  {\bf 67}, 465 (1992).
\bibitem{ariadne}  L. L\"onnblad, Comp. Phys. Com. {\bf 71}, 15 (1992).
\bibitem{lepto}    G. Ingelman, A. Edin, and J. Rathsman, 
                   preprint DESY-96-057 (1996), hep-ph/9605286.
\bibitem{isajet}   F. Paige and S. Protopopescu, in {\sl Supercollider
                   Physics}, edited by D. Soper (World Scientific, 1986);
                   H. Baer, F. Paige, S. Protopopescu and X. Tata, in
                   {\sl Proceedings of the Workshop on Physics at 
                   Current Accelerators and Supercolliders}, edited by 
                   J. Hewett, A. White and D. Zeppenfeld 
                   (Argonne National Laboratory 1993).
\bibitem{fritiof}  B. Andersson, G. Gustafson and B. Nilsson-Almquist,
                   Nucl. Phys. {\bf B281}, 289 (1987);
                   B. Nilsson-Almquist and E. Stenlund,
                   Computer Phys. Comm. {\bf 43} (1987), 387;
                   B. L\"orstad,  Int. J. Mod. Phys. {\bf A12}, 2861 (1989).
\bibitem{dpm}      A. Capella, U. Sukhatme, C.-I. Tan, and 
                   J. Tran Tranh Van, Phys. Rep. {\bf 236}, 225 (1994)
\bibitem{venus}    K. Werner,  Z. Phys. {\bf C42}, 85 (1989);
                   Phys. Rep. {\bf 232}, 87 (1993).
\bibitem{rqmd}     H. Sorge, H. St\"ocker, and W. Greiner, 
                   Nucl. Phys. {\bf A498}, 567c (1989); 
                   Ann. Phys. {\bf 192}, 266 (1989).
\bibitem{hijet}    A. Shor and R. Longacre, 
                   Phys. Lett. {\bf B218}, 100 (1989).
\bibitem{hijing}   X. N. Wang and M. Gyulassy, 
                   Phys. Rev. {\bf D44}, 3501 (1991);
                   Comp. Phys. Com. {\bf 83}, 307 (1994).
\bibitem{ms2}      K. Geiger, Phys. Rev. {\bf D46}, 4986 (1992).
\bibitem{bengt88}  H. U. Bengtsson,  Comp. Phys. Com. {\bf 31}, 323 (1984).


\end{thebibliography}
\end{document}